\documentclass[preprint,aps,pre,superscriptaddress]{revtex4-2}

\usepackage[utf8]{inputenc}
\usepackage{epsfig}
\usepackage{subfigure}
\usepackage{latexsym}
\usepackage{amsmath}
\usepackage{amsthm}
\usepackage{psfrag}
\usepackage{fullpage}
\usepackage{graphicx}
\usepackage{amssymb}
\usepackage{color}
\usepackage{placeins}
\usepackage{array}
\usepackage{hyperref}
\usepackage{ulem}

\newcolumntype{x}[1]{>{\centering\hspace{0pt}}p{#1}}

\usepackage{ulem}

\newcommand{\avg}[1]{\left\langle #1\right\rangle}

\newcommand{\T}[1]{\textrm{#1}}

\newcommand{\V}[1]{{\bf #1}}

\theoremstyle{definition}

\theoremstyle{remark}

\begin{document}
\title{Impact of physicality on network structure}

\author{M\'arton P\'osfai}
\email{Contributed equally.}
\affiliation{Department of Network and Data Science, Central European University, Vienna, Austria} 
\author{Bal\'azs Szegedy}
\email{Contributed equally.}
\affiliation{Alfr\'ed R\'enyi Institute of Mathematics, Budapest, Hungary}
\author{Iva Ba\v ci\'c}
\affiliation{Department of Network and Data Science, Central European University, Vienna, Austria} 
\affiliation{Institute of Physics, Belgrade, Serbia}
\author{Luka Blagojevi\'c}
\affiliation{Department of Network and Data Science, Central European University, Vienna, Austria} 
\author{Mikl\'os Ab\'ert}
\affiliation{Alfr\'ed R\'enyi Institute of Mathematics, Budapest, Hungary}
\author{J\'anos Kert\'esz}
\affiliation{Department of Network and Data Science, Central European University, Vienna, Austria} 
\author{L\'aszl\'o Lov\'asz}
\affiliation{Alfr\'ed R\'enyi Institute of Mathematics, Budapest, Hungary}
\author{Albert-L\'aszl\'o Barab\'asi}

\email{barabasi@gmail.com}
\affiliation{Department of Network and Data Science, Central European University, Vienna, Austria} 
\affiliation{Network Science Institute, Northeastern University, Boston, MA, USA}
\affiliation{Department of Medicine, Brigham and Women's Hospital, Harvard Medical School, Boston, MA, USA}

\date{\today}

\begin{abstract}
The emergence of detailed maps of physical networks, like the brain connectome, vascular networks, or composite networks in metamaterials, whose nodes and links are physical entities, have demonstrated the limits of the current network science toolset. Link physicality imposes a non-crossing condition that affects both the evolution and the structure of a network, in a way that the adjacency matrix alone – the starting point of all graph-based approaches – cannot capture. Here, we introduce a meta-graph that helps us discover an exact mapping between linear physical networks and independent sets, a central concept in graph theory. The mapping allows us to analytically derive both the onset of physical effects and the emergence of a jamming transition, and show that physicality impacts the network structure even when the total volume of the links is negligible. Finally, we construct the meta-graphs of several real physical networks, which allows us to predict functional features, such as synapse formation in the brain connectome, that agree with empirical data. Overall, our results show that, in order to understand the evolution and behavior of real complex networks, the role of physicality must be fully quantified.
\end{abstract}

\maketitle
\clearpage

Physical networks, describing molecular and composite networks in metamaterials~\cite{kadic20193d}, the hard-wiring of transistors in a computer chip, the brain connectome~\cite{scheffer2020connectome}, or the vascular system~\cite{banavar1999size,gagnon2015quantifying}, are networks whose nodes and links are physical entities with defined shape and volume, that cannot cross each other~\cite{dehmamy2018structural, liu2021isotopy}.
While network science offers a suite of tools to quantify abstract networks, whose structure is fully defined by their adjacency matrix~\cite{dorogovtsev2003evolution, caldarelli2007scale, cohen2010complex, newman2010networks, barabasi2016network}, these tools are insufficient if we wish to account for the impact of physicality.
Indeed, physical networks differ from the abstract networks in two key aspects. 
First, the nodes and links are embedded in a 3D Euclidean space, hence, similar to spatial networks~\cite{barthelemy2011spatial,horvat2016spatial}, we must specify node positions and link routing.
The second and the most defining feature of physicality is volume exclusion, i.e., the fact that the nodes and the links are not allowed to overlap~\cite{rubinstein2003polymer,song2008phase}. 
While recent experimental advances have provided increasingly accurate maps of physical networks, we lack a formalism to expand the toolset of network science to these systems, and understand how physicality affects the structure and the evolution of physical networks.

Here we unveil the impact of physicality through the discovery of an exact mapping of a physical network into the independent sets of a deterministic meta-graph~\cite{west2001introduction}, allowing us to analytically predict the onset of physicality and the emergence of a jamming transition.
The formalism allows us to construct the meta-graph for real physical networks, and to predict their functional features, like synapse formation in the brain.

\section*{Linear physical network (LPN) model}

To understand the impact of physicality on network structure, we aim to construct the simplest possible model that captures the role of volume exclusion.
For this reason, we eliminated all complexities that affect our ability to explore the role of volume exclusion, like the curving of the links or the impact of the node volume, prompting us to focus on linear physical networks (LPN), whose nodes are zero-volume points in the three-dimensional space and links are straight cylinders with diameter $\lambda$.

To generate a random $\lambda$-physical network $\mathcal G(\lambda,\mathcal P)$, we start from a random finite point set $\mathcal P$ placed in $\mathbb{R}^3$.
The points in $\mathcal P$, that serve as nodes, are placed uniformly randomly within the unit cube with the constraint that any two nodes must be at least at $\lambda$ distance from each other.
To construct the network, we first choose two unconnected nodes in $\mathcal P$ at random and connect them by a capped cylinder of thickness $\lambda$.
If the cylinder does not cross any preexisting link (except those connected to the two end nodes), we add it to the network; if, however, the proposed link overlaps with a previously added link, we delete the proposed link and select another random node pair to connect.

For $\lambda=0$ we lack physical constraints, and any point pair can be connected by a link.
Consequently $\mathcal G(0,\mathcal P)$ maps exactly into the Erd{\H o}s-R\'enyi model and leads to a fully connected network at $M=N(N-1)/2$.
For any $\lambda>0$, however, physicality induces a jammed state, implying that once we reach a maximal number of links $M_\T{max}(\lambda)$, no further links can be added, without violating volume exclusion.
To characterize the jamming process, we measured $\avg{l}$, the average length of the successfully added links, for different $\lambda$ values (Figure~\ref{fig:example-jamming}e). 
For $\lambda=0$ all links are accepted, hence the average length of the observed links is $l_\T{rs}\approx 0.662$, which is the expected length of a randomly selected segment from the unit cube (dashed line in Fig.~\ref{fig:example-jamming}e). 
For $\lambda>0$ the link length $\avg{l}$ deviates from $l_\T{rs}$ for large $M$, indicating that long links often violate physicality, and the larger $\lambda$ is, the earlier physicality manifests itself.

As Fig.~\ref{fig:example-jamming}e indicates, both the onset of physicality ($M_\T{phys}$, that captures the moment at which volume exclusion starts to play a role) and the jammed state ($M_\T{max}$) decrease with increasing link diameter $\lambda$.
Although $M_\T{max}$ and $M_\T{phys}$ are driven by random processes, we find that their value obtained for multiple independent networks generated with the same $(\lambda, \mathcal P)$ parameters is narrowly distributed (Fig.~\ref{fig:example-jamming}f,g),
indicating that $M_\T{phys}$ and $M_\T{max}$ are self-averaging (Supplementary Information Sec.~S3.2).
Our goal, therefore, is to unveil the processes that govern these variables, helping us understand the impact of physicality on the network structure.

\section*{Meta-graph and independent sets}

To uncover the dependence of the onset of physicality ($M_\T{phys})$ and the jamming transition ($M_\T{max}$) on the link thickness $\lambda$ and the number of nodes $N=\lvert\mathcal P\rvert$, we introduce the {\it meta-graph} $\mathcal M(\mathcal P,\lambda)$, designed to capture the physical constraints among the link candidates.
The meta-graph has $N(N-1)/2$ vertices, each corresponding to a possible link $(p_i,p_j)$ between the $N$ nodes.
Two links $(p_1,p_2)$ and $(p_3,p_4)$, corresponding to two vertices of the meta-graph, are connected if they violate physicality, i.e., if the distance between the line segments $(p_1,p_2)$ and $(p_3,p_4)$ is below $\lambda$ (Fig.~\ref{fig:example-jamming}a-d).
  Note that for a given point set $\mathcal P$ and link thickness $\lambda$ the construction of the meta-graph $\mathcal M(\mathcal P,\lambda)$ is fully deterministic.

 The value of the meta-graph stems from the discovery that any linear physical network $\mathcal G(\mathcal P,\lambda)$ corresponds to an independent set of vertices in $\mathcal M(\mathcal P,\lambda)$ and vice versa.
  A set of vertices is called {\it independent} if there are no edges between the elements of the set (Fig.~\ref{fig:Mmax}a,b).
For example, each vertex of the meta-graph of Fig.~\ref{fig:example-jamming}b,d corresponds to a potential link of the physical networks of Fig.~\ref{fig:example-jamming}a,c.
The meta-vertices shown in red on Fig.~\ref{fig:example-jamming}b,d form independent sets, as there are no direct edges between them.
Therefore, each link in the physical network that corresponds to a red meta-vertex can coexist with any other link corresponding to another red meta-vertex, as they do not violate physicality.

 Independent sets are extensively studied in combinatorics~\cite{west2001introduction}, computer science~\cite{tarjan1977finding}, probability theory and statistical physics~\cite{flory1939intramolecular, hartmann2006phase}.
The exact mapping between a $\lambda$-physical network $\mathcal G(\mathcal P, \lambda)$ and the independent vertex sets of the $\mathcal M(\mathcal P,\lambda)$ is our key result that, as we show next, allows us to develop an analytically solvable formalism to explore the structure and the evolution of physical networks.

\section*{Predicting the evolution of physical networks}

We rely on the mapping between $\lambda$-physical networks and the independent sets of meta-graphs to derive  $M_\T{phys}$ and $M_\T{max}$, and understand the role of physicality.
We must account for two limits as we proceed: 
 (i)~With fixed $\lambda$ we cannot take the large network limit ($N\rightarrow \infty$) because the total volume of the links, whose lower-bound scales as $N \lambda^3$ for networks with non-vanishing average degree, exceeds the available volume for large $N$, resulting in a disconnected network.
We therefore must decrease $\lambda$ as we increase $N$ to ensure that $\lambda\lesssim N^{-1/3}$.
(ii)~If $\lambda$ decreases too fast with $N$, the average meta-degree $\avg{k_\T{meta}}~\sim\lambda N^2$ (Supplementary Information Sec.~S1.4) converges to zero and physicality will stop playing a role, implying that $\lambda\gtrsim N^{-2}$.
To satisfy~(i) and (ii), we set
\begin{equation}
\lambda =\frac{C}{N^\alpha},
\end{equation}  
where $C$ is an arbitrary constant and the control parameter $\alpha$  interpolates between the crowded state (i, $\alpha=1/3$) and the loss of physicality (ii, $\alpha=2$).

We begin by observing that the construction of a random LPN is equivalent to building a greedy independent set $\mathcal I$ of $\mathcal M(\mathcal P, \lambda)$ by sequentially selecting the meta-vertices in random order and adding  the $t$th meta-vertex to $\mathcal I$ if none of its neighbors are in $\mathcal I$.
To analytically characterize this process, we introduce a randomized reference meta-graph $\mathcal M_\T{rr}(\mathcal P, \lambda)$ in which two link candidates with length $l_1$ and $l_2$ are connected independently with probability $(\pi/2)\lambda l_{1}l_{2}$, representing the approximate probability that the distance between two randomly selected segments of lengths $l_1$ and $l_2$ is at most $\lambda$.
This construction provides a first order approximation of $\mathcal M(\mathcal P, \lambda)$ in $\lambda$: the expected degree of a vertex with length $l$ is $\sim \lambda l$ in both $\mathcal M(\mathcal P, \lambda)$ and $\mathcal M_\T{rr}(\mathcal P, \lambda)$ in leading order, while higher-order structures, such as the number of triangles a vertex participates in, are not captured by $\mathcal M_\T{rr}(\mathcal P, \lambda)$.
We then leverage $\mathcal M_\T{rr}(\mathcal P, \lambda)$ to derive a differential equation that governs the temporal evolution of the total length $L_\T{total}(t)$ of links in $\mathcal I$~\cite{brightwell2017greedy}
(Supplementary Information Sec.~S3)
\begin{equation}\label{eq:Ltot_dyn}
\dot L_\T{total}(\tau) = \frac{N^2}{2}\int^{\sqrt{3}}_0 l\exp\left[-\frac{\pi}{2}\lambda L_\T{total}(\tau)l\right]p_\T{LC}(l)dl,
\end{equation}
where $\tau=2t/N(N-1)$ is the rescaled time, and $p_\T{LC}(l)$ is the length distribution of the link candidates. 
The expression $\exp\left[-\frac{\pi}{2}\lambda L_\T{total}(\tau)l\right]$ is the probability that a meta-vertex with length $l$ has no connections leading to the independent set $\mathcal I$ at $\tau$.
This means that the acceptance of link candidates decays exponentially with $l$, suggesting that $p(l)$, the distribution of the length of the accepted links, also decays exponentially,
in line with results on brain architecture finding an exponential law~\cite{ercsey2013predictive,bassett2017small}.
The total link length $L_\T{total}(\tau)$, however, increases over time, meaning that longer links are added earlier during the evolution of the network than in later stages.
In fact, we predict that this source of heterogeneity leads to a power law link length distribution $p(l)\sim l^{-3}$ with an upper cutoff dictated by the finite size of the system (Supplementary Information Sec.~S4).

The onset of physicality happens when link candidates get rejected with positive probability, i.e., $\lambda L_\T{total}(\tau)\rightarrow \T{const.}$ for $N\rightarrow\infty$, predicting
\begin{equation}\label{eq:Mphys_assym}
M_\T{phys}\sim N^\alpha,
\end{equation}
unveiling that the control parameter $\alpha$ directly governs the onset of physicality $M_\T{phys}$.
Equation (\ref{eq:Ltot_dyn}) also predicts that the jamming point scales as 
\begin{equation}\label{eq:Mmax_assym}
M_\T{max}\sim N^{\frac{3\alpha+4}{5}},
\end{equation}
 and the total link length in the jammed state scales as
\begin{equation}\label{eq:Ltot_assym}
L_\T{total}\sim N^{\frac{4\alpha+2}{5}}.
\end{equation}

Equation~\eqref{eq:Ltot_dyn} also allows us to study the fluctuations of these quantities,
yielding that their variances scale as $\sigma^2(M_\T{phys})\sim N^\alpha$, $\sigma^2_2\left(M_\T{max}\right)\sim N^{\frac{3\alpha+4}{5}}$ and $\sigma^2\left(L_\T{total}\right)\sim N^\alpha\ln N$.
Consequently, in the large network limit the typical fluctuations converge to zero compared to their expectation; indicating that Eqs.~(\ref{eq:Mphys_assym})-(\ref{eq:Ltot_assym}) correctly capture the typical behavior of LPNs.

Our analytical predictions are directly testable by simulations.
We begin by numerically solving Eq.~(\ref{eq:Ltot_dyn}) to determine the dependence of $M_\T{max}$ on $\lambda$ (Fig.~\ref{fig:Mmax}c), finding excellent agreement, particularly for small $\lambda$.
We next tested the predicted scaling behavior (\ref{eq:Mphys_assym})-(\ref{eq:Ltot_assym}) by constructing LPNs of increasing sizes, finding that they offer an accurate description of the onset of physicality~(Figs.~\ref{fig:Mmax}c-e).
 The predictive accuracy of Eq.~(\ref{eq:Ltot_dyn}) indicates that the likelihood of adding a physical link to the network is driven primarily by the length of the link, hence higher order effects, like the formation of triangles, that are ignored by $\mathcal M_\T{rr}(\mathcal P, \lambda)$, play a negligible role.

Figure~\ref{fig:Mmax}g summarizes the behavior of physical networks as predicted by Eqs.~(\ref{eq:Ltot_dyn}-\ref{eq:Ltot_assym}), documenting the vanishing role of physicality with increasing $\alpha$: 

For $\alpha<1/3$ the link widths are larger than the typical distance between adjacent nodes, hence the network remains disconnected.
The first realizable network emerges for $\alpha=1/3$, in which case Eq.~(\ref{eq:Mphys_assym}) predicts $M_\T{phys}\sim N^{1/3}$, meaning that physicality plays a role even when the network is ultra sparse ($\avg{k}=2M_\T{phys}/N\sim N^{-2/3}\rightarrow 0$).
For $\alpha=1/3$, we have $M_\T{max}\sim N$, indicating that the jammed network is also sparse ($\avg{k}=O(1)$).
The link length in the jammed network $l^*\sim L_\T{total}/M_\T{max}$ is of the order of the distance between physically adjacent nodes $\sim N^{-1/3}$.

Between $1/3\leq\alpha\leq 2$, we are in the physical regime, with two sub-regimes:
In the sub-linear regime ($1/3<\alpha\leq 1$), physicality plays a role even in sparse networks ($\avg{k}=O(1)$).
In contrast, in the super-linear regime ($1<\alpha\leq 2$) sparse LPNs are not affected by physicality, hence we need a super-linear number of links before physicality affects network formation. 

Finally, for $\alpha=2$, the onset of physicality scales as $M_\T{phys}\sim N^2$, and $M_\T{max}\sim N^2$, meaning that physical interactions are only important in dense networks.
 The average physical link length $\avg l$ in this regime is of the order of the system size, indicating the links can span the entire system, a consequence of the vanishing role of physical effects.

\section*{The adjacency matrix in the jammed state is predictive of node positions}\label{sec:jammed_state}

A key prediction of our formalism is that physicality impacts the structure of networks even with vanishing volume.
Indeed, according to (\ref{eq:Ltot_assym}), the network volume scales as $V\sim\lambda^2 L_\T{total}\sim N^{-\frac{6\alpha-2}{5}}$, hence in the $N\rightarrow \infty$ limit for any $\alpha>1/3$ the jammed network occupies a zero fraction of the available space.

As the number of links increases, so does the number of physical constraints that each new link must satisfy.
Hence our ability to place a new link becomes increasingly dependent on the existing links, which in turn leads to correlations between the adjacency matrix and the physical layout of the network. 
Indeed, before physicality turns on ($M<M_\T{phys}$), the distribution of eigenvalues follows Wigner's semicircle law (Fig.~\ref{fig:eigvec}a)\cite{van2010graph}.
When, however, the number of links $M$ approaches the jammed state $M_\T{max}$, three additional eigenvalues $\mu_2$, $\mu_3$ and $\mu_4$ separate from the bulk (Fig.~\ref{fig:eigvec}b).
In non-physical networks, such eigenvalues often indicate the presence of a large-scale organization in the network, for example, leading eigenvectors may correspond to a macroscopic community structure~\cite{abbe2017community}.
In contrast, in physical networks the leading eigenvectors are induced by physicality, signaling a strong spatial dependence: the typical link length $l^*\sim L_\T{total}/M_\T{max}\sim N^{-\frac{2-\alpha}{5}}$ decreases with increasing $N$ and nodes are densely connected in their $l^*$ neighborhood, while the majority of long-range links are suppressed.
The resulting large-scale three-dimensional spatial organization is captured by the eigenvalues $\mu_2$, $\mu_3$ and $\mu_4$ and the corresponding eigenvectors $\V v^{(2)}$, $\V v^{(3)}$ and $\V v^{(4)}$:
the eigenvectors at node $i$ become strongly correlated with the spatial coordinates of $i$.
In other words, as we approach the jammed state, the adjacency matrix becomes predictive of the position of individual nodes.
To quantify the predictive power of the leading eigenvectors, we rotate $\V v^{(2)}$, $\V v^{(3)}$ and $\V v^{(4)}$ to best align with the node coordinates and calculate the coefficient of determination, i.e., we measure how well $(v^{(2)}_i,v^{(3)}_i,v^{(4)}_i)$, the eigenvectors at node $i$, predict its coordinates $(x_i,y_i,z_i)$ (Supplementary Information~Sec.~S5.3).
We find that the coefficient of determination increases rapidly as we approach the jammed state, indicating that the node positions predicted by the adjacency matrix converge towards their true values (Fig.~\ref{fig:eigvec}c,d).
In other words, while a complete description of physical networks requires simultaneous information on the adjacency matrix, link routing and node layout, we find that unexpectedly in the jammed state, where physicality is the strongest, these features become so intertwined, that the adjacency matrix alone offers a complete description of the system.
Note that having reduced space around nodes does not by itself imply predictability: for example, the adjacency matrix of a densely packed tree does not carry any information about the physical location of the nodes, as such networks can be folded into a volume in many different ways, changing the layout but preserving the abstract network.

\section*{The meta-graph of real networks}

While the LPN model conceptualizes physical networks as nodes connected by straight links, in real physical networks, like the brain connectome or the vascular network, the links curve.
As we show next, the meta-graph offers a quantitative framework to characterize the impact of physicality for networks with  arbitrary link shapes and structure.

In their native state, neurons in the brain or the vessels of a vascular network obey volume exclusion.
If, however, we increase the thickness $\lambda$ of all links by a $\Delta \lambda$, conflicts can emerge. 
We therefore defined a generalized meta-graph $\mathcal M_\T g(\Delta \lambda)$, in which we connect two vertices of the meta-graph if the corresponding physical objects (links or nodes) overlap for a given $\Delta \lambda$, concisely capturing the spatial organization of a physical network (Supplementary Information~Sec.~S7).

We constructed $\mathcal M_\T g(\Delta \lambda)$ for multiple real physical networks, including the fruit fly's brain~\cite{scheffer2020connectome}, the vascular network of a mouse~\cite{gagnon2015quantifying} and a mitochondrial network~\cite{viana2020mitochondrial}.
We illustrate the process in Fig.~\ref{fig:realmeta_vis}a, which shows the meta-graph of the fruit fly connectome, consisting of $N=2,970$ neurons and $M=35,707$ synapses serving as links.

According to Peter's rule, neurons can only form synapses if their axons and dendrites are in close physical proximity~\cite{rees2017weighing, udvary2022impact}. 
Hence we expect and find a strong correlation between the meta-degree and the number of synapses (Supplementary Information Fig.~S21).
To abstract from these obvious correlations between the generalized meta-graph and synapse formation, we focus only on conflicts between neurons that are not connected by synapses and therefore are the result of the packing of the neurons in the brain.
We achieve this by building a restricted meta-graph, where we remove the synaptically connected links from the meta-graph. 
Figure~\ref{fig:realmeta_vis}a highlights the vertex with the highest restricted meta-graph degree $k_\T A=13$, corresponding to the most physically confined neuron, bordered by 13 other neurons that it does not synapse with (Fig.~\ref{fig:realmeta_vis}b,c).
This prompts the question: Is the most confined neuron also the most central in the synaptic network?
To find an answer, we performed a linear regression between the restricted meta-degree and the logarithm of synapses, revealing a positive association between the physical confinement and the functional role of neurons (Fig.~\ref{fig:realmeta_vis}d, slope $a=0.356\pm 0.022$ and $R^2=0.26$). 
Our result indicates that synaptically central neurons in the connectome are tightly confined in the brain by non-synaptic partners.
This is non-obvious, as we can construct physical networks that have negative correlations between the number of synapses and the restricted meta-degree: consider a physical network with $N$ nodes where each neuron is physically adjacent to all $N-1$ other neurons (Supplementary Information~Fig.~S23).
If a neuron $i$ has $k$ synaptic partners, it has $N-1-k$ meta-degree, resulting in a perfect anticorrelation between the number of synapses and the meta-degree.
Overall, the observed correlations confirm that the meta-graph captures important properties of the physical layout and can be used to systematically study the connection between physical and abstract network structure.

As connectome mapping aspires to scale up to the $10^9$ neurons of the human brain, new mathematical and computational formalisms, like the one offered by the meta-graph, are needed to unveil the predictive power of these exceptionally large physical network maps.
Full description of the layout of a physical network requires copious amounts of data that is difficult to handle computationally and also limits analytical advances.
For example the Hemibrain dataset describes the 3D trajectory of approximately 25,000 neurons of a fruit fly using 117 million linear segments~\cite{scheffer2020connectome}.
Na\"ive identification of physical conflicts, therefore, requires $10^{16}$ distance computations, a prohibitive computational burden for most researchers.
In contrast, the $25,000\times 25,000$ adjacency matrix of the generalized meta-graph can be represented using a few hundred MB of data,
 hence publishing it together with the adjacency of the connectome would allow the computationally efficient study of the relation between physical and abstract network structure.

\section*{Discussion}

Recent experimental advances, driven by connectomics and high resolution MRI, have offered detailed and accurate maps of a wide range of physical networks, from the structure of individual neurons in a brain, to 3D maps of large vascular systems.
These advances unveiled an important gap in network science: the lack of understanding of how physicality affects the network structure.
The need for a quantitative and conceptual framework go beyond biology: complex metamaterials, combining random and repetitive local structures~\cite{nicolaou2012mechanical, kadic20193d, kim2022nonlinear}, offer other manifestations of physical networks, and so do computer chips that pack billions of transistors.
Here we introduced a formalism designed to systematically explore the structure of physical networks.
We show that the impact of physicality is not limited to dense networks -- to the contrary, in their jammed state physical networks are sparse, with the relative volume of their links converging to zero for large systems.
In other words, physicality is not a simple manifestation of crowding, but has subtle and nontrivial consequences on the network structure.
The advances presented here raise multiple open questions, many of which can be addressed using meta-graphs.
For example, many real networks are characterized by non-uniform node density, heterogeneous link diameters, bent links, and are potentially affected by the order the nodes and the links are added to the network.
The impact of these features can be studied by extending the random LPN model or using the generalized meta-graphs.
These variants of LPNs may serve as null models to understand the features of the physical layout and network structure of real systems (Supplementary Information Sec.~S7.6).  
Other issues are less straightforward extensions of our work, but may benefit from the meta-graph framework, such as understanding the effect of the physical architectures on network robustness~\cite{dsouza2019explosive,heroy2022rigidity} or on dynamics on networks~\cite{barrat2008dynamical, ulloa2016role, bianconi2018multilayer,case2019braess, kim2019conformational}.

A quantitative understanding of physicality can directly impact multiple areas of science.
For example, at this point it is unclear to what degree the observed brain connectomes are driven by the genetic processes that govern their developmental biology~\cite{kovacs2020uncovering}, or by physical constraints that the neurons and their interactions must obey, limiting a neuron's ability to synapse with desired target neurons in a very dense environment.
Answers require a modeling and analytical platform that helps us systematically explore the competing role of genetics and physicality.

\subsection*{Acknowledgments}

This research was funded by ERC grant No.~810115-DYNASNET.

\subsection*{Author Contributions Statement}

M.P. developed and performed the simulations. M.P. and B.Sz. derived the analytical results. M.P. and L.B. analyzed the empirical data. All authors contributed to the conceptual design of the study and writing the manuscript. A.-L.B. was the lead writer of the manuscript.

\subsection*{Competing Interests Statement}

A.-L.B. is the founder of Foodome and ScipherMedicine companies that explore the role of networks in health and urban environments.
The other authors declare no competing interests.

\subsection*{Data and Code Availability}

Code and data to generate random LPNs and reproduce the figures are available at {\it https://github.com/posfaim/randLPN}. 

\clearpage

\begin{figure*}[ht]
	\centering
	\includegraphics[width=1.\textwidth]{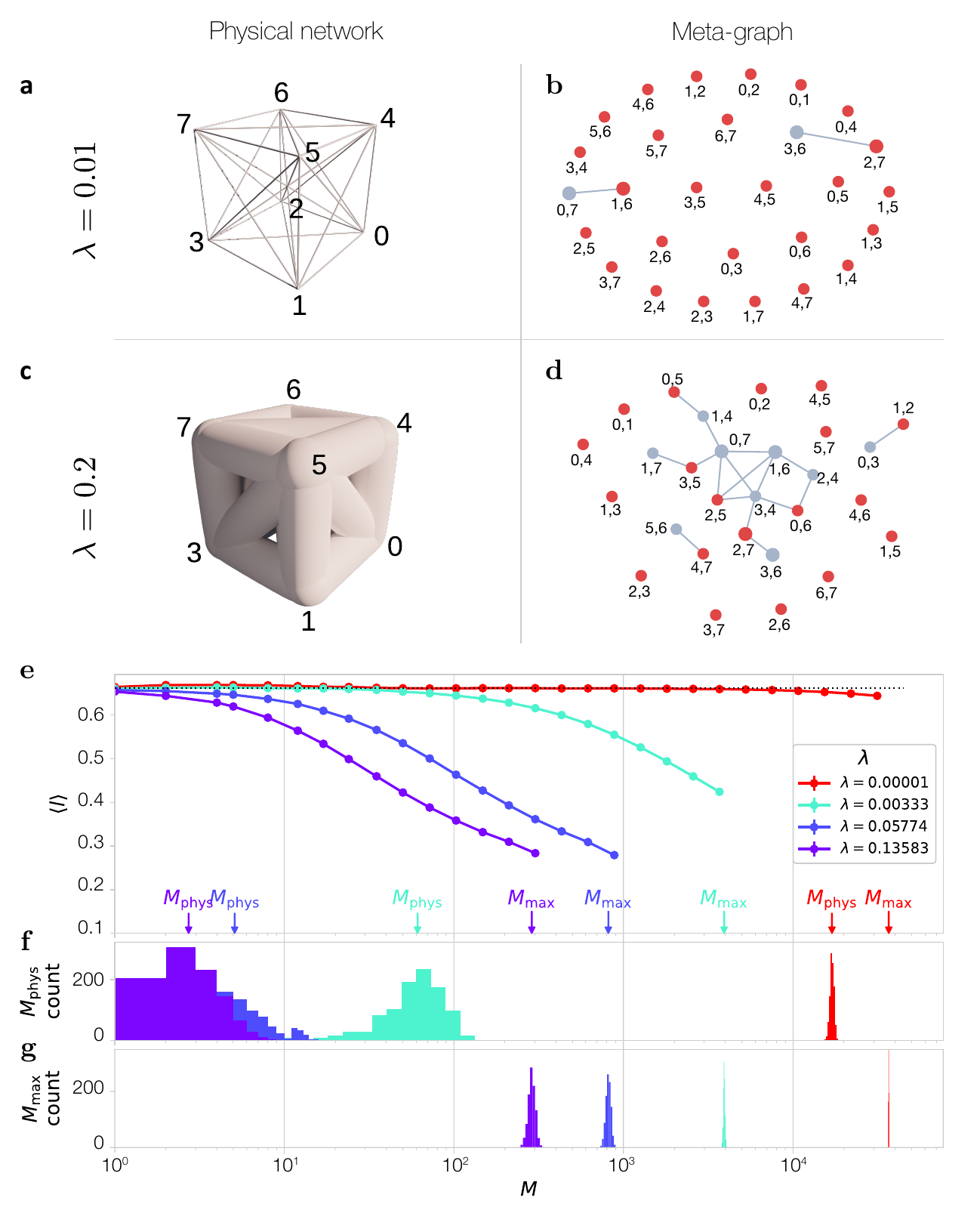}
\end{figure*}

\begin{figure}[ht]
	\centering
	\caption{{\bf Linear Physical Networks (LPN)} {\bf (a,c)}~A linear physical network (LPN) with eight nodes, showing its structure for two different $\lambda$ values.
While for the small $\lambda$ in (a) most links are allowed, for $\lambda=0.2$ in (b) many links are forbidden, as they would overlap with other links.
{\bf(b,d)}~The 28 vertices of the meta-graph represent the candidate links of the physical network, each labeled by the node numbers they attempt to connect.
Two vertices are connected if the corresponding links overlap, hence they cannot coexist in a physical network.
Each independent vertex set of these meta-graphs corresponds to a valid physical network: the independent set formed by the red nodes represent the physical network shown in (a,c).
{\bf (e-g)}~To model the evolution of a LPN, we generate a point set $\mathcal P$ with $N=300$ nodes, randomly adding links if they do not violate $\lambda$-physicality, repeating the process 1000 times for the same $\mathcal P$.
{\bf (e)}~The observed length of the links after the addition of $M$ links.
The data points are logarithmically spaced, and the dashed line corresponds to $l_\T{rs}\approx 0.662$, the expected length of a random segment chosen from the unit cube (expected for $\lambda=0$).
The higher the $\lambda$, the more conflicts links have, hence the more the observed $l$ deviates from $l_\T{rs}$.
Error bars representing the SEM are smaller than the marker size.
{\bf (f,g)}~Histogram of $M_\T{phys}$ and $M_\T{max}$ for different realizations, showing that $M_\T{phys}$ and $M_\T{max}$ are concentrated on a narrow range, being largely independent of the order the links are added. Due to the logarithmic scale the histograms for low $M_\T{phys}$ and $M_\T{max}$ appear to be wider. In simulations, we measure $M_\T{phys}$ as the number of links above which at least 1/6th of the link candidates are rejected. (Supplementary Information Sec.~S1.5.5)}
	\label{fig:example-jamming}
\end{figure}

\begin{figure*}[ht]
	\centering
	\includegraphics[width=1.\textwidth]{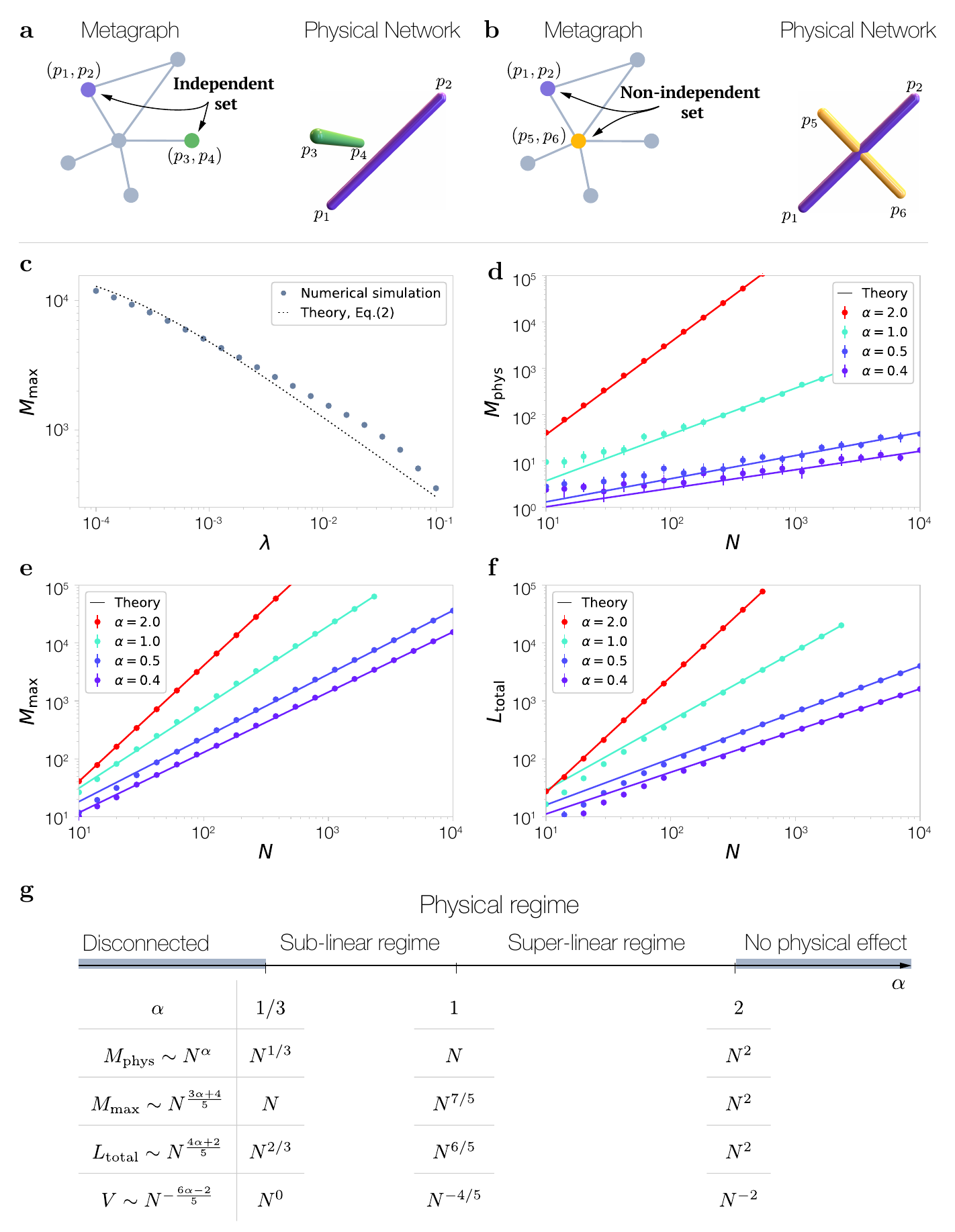}
\end{figure*}

\begin{figure}[ht]
	\centering
	\caption{{ \bf Meta-graph and Independent Sets.} 
{\bf (a)}~The purple and green vertices of the meta-graph have no direct edges between them, forming an independent set, indicating that the corresponding physical links $(p_1,p_2)$ and $(p_3,p_4)$ are non-overlapping (conflict free).
{\bf (b)}~The purple and yellow vertices do not form an independent set, as they are connected by a direct edge, indicating that the physical links corresponding to them overlap, hence they cannot be added simultaneously.
{\bf (c-g)}~To test the analytical predictions provided by the meta-graph, we simulated LPNs and numerically measured the number of links at the onset of physicality ($M_\T{phys}$), the maximal number of links ($M_\T{max}$), and their length ($L_\T{tot}$) in the jammed state.
{\bf (c)}~Comparing the prediction of Eq.~(\ref{eq:Ltot_dyn}) with the numerical estimate of $M_\T{max}$ from simulations of LPNs with $N=200$.
Markers represent the average of 50 independent networks, the error bars representing the SEM are smaller than the marker size.
{\bf (d-f)}~We generated LPNs with increasing $N$ and link thickness scaling as $\lambda=C/N^\alpha$.
The symbols indicate the numerical results and the slope of the continuous lines correspond to the scaling exponents predicted by Eqs.~(\ref{eq:Mphys_assym})-(\ref{eq:Ltot_assym}).
The data points represent an average of 10 independent runs, and the error bars representing the SEM are typically smaller than the marker size.
{\bf (g)}~The behavior of physical networks in function of $\alpha$.
For $\alpha<1/3$ the links are wider than the typical distance between physically adjacent nodes, leading to disconnected physical networks with zero average degree in the $N\rightarrow \infty$ limit.
In contrast, for $\alpha>2$ the role of physicality vanishes.
Between these two limits, physicality effects the formation of networks with more than $M_\T{phys}\sim N^\alpha$ links.
For $1/3\leq\alpha\leq 1$ physicality turns on after the addition of a sub-linear number of links and therefore even sparse networks ($M\sim N$) are affected by physicality.
In contrast, for $1\leq\alpha\leq 2$ volume exclusion has an effect only after the addition of a super-linear number of links, hence only dense networks are affected by physicality.
Overall, the role of physicality weakens for increasing $\alpha$.}
	\label{fig:Mmax}
\end{figure}

\begin{figure}[ht]
	\centering
	\includegraphics[width=1.\textwidth]{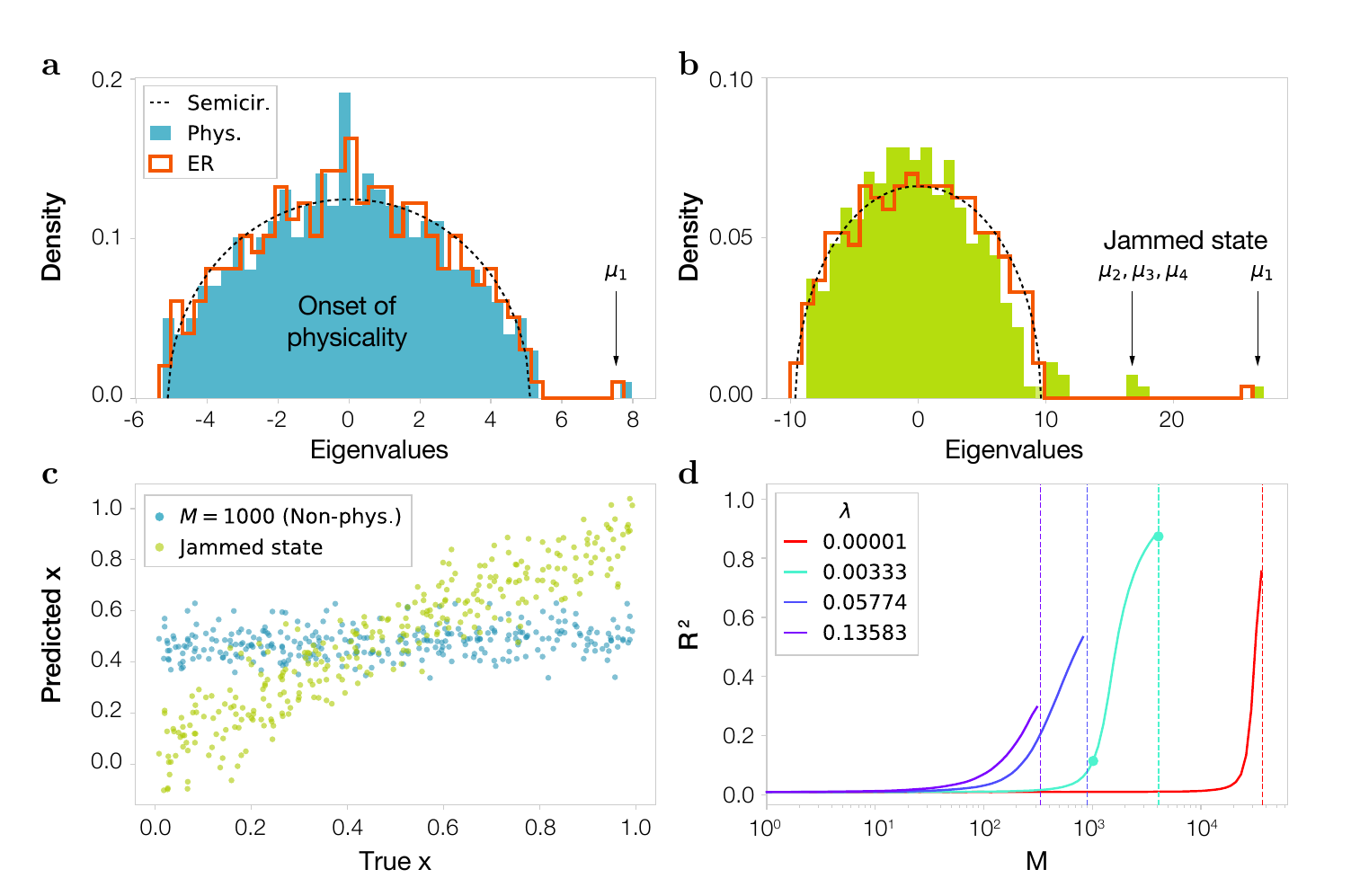}
\end{figure}

\begin{figure}[ht]
	\centering
	\caption{{ \bf Predicting Node Positions in the Jammed State.}
	{\bf (a)}~The spectral density of the adjacency matrix for $N=300$ and $\lambda=1/N$ after the addition of $M=1000$ links (blue).
Before the onset of physicality the eigenvalues of the LPN are consistent with an ER network of the same size (red outline): the bulk of the spectral density is well approximated by Wigner's semicircle law (dashed line), and the largest eigenvalue separated from the bulk ($\mu_1$) corresponds to the average degree of the network.
{\bf (b)}~In the jammed state, the eigenspectrum of the LPN (green) deviates from the spectrum of an ER network (red outline): three additional eigenvalues $\mu_2$, $\mu_3$ and $\mu_4$ become separated from the bulk as a consequence of physical interactions.
{\bf (c)}~Comparing the predicted and the true $x$ coordinate of each node shows that while for small link density the adjacency matrix has no predictive power (blue symbols), the adjacency matrix can reliably predict the position of nodes in the jammed state (green symbols).
{\bf (d)}~The coefficient of determination $R^2$ increases as we add links to LPNs, indicating that as we approach the jammed state the predictive power of the adjacency matrix increases. The circles highlight the $R^2$ values corresponding to the $\lambda$ shown in (a) and (b).
Subplots (a-c) show results for a single LPN, while (d) shows the average of 1000 independent networks.
}
	\label{fig:eigvec}
\end{figure}

\begin{figure*}[ht]
	\centering
	\includegraphics[width=1.\textwidth]{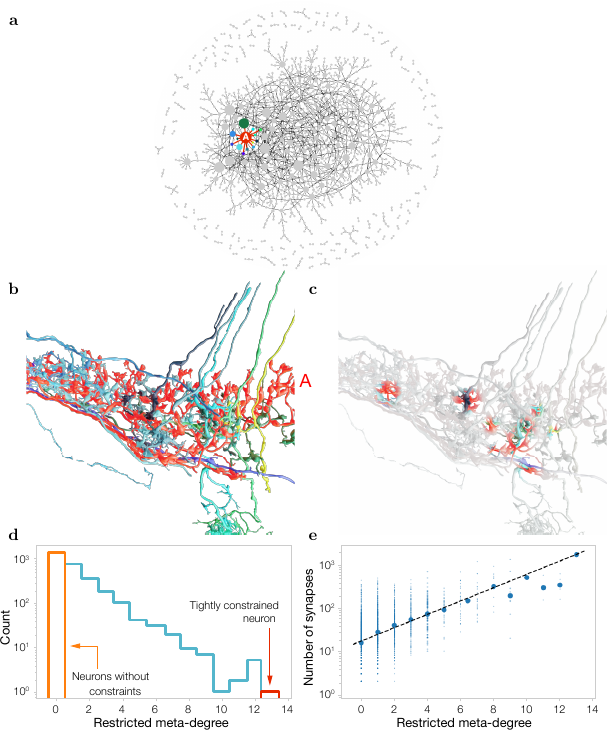}
\end{figure*}

\begin{figure}[ht]
	\centering

	\caption{{ \bf Meta-graph of Real Networks.}
{\bf (a)}~Each vertex of the restricted meta-graph represents a neuron in the fruit fly connectome~\cite{scheffer2020connectome}.
A link between two vertices of the restricted meta-graph implies that the corresponding neurons overlap if we increase their thickness by $\Delta \lambda\approx 0.028$ but they are not connected by synapses.
The neuron with the highest restricted meta-degree, A, has 13 connections, while  $1469$ isolated vertices (not shown) correspond to neurons that are conflict-free for $\Delta \lambda\approx 0.028$.
{\bf (b)}~Neuron A (red) has the most excess confinement. We show it together with the 13 neurons that are within distance $\Delta \lambda$ of A, hence are connected to A in the restricted meta-graph and highlighted in (a).
The neuron colors match the colors of the meta-vertices in (a).
{\bf (c)}~Neuron A is an extended, physical object, whose physical conflicts with other neurons are localized in specific neighborhoods of the physical network, highlighted in the figure.
{\bf (d)}~The degree distribution of the restricted meta-graph for $\Delta \lambda\approx 0.028$.
Vertices with degree zero correspond to conflict-free neurons, i.e., lack proximity within $\Delta \lambda$ with other neurons that are not connected to them via synapses.
Physically confined neurons have high restricted meta-degree, indicative of a large number of physical conflicts.
{\bf (e)}~The dependence of the restricted meta-degree on the number of synapses of each neuron indicates that the restricted meta-degree is predictive of synapse formation. The dashed line corresponds to linear regression between the restricted meta-degree and the logarithm of the number of synapses for each neuron.
Small markers represent individual neurons, large markers are binned averages.
For illustration purposes we chose $\Delta \lambda$ such that the meta-graph is sparse.
In Sec.~S7 of the Supplementary Information we repeat the above analysis for various $\Delta \lambda$ values, finding that the positive association between the restricted meta-graph and the synaptic network is robust.}
	\label{fig:realmeta_vis}
\end{figure}


\clearpage

\begin{thebibliography}{37}%
\makeatletter
\providecommand \@ifxundefined [1]{%
 \@ifx{#1\undefined}
}%
\providecommand \@ifnum [1]{%
 \ifnum #1\expandafter \@firstoftwo
 \else \expandafter \@secondoftwo
 \fi
}%
\providecommand \@ifx [1]{%
 \ifx #1\expandafter \@firstoftwo
 \else \expandafter \@secondoftwo
 \fi
}%
\providecommand \natexlab [1]{#1}%
\providecommand \enquote  [1]{``#1''}%
\providecommand \bibnamefont  [1]{#1}%
\providecommand \bibfnamefont [1]{#1}%
\providecommand \citenamefont [1]{#1}%
\providecommand \href@noop [0]{\@secondoftwo}%
\providecommand \href [0]{\begingroup \@sanitize@url \@href}%
\providecommand \@href[1]{\@@startlink{#1}\@@href}%
\providecommand \@@href[1]{\endgroup#1\@@endlink}%
\providecommand \@sanitize@url [0]{\catcode `\\12\catcode `\$12\catcode
  `\&12\catcode `\#12\catcode `\^12\catcode `\_12\catcode `\%12\relax}%
\providecommand \@@startlink[1]{}%
\providecommand \@@endlink[0]{}%
\providecommand \url  [0]{\begingroup\@sanitize@url \@url }%
\providecommand \@url [1]{\endgroup\@href {#1}{\urlprefix }}%
\providecommand \urlprefix  [0]{URL }%
\providecommand \Eprint [0]{\href }%
\providecommand \doibase [0]{https://doi.org/}%
\providecommand \selectlanguage [0]{\@gobble}%
\providecommand \bibinfo  [0]{\@secondoftwo}%
\providecommand \bibfield  [0]{\@secondoftwo}%
\providecommand \translation [1]{[#1]}%
\providecommand \BibitemOpen [0]{}%
\providecommand \bibitemStop [0]{}%
\providecommand \bibitemNoStop [0]{.\EOS\space}%
\providecommand \EOS [0]{\spacefactor3000\relax}%
\providecommand \BibitemShut  [1]{\csname bibitem#1\endcsname}%
\let\auto@bib@innerbib\@empty
\bibitem [{\citenamefont {Kadic}\ \emph {et~al.}(2019)\citenamefont {Kadic},
  \citenamefont {Milton}, \citenamefont {van Hecke},\ and\ \citenamefont
  {Wegener}}]{kadic20193d}%
  \BibitemOpen
  \bibfield  {author} {\bibinfo {author} {\bibfnamefont {M.}~\bibnamefont
  {Kadic}}, \bibinfo {author} {\bibfnamefont {G.~W.}\ \bibnamefont {Milton}},
  \bibinfo {author} {\bibfnamefont {M.}~\bibnamefont {van Hecke}},\ and\
  \bibinfo {author} {\bibfnamefont {M.}~\bibnamefont {Wegener}},\ }\bibfield
  {title} {\bibinfo {title} {3d metamaterials},\ }\href@noop {} {\bibfield
  {journal} {\bibinfo  {journal} {Nature Reviews Physics}\ }\textbf {\bibinfo
  {volume} {1}},\ \bibinfo {pages} {198} (\bibinfo {year} {2019})}\BibitemShut
  {NoStop}%
\bibitem [{\citenamefont {Scheffer}\ \emph {et~al.}(2020)\citenamefont
  {Scheffer}, \citenamefont {Xu}, \citenamefont {Januszewski}, \citenamefont
  {Lu}, \citenamefont {Takemura}, \citenamefont {Hayworth}, \citenamefont
  {Huang}, \citenamefont {Shinomiya}, \citenamefont {Maitlin-Shepard},
  \citenamefont {Berg} \emph {et~al.}}]{scheffer2020connectome}%
  \BibitemOpen
  \bibfield  {author} {\bibinfo {author} {\bibfnamefont {L.~K.}\ \bibnamefont
  {Scheffer}}, \bibinfo {author} {\bibfnamefont {C.~S.}\ \bibnamefont {Xu}},
  \bibinfo {author} {\bibfnamefont {M.}~\bibnamefont {Januszewski}}, \bibinfo
  {author} {\bibfnamefont {Z.}~\bibnamefont {Lu}}, \bibinfo {author}
  {\bibfnamefont {S.-y.}\ \bibnamefont {Takemura}}, \bibinfo {author}
  {\bibfnamefont {K.~J.}\ \bibnamefont {Hayworth}}, \bibinfo {author}
  {\bibfnamefont {G.~B.}\ \bibnamefont {Huang}}, \bibinfo {author}
  {\bibfnamefont {K.}~\bibnamefont {Shinomiya}}, \bibinfo {author}
  {\bibfnamefont {J.}~\bibnamefont {Maitlin-Shepard}}, \bibinfo {author}
  {\bibfnamefont {S.}~\bibnamefont {Berg}}, \emph {et~al.},\ }\bibfield
  {title} {\bibinfo {title} {A connectome and analysis of the adult drosophila
  central brain},\ }\href@noop {} {\bibfield  {journal} {\bibinfo  {journal}
  {Elife}\ }\textbf {\bibinfo {volume} {9}},\ \bibinfo {pages} {e57443}
  (\bibinfo {year} {2020})}\BibitemShut {NoStop}%
\bibitem [{\citenamefont {Banavar}\ \emph {et~al.}(1999)\citenamefont
  {Banavar}, \citenamefont {Maritan},\ and\ \citenamefont
  {Rinaldo}}]{banavar1999size}%
  \BibitemOpen
  \bibfield  {author} {\bibinfo {author} {\bibfnamefont {J.~R.}\ \bibnamefont
  {Banavar}}, \bibinfo {author} {\bibfnamefont {A.}~\bibnamefont {Maritan}},\
  and\ \bibinfo {author} {\bibfnamefont {A.}~\bibnamefont {Rinaldo}},\
  }\bibfield  {title} {\bibinfo {title} {Size and form in efficient
  transportation networks},\ }\href@noop {} {\bibfield  {journal} {\bibinfo
  {journal} {Nature}\ }\textbf {\bibinfo {volume} {399}},\ \bibinfo {pages}
  {130} (\bibinfo {year} {1999})}\BibitemShut {NoStop}%
\bibitem [{\citenamefont {Gagnon}\ \emph {et~al.}(2015)\citenamefont {Gagnon},
  \citenamefont {Sakad{\v{z}}i{\'c}}, \citenamefont {Lesage}, \citenamefont
  {Musacchia}, \citenamefont {Lefebvre}, \citenamefont {Fang}, \citenamefont
  {Y{\"u}cel}, \citenamefont {Evans}, \citenamefont {Mandeville}, \citenamefont
  {Cohen-Adad} \emph {et~al.}}]{gagnon2015quantifying}%
  \BibitemOpen
  \bibfield  {author} {\bibinfo {author} {\bibfnamefont {L.}~\bibnamefont
  {Gagnon}}, \bibinfo {author} {\bibfnamefont {S.}~\bibnamefont
  {Sakad{\v{z}}i{\'c}}}, \bibinfo {author} {\bibfnamefont {F.}~\bibnamefont
  {Lesage}}, \bibinfo {author} {\bibfnamefont {J.~J.}\ \bibnamefont
  {Musacchia}}, \bibinfo {author} {\bibfnamefont {J.}~\bibnamefont {Lefebvre}},
  \bibinfo {author} {\bibfnamefont {Q.}~\bibnamefont {Fang}}, \bibinfo {author}
  {\bibfnamefont {M.~A.}\ \bibnamefont {Y{\"u}cel}}, \bibinfo {author}
  {\bibfnamefont {K.~C.}\ \bibnamefont {Evans}}, \bibinfo {author}
  {\bibfnamefont {E.~T.}\ \bibnamefont {Mandeville}}, \bibinfo {author}
  {\bibfnamefont {J.}~\bibnamefont {Cohen-Adad}}, \emph {et~al.},\ }\bibfield
  {title} {\bibinfo {title} {Quantifying the microvascular origin of bold-fmri
  from first principles with two-photon microscopy and an oxygen-sensitive
  nanoprobe},\ }\href@noop {} {\bibfield  {journal} {\bibinfo  {journal}
  {Journal of Neuroscience}\ }\textbf {\bibinfo {volume} {35}},\ \bibinfo
  {pages} {3663} (\bibinfo {year} {2015})}\BibitemShut {NoStop}%
\bibitem [{\citenamefont {Dehmamy}\ \emph {et~al.}(2018)\citenamefont
  {Dehmamy}, \citenamefont {Milanlouei},\ and\ \citenamefont
  {Barab{\'a}si}}]{dehmamy2018structural}%
  \BibitemOpen
  \bibfield  {author} {\bibinfo {author} {\bibfnamefont {N.}~\bibnamefont
  {Dehmamy}}, \bibinfo {author} {\bibfnamefont {S.}~\bibnamefont
  {Milanlouei}},\ and\ \bibinfo {author} {\bibfnamefont {A.-L.}\ \bibnamefont
  {Barab{\'a}si}},\ }\bibfield  {title} {\bibinfo {title} {A structural
  transition in physical networks},\ }\href@noop {} {\bibfield  {journal}
  {\bibinfo  {journal} {Nature}\ }\textbf {\bibinfo {volume} {563}},\ \bibinfo
  {pages} {676} (\bibinfo {year} {2018})}\BibitemShut {NoStop}%
\bibitem [{\citenamefont {Liu}\ \emph {et~al.}(2021)\citenamefont {Liu},
  \citenamefont {Dehmamy},\ and\ \citenamefont
  {Barab{\'a}si}}]{liu2021isotopy}%
  \BibitemOpen
  \bibfield  {author} {\bibinfo {author} {\bibfnamefont {Y.}~\bibnamefont
  {Liu}}, \bibinfo {author} {\bibfnamefont {N.}~\bibnamefont {Dehmamy}},\ and\
  \bibinfo {author} {\bibfnamefont {A.-L.}\ \bibnamefont {Barab{\'a}si}},\
  }\bibfield  {title} {\bibinfo {title} {Isotopy and energy of physical
  networks},\ }\href@noop {} {\bibfield  {journal} {\bibinfo  {journal} {Nature
  Physics}\ }\textbf {\bibinfo {volume} {17}},\ \bibinfo {pages} {216}
  (\bibinfo {year} {2021})}\BibitemShut {NoStop}%
\bibitem [{\citenamefont {Dorogovtsev}\ and\ \citenamefont
  {Mendes}(2003)}]{dorogovtsev2003evolution}%
  \BibitemOpen
  \bibfield  {author} {\bibinfo {author} {\bibfnamefont {S.~N.}\ \bibnamefont
  {Dorogovtsev}}\ and\ \bibinfo {author} {\bibfnamefont {J.~F.}\ \bibnamefont
  {Mendes}},\ }\href@noop {} {\emph {\bibinfo {title} {Evolution of networks:
  From biological nets to the Internet and WWW}}}\ (\bibinfo  {publisher}
  {Oxford University Press},\ \bibinfo {year} {2003})\BibitemShut {NoStop}%
\bibitem [{\citenamefont {Caldarelli}(2007)}]{caldarelli2007scale}%
  \BibitemOpen
  \bibfield  {author} {\bibinfo {author} {\bibfnamefont {G.}~\bibnamefont
  {Caldarelli}},\ }\href@noop {} {\emph {\bibinfo {title} {Scale-free networks:
  complex webs in nature and technology}}}\ (\bibinfo  {publisher} {Oxford
  University Press},\ \bibinfo {year} {2007})\BibitemShut {NoStop}%
\bibitem [{\citenamefont {Cohen}\ and\ \citenamefont
  {Havlin}(2010)}]{cohen2010complex}%
  \BibitemOpen
  \bibfield  {author} {\bibinfo {author} {\bibfnamefont {R.}~\bibnamefont
  {Cohen}}\ and\ \bibinfo {author} {\bibfnamefont {S.}~\bibnamefont {Havlin}},\
  }\href@noop {} {\emph {\bibinfo {title} {Complex networks: structure,
  robustness and function}}}\ (\bibinfo  {publisher} {Cambridge University
  Press},\ \bibinfo {year} {2010})\BibitemShut {NoStop}%
\bibitem [{\citenamefont {Newman}(2010)}]{newman2010networks}%
  \BibitemOpen
  \bibfield  {author} {\bibinfo {author} {\bibfnamefont {M.}~\bibnamefont
  {Newman}},\ }\href@noop {} {\emph {\bibinfo {title} {Networks: An
  introduction}}}\ (\bibinfo  {publisher} {Oxford University Press},\ \bibinfo
  {year} {2010})\BibitemShut {NoStop}%
\bibitem [{\citenamefont {Barab{\'a}si}(2016)}]{barabasi2016network}%
  \BibitemOpen
  \bibfield  {author} {\bibinfo {author} {\bibfnamefont {A.-L.}\ \bibnamefont
  {Barab{\'a}si}},\ }\href@noop {} {\emph {\bibinfo {title} {Network
  Science}}}\ (\bibinfo  {publisher} {Cambridge University Press},\ \bibinfo
  {year} {2016})\BibitemShut {NoStop}%
\bibitem [{\citenamefont {Barth{\'e}lemy}(2011)}]{barthelemy2011spatial}%
  \BibitemOpen
  \bibfield  {author} {\bibinfo {author} {\bibfnamefont {M.}~\bibnamefont
  {Barth{\'e}lemy}},\ }\bibfield  {title} {\bibinfo {title} {Spatial
  networks},\ }\href@noop {} {\bibfield  {journal} {\bibinfo  {journal}
  {Physics Reports}\ }\textbf {\bibinfo {volume} {499}},\ \bibinfo {pages} {1}
  (\bibinfo {year} {2011})}\BibitemShut {NoStop}%
\bibitem [{\citenamefont {Horv{\'a}t}\ \emph {et~al.}(2016)\citenamefont
  {Horv{\'a}t}, \citenamefont {G{\u{a}}m{\u{a}}nu\textcommabelow{t}},
  \citenamefont {Ercsey-Ravasz}, \citenamefont {Magrou}, \citenamefont
  {G{\u{a}}m{\u{a}}nu\textcommabelow{t}}, \citenamefont {Van~Essen},
  \citenamefont {Burkhalter}, \citenamefont {Knoblauch}, \citenamefont
  {Toroczkai},\ and\ \citenamefont {Kennedy}}]{horvat2016spatial}%
  \BibitemOpen
  \bibfield  {author} {\bibinfo {author} {\bibfnamefont {S.}~\bibnamefont
  {Horv{\'a}t}}, \bibinfo {author} {\bibfnamefont {R.}~\bibnamefont
  {G{\u{a}}m{\u{a}}nu\textcommabelow{t}}}, \bibinfo {author} {\bibfnamefont
  {M.}~\bibnamefont {Ercsey-Ravasz}}, \bibinfo {author} {\bibfnamefont
  {L.}~\bibnamefont {Magrou}}, \bibinfo {author} {\bibfnamefont
  {B.}~\bibnamefont {G{\u{a}}m{\u{a}}nu\textcommabelow{t}}}, \bibinfo {author}
  {\bibfnamefont {D.~C.}\ \bibnamefont {Van~Essen}}, \bibinfo {author}
  {\bibfnamefont {A.}~\bibnamefont {Burkhalter}}, \bibinfo {author}
  {\bibfnamefont {K.}~\bibnamefont {Knoblauch}}, \bibinfo {author}
  {\bibfnamefont {Z.}~\bibnamefont {Toroczkai}},\ and\ \bibinfo {author}
  {\bibfnamefont {H.}~\bibnamefont {Kennedy}},\ }\bibfield  {title} {\bibinfo
  {title} {Spatial embedding and wiring cost constrain the functional layout of
  the cortical network of rodents and primates},\ }\href@noop {} {\bibfield
  {journal} {\bibinfo  {journal} {PLoS biology}\ }\textbf {\bibinfo {volume}
  {14}},\ \bibinfo {pages} {e1002512} (\bibinfo {year} {2016})}\BibitemShut
  {NoStop}%
\bibitem [{\citenamefont {Rubinstein}\ \emph {et~al.}(2003)\citenamefont
  {Rubinstein}, \citenamefont {Colby} \emph {et~al.}}]{rubinstein2003polymer}%
  \BibitemOpen
  \bibfield  {author} {\bibinfo {author} {\bibfnamefont {M.}~\bibnamefont
  {Rubinstein}}, \bibinfo {author} {\bibfnamefont {R.~H.}\ \bibnamefont
  {Colby}}, \emph {et~al.},\ }\href@noop {} {\emph {\bibinfo {title} {Polymer
  physics}}},\ Vol.~\bibinfo {volume} {23}\ (\bibinfo  {publisher} {Oxford
  university press New York},\ \bibinfo {year} {2003})\BibitemShut {NoStop}%
\bibitem [{\citenamefont {Song}\ \emph {et~al.}(2008)\citenamefont {Song},
  \citenamefont {Wang},\ and\ \citenamefont {Makse}}]{song2008phase}%
  \BibitemOpen
  \bibfield  {author} {\bibinfo {author} {\bibfnamefont {C.}~\bibnamefont
  {Song}}, \bibinfo {author} {\bibfnamefont {P.}~\bibnamefont {Wang}},\ and\
  \bibinfo {author} {\bibfnamefont {H.~A.}\ \bibnamefont {Makse}},\ }\bibfield
  {title} {\bibinfo {title} {A phase diagram for jammed matter},\ }\href@noop
  {} {\bibfield  {journal} {\bibinfo  {journal} {Nature}\ }\textbf {\bibinfo
  {volume} {453}},\ \bibinfo {pages} {629} (\bibinfo {year}
  {2008})}\BibitemShut {NoStop}%
\bibitem [{\citenamefont {West}\ \emph {et~al.}(2001)\citenamefont {West} \emph
  {et~al.}}]{west2001introduction}%
  \BibitemOpen
  \bibfield  {author} {\bibinfo {author} {\bibfnamefont {D.~B.}\ \bibnamefont
  {West}} \emph {et~al.},\ }\href@noop {} {\emph {\bibinfo {title}
  {Introduction to graph theory}}},\ Vol.~\bibinfo {volume} {2}\ (\bibinfo
  {publisher} {Prentice hall Upper Saddle River},\ \bibinfo {year}
  {2001})\BibitemShut {NoStop}%
\bibitem [{\citenamefont {Tarjan}\ and\ \citenamefont
  {Trojanowski}(1977)}]{tarjan1977finding}%
  \BibitemOpen
  \bibfield  {author} {\bibinfo {author} {\bibfnamefont {R.~E.}\ \bibnamefont
  {Tarjan}}\ and\ \bibinfo {author} {\bibfnamefont {A.~E.}\ \bibnamefont
  {Trojanowski}},\ }\bibfield  {title} {\bibinfo {title} {Finding a maximum
  independent set},\ }\href@noop {} {\bibfield  {journal} {\bibinfo  {journal}
  {SIAM Journal on Computing}\ }\textbf {\bibinfo {volume} {6}},\ \bibinfo
  {pages} {537} (\bibinfo {year} {1977})}\BibitemShut {NoStop}%
\bibitem [{\citenamefont {Flory}(1939)}]{flory1939intramolecular}%
  \BibitemOpen
  \bibfield  {author} {\bibinfo {author} {\bibfnamefont {P.~J.}\ \bibnamefont
  {Flory}},\ }\bibfield  {title} {\bibinfo {title} {Intramolecular reaction
  between neighboring substituents of vinyl polymers},\ }\href@noop {}
  {\bibfield  {journal} {\bibinfo  {journal} {Journal of the American Chemical
  Society}\ }\textbf {\bibinfo {volume} {61}},\ \bibinfo {pages} {1518}
  (\bibinfo {year} {1939})}\BibitemShut {NoStop}%
\bibitem [{\citenamefont {Hartmann}\ and\ \citenamefont
  {Weigt}(2006)}]{hartmann2006phase}%
  \BibitemOpen
  \bibfield  {author} {\bibinfo {author} {\bibfnamefont {A.~K.}\ \bibnamefont
  {Hartmann}}\ and\ \bibinfo {author} {\bibfnamefont {M.}~\bibnamefont
  {Weigt}},\ }\href@noop {} {\emph {\bibinfo {title} {Phase transitions in
  combinatorial optimization problems: basics, algorithms and statistical
  mechanics}}}\ (\bibinfo  {publisher} {John Wiley \& Sons},\ \bibinfo {year}
  {2006})\BibitemShut {NoStop}%
\bibitem [{\citenamefont {Brightwell}\ \emph {et~al.}(2017)\citenamefont
  {Brightwell}, \citenamefont {Janson},\ and\ \citenamefont
  {Luczak}}]{brightwell2017greedy}%
  \BibitemOpen
  \bibfield  {author} {\bibinfo {author} {\bibfnamefont {G.}~\bibnamefont
  {Brightwell}}, \bibinfo {author} {\bibfnamefont {S.}~\bibnamefont {Janson}},\
  and\ \bibinfo {author} {\bibfnamefont {M.}~\bibnamefont {Luczak}},\
  }\bibfield  {title} {\bibinfo {title} {The greedy independent set in a random
  graph with given degrees},\ }\href@noop {} {\bibfield  {journal} {\bibinfo
  {journal} {Random Structures \& Algorithms}\ }\textbf {\bibinfo {volume}
  {51}},\ \bibinfo {pages} {565} (\bibinfo {year} {2017})}\BibitemShut
  {NoStop}%
\bibitem [{\citenamefont {Ercsey-Ravasz}\ \emph {et~al.}(2013)\citenamefont
  {Ercsey-Ravasz}, \citenamefont {Markov}, \citenamefont {Lamy}, \citenamefont
  {Van~Essen}, \citenamefont {Knoblauch}, \citenamefont {Toroczkai},\ and\
  \citenamefont {Kennedy}}]{ercsey2013predictive}%
  \BibitemOpen
  \bibfield  {author} {\bibinfo {author} {\bibfnamefont {M.}~\bibnamefont
  {Ercsey-Ravasz}}, \bibinfo {author} {\bibfnamefont {N.~T.}\ \bibnamefont
  {Markov}}, \bibinfo {author} {\bibfnamefont {C.}~\bibnamefont {Lamy}},
  \bibinfo {author} {\bibfnamefont {D.~C.}\ \bibnamefont {Van~Essen}}, \bibinfo
  {author} {\bibfnamefont {K.}~\bibnamefont {Knoblauch}}, \bibinfo {author}
  {\bibfnamefont {Z.}~\bibnamefont {Toroczkai}},\ and\ \bibinfo {author}
  {\bibfnamefont {H.}~\bibnamefont {Kennedy}},\ }\bibfield  {title} {\bibinfo
  {title} {A predictive network model of cerebral cortical connectivity based
  on a distance rule},\ }\href@noop {} {\bibfield  {journal} {\bibinfo
  {journal} {Neuron}\ }\textbf {\bibinfo {volume} {80}},\ \bibinfo {pages}
  {184} (\bibinfo {year} {2013})}\BibitemShut {NoStop}%
\bibitem [{\citenamefont {Bassett}\ and\ \citenamefont
  {Bullmore}(2017)}]{bassett2017small}%
  \BibitemOpen
  \bibfield  {author} {\bibinfo {author} {\bibfnamefont {D.~S.}\ \bibnamefont
  {Bassett}}\ and\ \bibinfo {author} {\bibfnamefont {E.~T.}\ \bibnamefont
  {Bullmore}},\ }\bibfield  {title} {\bibinfo {title} {Small-world brain
  networks revisited},\ }\href@noop {} {\bibfield  {journal} {\bibinfo
  {journal} {The Neuroscientist}\ }\textbf {\bibinfo {volume} {23}},\ \bibinfo
  {pages} {499} (\bibinfo {year} {2017})}\BibitemShut {NoStop}%
\bibitem [{\citenamefont {Van~Mieghem}(2010)}]{van2010graph}%
  \BibitemOpen
  \bibfield  {author} {\bibinfo {author} {\bibfnamefont {P.}~\bibnamefont
  {Van~Mieghem}},\ }\href@noop {} {\emph {\bibinfo {title} {Graph spectra for
  complex networks}}}\ (\bibinfo  {publisher} {Cambridge University Press},\
  \bibinfo {year} {2010})\BibitemShut {NoStop}%
\bibitem [{\citenamefont {Abbe}(2017)}]{abbe2017community}%
  \BibitemOpen
  \bibfield  {author} {\bibinfo {author} {\bibfnamefont {E.}~\bibnamefont
  {Abbe}},\ }\bibfield  {title} {\bibinfo {title} {Community detection and
  stochastic block models: recent developments},\ }\href@noop {} {\bibfield
  {journal} {\bibinfo  {journal} {The Journal of Machine Learning Research}\
  }\textbf {\bibinfo {volume} {18}},\ \bibinfo {pages} {6446} (\bibinfo {year}
  {2017})}\BibitemShut {NoStop}%
\bibitem [{\citenamefont {Viana}\ \emph {et~al.}(2020)\citenamefont {Viana},
  \citenamefont {Brown}, \citenamefont {Mueller}, \citenamefont {Goul},
  \citenamefont {Koslover},\ and\ \citenamefont
  {Rafelski}}]{viana2020mitochondrial}%
  \BibitemOpen
  \bibfield  {author} {\bibinfo {author} {\bibfnamefont {M.~P.}\ \bibnamefont
  {Viana}}, \bibinfo {author} {\bibfnamefont {A.~I.}\ \bibnamefont {Brown}},
  \bibinfo {author} {\bibfnamefont {I.~A.}\ \bibnamefont {Mueller}}, \bibinfo
  {author} {\bibfnamefont {C.}~\bibnamefont {Goul}}, \bibinfo {author}
  {\bibfnamefont {E.~F.}\ \bibnamefont {Koslover}},\ and\ \bibinfo {author}
  {\bibfnamefont {S.~M.}\ \bibnamefont {Rafelski}},\ }\bibfield  {title}
  {\bibinfo {title} {Mitochondrial fission and fusion dynamics generate
  efficient, robust, and evenly distributed network topologies in budding yeast
  cells},\ }\href@noop {} {\bibfield  {journal} {\bibinfo  {journal} {Cell
  systems}\ }\textbf {\bibinfo {volume} {10}},\ \bibinfo {pages} {287}
  (\bibinfo {year} {2020})}\BibitemShut {NoStop}%
\bibitem [{\citenamefont {Rees}\ \emph {et~al.}(2017)\citenamefont {Rees},
  \citenamefont {Moradi},\ and\ \citenamefont {Ascoli}}]{rees2017weighing}%
  \BibitemOpen
  \bibfield  {author} {\bibinfo {author} {\bibfnamefont {C.~L.}\ \bibnamefont
  {Rees}}, \bibinfo {author} {\bibfnamefont {K.}~\bibnamefont {Moradi}},\ and\
  \bibinfo {author} {\bibfnamefont {G.~A.}\ \bibnamefont {Ascoli}},\ }\bibfield
   {title} {\bibinfo {title} {Weighing the evidence in peters\u2019 rule: does
  neuronal morphology predict connectivity?},\ }\href@noop {} {\bibfield
  {journal} {\bibinfo  {journal} {Trends in neurosciences}\ }\textbf {\bibinfo
  {volume} {40}},\ \bibinfo {pages} {63} (\bibinfo {year} {2017})}\BibitemShut
  {NoStop}%
\bibitem [{\citenamefont {Udvary}\ \emph {et~al.}(2022)\citenamefont {Udvary},
  \citenamefont {Harth}, \citenamefont {Macke}, \citenamefont {Hege},
  \citenamefont {de~Kock}, \citenamefont {Sakmann},\ and\ \citenamefont
  {Oberlaender}}]{udvary2022impact}%
  \BibitemOpen
  \bibfield  {author} {\bibinfo {author} {\bibfnamefont {D.}~\bibnamefont
  {Udvary}}, \bibinfo {author} {\bibfnamefont {P.}~\bibnamefont {Harth}},
  \bibinfo {author} {\bibfnamefont {J.~H.}\ \bibnamefont {Macke}}, \bibinfo
  {author} {\bibfnamefont {H.-C.}\ \bibnamefont {Hege}}, \bibinfo {author}
  {\bibfnamefont {C.~P.}\ \bibnamefont {de~Kock}}, \bibinfo {author}
  {\bibfnamefont {B.}~\bibnamefont {Sakmann}},\ and\ \bibinfo {author}
  {\bibfnamefont {M.}~\bibnamefont {Oberlaender}},\ }\bibfield  {title}
  {\bibinfo {title} {The impact of neuron morphology on cortical network
  architecture},\ }\href@noop {} {\bibfield  {journal} {\bibinfo  {journal}
  {Cell Reports}\ }\textbf {\bibinfo {volume} {39}},\ \bibinfo {pages} {110677}
  (\bibinfo {year} {2022})}\BibitemShut {NoStop}%
\bibitem [{\citenamefont {Nicolaou}\ and\ \citenamefont
  {Motter}(2012)}]{nicolaou2012mechanical}%
  \BibitemOpen
  \bibfield  {author} {\bibinfo {author} {\bibfnamefont {Z.~G.}\ \bibnamefont
  {Nicolaou}}\ and\ \bibinfo {author} {\bibfnamefont {A.~E.}\ \bibnamefont
  {Motter}},\ }\bibfield  {title} {\bibinfo {title} {Mechanical metamaterials
  with negative compressibility transitions},\ }\href@noop {} {\bibfield
  {journal} {\bibinfo  {journal} {Nature materials}\ }\textbf {\bibinfo
  {volume} {11}},\ \bibinfo {pages} {608} (\bibinfo {year} {2012})}\BibitemShut
  {NoStop}%
\bibitem [{\citenamefont {Kim}\ \emph {et~al.}(2022)\citenamefont {Kim},
  \citenamefont {Lu}, \citenamefont {Blevins},\ and\ \citenamefont
  {Bassett}}]{kim2022nonlinear}%
  \BibitemOpen
  \bibfield  {author} {\bibinfo {author} {\bibfnamefont {J.~Z.}\ \bibnamefont
  {Kim}}, \bibinfo {author} {\bibfnamefont {Z.}~\bibnamefont {Lu}}, \bibinfo
  {author} {\bibfnamefont {A.~S.}\ \bibnamefont {Blevins}},\ and\ \bibinfo
  {author} {\bibfnamefont {D.~S.}\ \bibnamefont {Bassett}},\ }\bibfield
  {title} {\bibinfo {title} {Nonlinear dynamics and chaos in conformational
  changes of mechanical metamaterials},\ }\href@noop {} {\bibfield  {journal}
  {\bibinfo  {journal} {Physical Review X}\ }\textbf {\bibinfo {volume} {12}},\
  \bibinfo {pages} {011042} (\bibinfo {year} {2022})}\BibitemShut {NoStop}%
\bibitem [{\citenamefont {D'Souza}\ \emph {et~al.}(2019)\citenamefont
  {D'Souza}, \citenamefont {G{\'o}mez-Gardenes}, \citenamefont {Nagler},\ and\
  \citenamefont {Arenas}}]{dsouza2019explosive}%
  \BibitemOpen
  \bibfield  {author} {\bibinfo {author} {\bibfnamefont {R.~M.}\ \bibnamefont
  {D'Souza}}, \bibinfo {author} {\bibfnamefont {J.}~\bibnamefont
  {G{\'o}mez-Gardenes}}, \bibinfo {author} {\bibfnamefont {J.}~\bibnamefont
  {Nagler}},\ and\ \bibinfo {author} {\bibfnamefont {A.}~\bibnamefont
  {Arenas}},\ }\bibfield  {title} {\bibinfo {title} {Explosive phenomena in
  complex networks},\ }\href@noop {} {\bibfield  {journal} {\bibinfo  {journal}
  {Advances in Physics}\ }\textbf {\bibinfo {volume} {68}},\ \bibinfo {pages}
  {123} (\bibinfo {year} {2019})}\BibitemShut {NoStop}%
\bibitem [{\citenamefont {Heroy}\ \emph {et~al.}(2022)\citenamefont {Heroy},
  \citenamefont {Taylor}, \citenamefont {Shi}, \citenamefont {Forest},\ and\
  \citenamefont {Mucha}}]{heroy2022rigidity}%
  \BibitemOpen
  \bibfield  {author} {\bibinfo {author} {\bibfnamefont {S.}~\bibnamefont
  {Heroy}}, \bibinfo {author} {\bibfnamefont {D.}~\bibnamefont {Taylor}},
  \bibinfo {author} {\bibfnamefont {F.}~\bibnamefont {Shi}}, \bibinfo {author}
  {\bibfnamefont {M.~G.}\ \bibnamefont {Forest}},\ and\ \bibinfo {author}
  {\bibfnamefont {P.~J.}\ \bibnamefont {Mucha}},\ }\bibfield  {title} {\bibinfo
  {title} {Rigidity percolation in disordered 3d rod systems},\ }\href@noop {}
  {\bibfield  {journal} {\bibinfo  {journal} {Multiscale Modeling \&
  Simulation}\ }\textbf {\bibinfo {volume} {20}},\ \bibinfo {pages} {250}
  (\bibinfo {year} {2022})}\BibitemShut {NoStop}%
\bibitem [{\citenamefont {Barrat}\ \emph {et~al.}(2008)\citenamefont {Barrat},
  \citenamefont {Barthelemy},\ and\ \citenamefont
  {Vespignani}}]{barrat2008dynamical}%
  \BibitemOpen
  \bibfield  {author} {\bibinfo {author} {\bibfnamefont {A.}~\bibnamefont
  {Barrat}}, \bibinfo {author} {\bibfnamefont {M.}~\bibnamefont {Barthelemy}},\
  and\ \bibinfo {author} {\bibfnamefont {A.}~\bibnamefont {Vespignani}},\
  }\href@noop {} {\emph {\bibinfo {title} {Dynamical processes on complex
  networks}}}\ (\bibinfo  {publisher} {Cambridge University Press},\ \bibinfo
  {year} {2008})\BibitemShut {NoStop}%
\bibitem [{\citenamefont {Ulloa~Severino}\ \emph {et~al.}(2016)\citenamefont
  {Ulloa~Severino}, \citenamefont {Ban}, \citenamefont {Song}, \citenamefont
  {Tang}, \citenamefont {Bianconi}, \citenamefont {Cheng},\ and\ \citenamefont
  {Torre}}]{ulloa2016role}%
  \BibitemOpen
  \bibfield  {author} {\bibinfo {author} {\bibfnamefont {F.~P.}\ \bibnamefont
  {Ulloa~Severino}}, \bibinfo {author} {\bibfnamefont {J.}~\bibnamefont {Ban}},
  \bibinfo {author} {\bibfnamefont {Q.}~\bibnamefont {Song}}, \bibinfo {author}
  {\bibfnamefont {M.}~\bibnamefont {Tang}}, \bibinfo {author} {\bibfnamefont
  {G.}~\bibnamefont {Bianconi}}, \bibinfo {author} {\bibfnamefont
  {G.}~\bibnamefont {Cheng}},\ and\ \bibinfo {author} {\bibfnamefont
  {V.}~\bibnamefont {Torre}},\ }\bibfield  {title} {\bibinfo {title} {The role
  of dimensionality in neuronal network dynamics},\ }\href@noop {} {\bibfield
  {journal} {\bibinfo  {journal} {Scientific reports}\ }\textbf {\bibinfo
  {volume} {6}},\ \bibinfo {pages} {1} (\bibinfo {year} {2016})}\BibitemShut
  {NoStop}%
\bibitem [{\citenamefont {Bianconi}(2018)}]{bianconi2018multilayer}%
  \BibitemOpen
  \bibfield  {author} {\bibinfo {author} {\bibfnamefont {G.}~\bibnamefont
  {Bianconi}},\ }\href@noop {} {\emph {\bibinfo {title} {Multilayer networks:
  structure and function}}}\ (\bibinfo  {publisher} {Oxford University Press},\
  \bibinfo {year} {2018})\BibitemShut {NoStop}%
\bibitem [{\citenamefont {Case}\ \emph {et~al.}(2019)\citenamefont {Case},
  \citenamefont {Liu}, \citenamefont {Kiss}, \citenamefont {Angilella},\ and\
  \citenamefont {Motter}}]{case2019braess}%
  \BibitemOpen
  \bibfield  {author} {\bibinfo {author} {\bibfnamefont {D.~J.}\ \bibnamefont
  {Case}}, \bibinfo {author} {\bibfnamefont {Y.}~\bibnamefont {Liu}}, \bibinfo
  {author} {\bibfnamefont {I.~Z.}\ \bibnamefont {Kiss}}, \bibinfo {author}
  {\bibfnamefont {J.-R.}\ \bibnamefont {Angilella}},\ and\ \bibinfo {author}
  {\bibfnamefont {A.~E.}\ \bibnamefont {Motter}},\ }\bibfield  {title}
  {\bibinfo {title} {Braess's paradox and programmable behaviour in
  microfluidic networks},\ }\href@noop {} {\bibfield  {journal} {\bibinfo
  {journal} {Nature}\ }\textbf {\bibinfo {volume} {574}},\ \bibinfo {pages}
  {647} (\bibinfo {year} {2019})}\BibitemShut {NoStop}%
\bibitem [{\citenamefont {Kim}\ \emph {et~al.}(2019)\citenamefont {Kim},
  \citenamefont {Lu}, \citenamefont {Strogatz},\ and\ \citenamefont
  {Bassett}}]{kim2019conformational}%
  \BibitemOpen
  \bibfield  {author} {\bibinfo {author} {\bibfnamefont {J.~Z.}\ \bibnamefont
  {Kim}}, \bibinfo {author} {\bibfnamefont {Z.}~\bibnamefont {Lu}}, \bibinfo
  {author} {\bibfnamefont {S.~H.}\ \bibnamefont {Strogatz}},\ and\ \bibinfo
  {author} {\bibfnamefont {D.~S.}\ \bibnamefont {Bassett}},\ }\bibfield
  {title} {\bibinfo {title} {Conformational control of mechanical networks},\
  }\href@noop {} {\bibfield  {journal} {\bibinfo  {journal} {Nature Physics}\
  }\textbf {\bibinfo {volume} {15}},\ \bibinfo {pages} {714} (\bibinfo {year}
  {2019})}\BibitemShut {NoStop}%
\bibitem [{\citenamefont {Kov{\'a}cs}\ \emph {et~al.}(2020)\citenamefont
  {Kov{\'a}cs}, \citenamefont {Barab{\'a}si},\ and\ \citenamefont
  {Barab{\'a}si}}]{kovacs2020uncovering}%
  \BibitemOpen
  \bibfield  {author} {\bibinfo {author} {\bibfnamefont {I.~A.}\ \bibnamefont
  {Kov{\'a}cs}}, \bibinfo {author} {\bibfnamefont {D.~L.}\ \bibnamefont
  {Barab{\'a}si}},\ and\ \bibinfo {author} {\bibfnamefont {A.-L.}\ \bibnamefont
  {Barab{\'a}si}},\ }\bibfield  {title} {\bibinfo {title} {Uncovering the
  genetic blueprint of the c. elegans nervous system},\ }\href@noop {}
  {\bibfield  {journal} {\bibinfo  {journal} {Proceedings of the National
  Academy of Sciences}\ }\textbf {\bibinfo {volume} {117}},\ \bibinfo {pages}
  {33570} (\bibinfo {year} {2020})}\BibitemShut {NoStop}%
\end{thebibliography}
\end{document}


\title{Supplementary Information -- Understanding the impact of physicality on network structure}

\author{M\'arton P\'osfai}
\email{Contributed equally.}
\affiliation{Department of Network and Data Science, Central European University, Vienna, Austria} 
\author{Bal\'azs Szegedy}
\email{Contributed equally.}
\affiliation{Alfr\'ed R\'enyi Institute of Mathematics, Budapest, Hungary}
\author{Iva Ba\v ci\'c}
\affiliation{Department of Network and Data Science, Central European University, Vienna, Austria} 
\affiliation{Institute of Physics, Belgrade, Serbia}
\author{Luka Blagojevi\'c}
\affiliation{Department of Network and Data Science, Central European University, Vienna, Austria} 
\author{Mikl\'os Ab\'ert}
\affiliation{Alfr\'ed R\'enyi Institute of Mathematics, Budapest, Hungary}
\author{J\'anos Kert\'esz}
\affiliation{Department of Network and Data Science, Central European University, Vienna, Austria} 
\author{L\'aszl\'o Lov\'asz}
\affiliation{Alfr\'ed R\'enyi Institute of Mathematics, Budapest, Hungary}
\author{Albert-L\'aszl\'o Barab\'asi}

\affiliation{Department of Network and Data Science, Central European University, Vienna, Austria} 
\affiliation{Network Science Institute, Northeastern University, Boston, MA, USA}
\affiliation{Department of Medicine, Brigham and Women's Hospital, Harvard Medical School, Boston, MA, USA}


\tableofcontents

\newpage

\section{Linear physical networks and the meta-graph}

In this section we provide formal definitions for the objects and quantities that define physical networks.

\subsection{Linear physical networks}

A linear physical network (LPN) is a network embedded in three-dimensional Euclidean space such that each node in the network is a sphere and each link is a capped cylinder with diameter $\lambda$.
The nodes and links satisfy volume exclusion, meaning that they cannot overlap in space.
To avoid restricting the maximum node degree, we allow a node to overlap with the links that are connected to it, and we allow links to overlap with each other if they share an endpoint, leading to the following formal definition:

\begin{definition}\label{def:LPN}
A $\lambda$-linear physical network (LPN) in its strictest sense is a graph $\mathcal G$ such that the vertex set of $\s G$ is a point set $\mathcal P\subset\mathbb{R}^3$ and the edges $(p_1,p_2)\in \s E\subset \s P_2$ are straight segments connecting these points, where $\s P_2$ is every unordered pair formed of elements of $\s P$.
We require that the distance is at least $\lambda$ between
\begin{enumerate}[(i)]
\item every point pair $p_1,p_2\in \s P$, with $p_1\neq p_2$ (node-node interaction);
\item every point $p_1$ and every edge $(p_2,p_3)\in \s E$, with $\{p_1\}\cap\{p_2,p_3\}=\emptyset$ (node-link interaction);
\item every pair of edges $(p_1,p_2),(p_3,p_4)$, respectively, with $\{p_1,p_2\}\cap\{p_3,p_4\}=\emptyset$ (link-link interaction).
\end{enumerate}
\end{definition}
We can define a more permissive version of the LPNs by relaxing some of the (i-iii) conditions.
In the main text, for example, we study LPNs with only (iii) link-link interactions, offering a simpler exposition, while the behavior of key properties that we study, such as scaling of the jammed state and the space-dependent eigenvectors, remain similar to the strictest definition of LPNs, as we demonstrate in Secs.~\ref{sec:Mmax_bounds}, \ref{sec:rand_indep_set}, \ref{sec:link_length}, \ref{sec:degvar_clust} and \ref{sec:spectra}.

\subsection{Generating random LPNs}\label{sec:generating_LPNs}

Here we define a random LPN model.
We generate random LPNs in two stages: (i)~we place points $p\in \s P$ corresponding to nodes in $\mathbb R^3$ and then (ii)~we connect unordered node pairs $(p,q)\in \s P_2$ such that we ensure that we do not violate Def.~\ref{def:LPN}.
Throughout this paper we place nodes randomly in the unit cube and node pairs are connected in random order.

Specifically, for node placement, if node-node interactions are ignored, i.e., we allow nodes to overlap, the center of each node $p\in \s P$ is placed uniformly at random in the unit cube.
If node-node interactions are considered, nodes are placed one at time, always choosing uniformly at random from positions at least $\lambda$ distance away from the center of any existing node.
This process is equivalent to random sequential deposition (RSD) studied in the context of hard-particle packings.
It is known that the maximum density achievable via RSD is $\rho \approx 0.38$~\cite{zhang2013precise}, in our simulations we remain well below this limit.
Higher node density is achievable relying on non-random packings or simulated annealing-type algorithms~\cite{torquato2010jammed}.

For link placement, we randomly order the link candidates, i.e., every possible pair of nodes $(p,q)\in\s P_2$. 
We then sequentially add them to the network if they do not violate physicality, i.e., they do not be overlap with any already existing link and, if node-link interactions are also considered, they do not overlap with any node other than their endpoints.
We halt either after the addition of a predefined number of links $M$ or we continue until no more links can be added.

Note that the definition of LPNs is more general than the random model introduced here, other models, such as growing physical networks, are also possible.

\subsection{Meta-graph}\label{sec:meta-graph}

Here we define the meta-graph $\mathcal M(\mathcal P,\lambda)$, which is an auxiliary graph that captures the physical constraints between link candidates connecting point pairs $(p,q)\in\s P\times\s P$. 

\begin{definition}\label{def:meta}
The meta-graph  $\mathcal M(\mathcal P,\lambda)$ is a graph defined for a $\lambda>0$ and point set $\s P$, such that 
\begin{enumerate}[(i)]
\item the vertex set of $\mathcal M(\mathcal P,\lambda)$ is the set of link candidates that do not overlap with nodes, i.e., 
$$\s V_\T{meta}=\{(p,q)\in\s P_2: d((p,q),r)\geq\lambda \forall r\in P\setminus\{p,q\} \}\;$$
\item and the edges of $\mathcal M(\mathcal P,\lambda)$ connect link candidates that overlap in space, i.e.,
$$\s E_\T{meta}=\{((p,q),(r,s))\in\s V_\T{meta}\times\s V_\T{meta}: d((p,q),(r,s)) \forall (r,s)\in P \times\s P \T{and} \{p,q\}\cap\{r,s\}=\emptyset \}.$$
\end{enumerate}
\end{definition}

If link-node interaction is not considered, the vertex set of the meta-graph contains all possible point pairs, i.e., $\s V_\T{meta}=(p,q)\in\s P\times\s P$ and $N_\T{meta}=N(N-1)/2$, where $N_\T{meta}=\lvert \s V_\T{meta}\rvert$ and $N=\lvert \s P\rvert$ .

\subsubsection{Independent sets of the meta-graph and physical networks}

\begin{definition}
Given a graph $\s G(\s V, \s E)$ with vertex set $\s V$ and edge set $\s E$, a subset of vertices $\s I\subset \s V$ is independent if no two nodes in $\s I$ are connected in $\s G$, i.e., $\forall v,w\in I$   $(v,w)\notin E$.
\end{definition}

An independent set in the meta-graph represents to a conflict-free set of link candidates; therefore there is a one-to-one correspondence between $\lambda$-linear physical networks on $\s P$ and independent sets in $\s M(\s P,\lambda)$.
Independent sets are extensively studied in graph combinatorics, computer science, and physics~\cite{west2001introduction,tarjan1977finding,hartmann2006phase}.

Therefore mapping between independent sets and physical networks provides a range of tools to characterize LPNs.
For example, the random LPN generation introduced in Sec.~\ref{sec:generating_LPNs} corresponds to the greedy maximal independent set construction.
Similarly to the mapping between LPNs and independent sets of the meta-graph, greedy independent sets were used to study systems with volume exclusion in statistical physics and chemistry starting with the work of Flory~\cite{flory1939intramolecular,evans1993random}.

\subsubsection{Relation to the dual graph}

\begin{figure}[h]
	\centering
	\includegraphics[width=.85\textwidth]{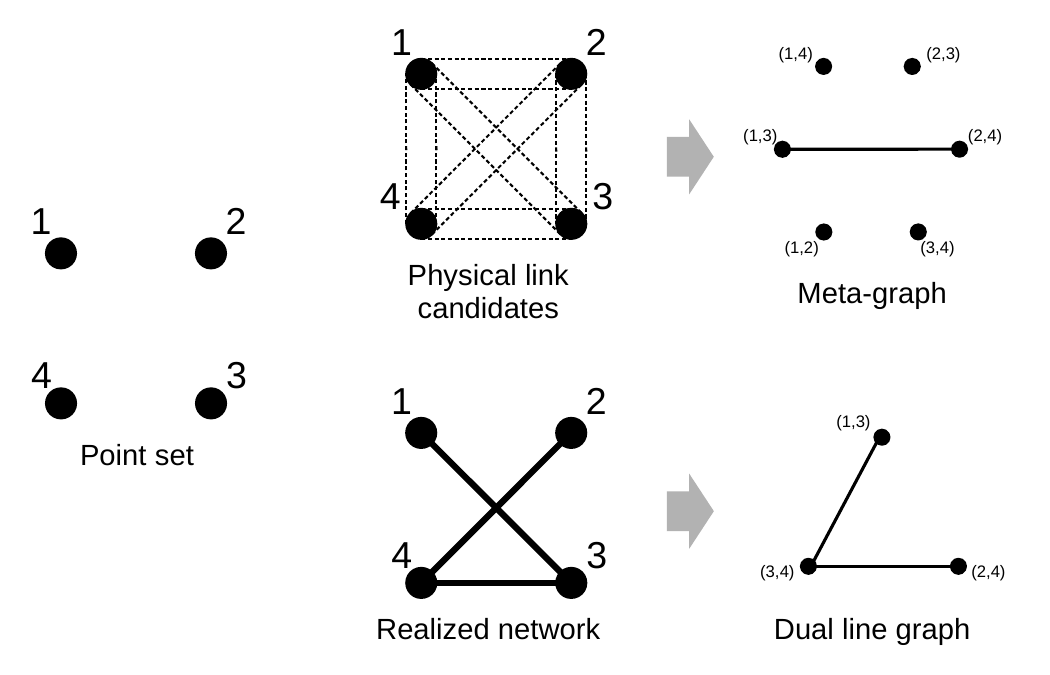}
	\caption{{ \bf Meta and line graphs.} The meta and line graphs are somewhat similar concepts as both meta-nodes and line graph-nodes represent links in the original network; however, the two graphs capture different relationships between these links. The meta-nodes are physical link candidates and the meta-links capture physical proximity: two meta-nodes are connected if the corresponding link candidates overlap and do not share endpoints, such as (1,3) and (2,4) in the example. While the nodes in the line graph are not link candidates but they represent links that are actually realized in a network, and two are connected if they are adjacent in the 	abstract network, for example (1,3) and (2,4) do not share endpoints and therefore are not connected in the line graph.}
	\label{fig:meta_vs_line}
\end{figure}

The dual line graph is a somewhat similar but distinct concept from the meta-graph~(Fig.~\ref{fig:meta_vs_line}).
The vertices of both the meta-graph and the line graph correspond to links in the original network.
The line graph, however, is associated to a realized network $\mathcal G$, while the meta-graph is associated to a physical point set $\mathcal P$.
Vertices of the line graph are links in $\mathcal G$ and connections between them represent adjacency in the network $\mathcal G$, i.e., they are connected if their associated edges share endpoints in $\mathcal G$. While the vertices of the meta-graph are all possible links connecting points in $\mathcal P$, and meta-edges represent physical proximity, i.e., two meta-nodes are connected if they overlap and do not share endpoints.

\subsection{Large $N$ limit and the $\alpha$ parameter}\label{sec:large_N}

We are interested in the large network limit $N\rightarrow\infty$ of LPNs.
Increasing $N$ while keeping $\lambda$ constant, however, is not possible, since the unit cube quickly fills up limiting the number of physical nodes it can hold.
Therefore, to obtain a useful thermodynamic limit, we must decrease diameter $\lambda$ while increasing $N$.
To find the precise relationship between $N$ and $\lambda$ that produces a non-trivial large network limit we estimate the average degree of the meta-graph $\avg{k_\T{meta}}$ as a function of $N$ and $\lambda$.

\subsubsection{Random link-link and node-link intersection}

To estimate $\avg{k_\T{meta}}$, we first we calculate the probability that two randomly placed links intersect.
Assume that the links have lengths $l_1$ and $l_2$, respectively, diameter $\lambda$, a given orientation, and that the links are rod-like, i.e., $l_1,l_2\gg\lambda$.
If we first place $l_1$ in the available volume, the excluded volume (i.e., the volume where we cannot place $l_2$ without violating physicality) in leading order of $\lambda$ is provided by a parallelepiped with sides $l_1$, $l_2$ and $2\lambda$, meaning that
\begin{equation}
V_\T{ll, exc}\sim \lambda l_1 l_2.
\end{equation}
The probability that the two links intersect is $p_\T{ll}=V_\T{ll, exc}/V_\T{total}$, where in case of placing links in the unit cube $V_\T{total}=1$.

The excluded volume between a randomly placed node and a link with length $l$ is simply the volume of a capped cylinder with length $l$ and diameter $2\lambda$, meaning that 
\begin{equation}
V_\T{nl, exc}\sim \lambda^2 l,
\end{equation}
in leading order of $\lambda$.
The fact that $V_\T{ll, exc}\sim \lambda$ and $V_\T{nl, exc}\sim \lambda^2$ hints at that as long as the typical link length is much longer than the diameter, i.e., $l^*\gg \lambda$, link-link interactions will dominate over node-link interactions in the large network limit.

\subsubsection{The number of vertices and edges in the meta-graph}

By definition, if node-link interactions are taken into account, the vertex set of the meta-graph is the set of physical link candidates that do not overlap with physical nodes.
There are $N=\lvert \s P \rvert$ physical nodes; therefore the probability that a random link with length $l$ does not intersect with any nodes is
\begin{equation}\label{eq:node_link_intersect_prob}
(1-c\cdot \lambda^2 l)^N\approx \exp(-c\lambda^2 lN),
\end{equation}
where $c$ is a constant.
There are $\sim N^2$ link candidates, therefore the total number of vertices in the meta-graph is
\begin{equation}\label{eq:Nmeta_fixed_N}
N_\T{meta}\sim N^2\int p_\T{LC}(l) \exp(-c\lambda^2 lN)dl,
\end{equation}
where $p_\T{LC}(l)$ is length distribution of the link candidates, and it is approximately equal to the distribution of the length of a randomly selected segment in the unit cube, also known as the cube line picking distribution~\cite{finch2003mathematical, philip2007probability}. 

The probability that two physical link candidates with length $l_1$ and $l_2$ overlap is $\sim \lambda l_1 l_2$.
Since we have $N_\T{meta}^2$ meta-vertex pairs, the number of edges in the meta-graph 
\begin{equation}\label{eq:Mmeta_fixed_N}
M_\T{meta}\sim\lambda N^4 \int p_\T{LC}(l_1)\exp(-c\lambda^2 l_1N)dl_1\int p_\T{LC}(l_2)\exp(-c\lambda^2 l_2N)dl_2,
\end{equation}
meaning that $M_\T{meta}$ grows linearly as a function of $\lambda$ with an exponential cutoff around $\lambda\sim \sqrt{N}$, a prediction supported by simulations.

If, as in the main text, we ignore node-link interactions, the exponential cutoff disappears, leading to
\begin{align}\label{eq:Nmeta_Mmeta_fixed_N_link-link}
N_\T{meta}&\sim N^2\\
M_\T{meta}&\sim \lambda N^4,
\end{align}
meaning that the average meta-degree is $\avg{k_\T{meta}}~\sim \lambda N^2$.

\subsubsection{The $\alpha$ exponent}

We now return to the question of how to take the $N\rightarrow\infty$ limit.
First note that if  
\begin{equation}
\lambda\lesssim N^{-2},
\end{equation}
then the average meta-degree $\avg{k_\T{meta}}\sim\lambda N^2$ tends to zero, meaning that physicality will have a diminishing effect and almost all links can be added to the physical network.
On the other hand if 
\begin{equation}
\lambda\gtrsim N^{-1/3},
\end{equation}
then the total volume of the nodes $V_\T{nodes}\sim \lambda^3 N$ will exceed the total available volume.
We introduce the parameter $\alpha$ to interpolate between the two limiting cases:
\begin{equation}\label{eq:alpha}
\lambda = C\cdot N^{-\alpha}.
\end{equation}

\subsubsection{Asymptotic scaling of the average meta-degree}

The definition of the $\alpha$ exponent ensures that the average meta-degree $\avg{k_\T{meta}}$ remains positive in the large network limit for $\alpha<2$.
Equations~(\ref{eq:Nmeta_fixed_N}) and (\ref{eq:Mmeta_fixed_N}) allow us to calculate the asymptotic scaling of $\avg{k_\T{meta}}$ for any $\alpha$.

We start with calculating $\avg{k_\T{meta}}$ for LPNs including node-link interactions.
We found in Eq.~(\ref{eq:Nmeta_fixed_N}) that for fixed $N$, the number of meta-vertices is constant with an exponential cutoff at $\lambda\sim N^{-1/2}$.
The characteristic length of the exponential cutoff is $(\lambda^2 N)^{-1}\sim N^{2\alpha-1}$.
For $\alpha>1/2$, this characteristic length tends to infinity in the large network limit, hence it becomes irrelevant and $N_\T{meta}\sim N^2$.
For $\alpha< 1/2$, however, the characteristic length tends to zero and the exponential cutoff matters.
For small $l$, the cube line picking distribution is $p_\T{LC}(l)\sim l^2$, leading to $N_\T{meta}\sim N^{6\alpha-1}$.
Therefore, if the LPN includes node-link interactions the meta-graph has
\begin{equation}\label{eq:Nmeta_scaling}
N_\T{meta}\sim
\begin{cases}
N^2 & \T{for } \alpha\geq 1/2,\\
N^{6\alpha-1} & \T{for } \alpha< 1/2\\
\end{cases}
\end{equation}
vertices.

Following similar considerations as for $N_\T{meta}$, we find that in the $N\rightarrow\infty$ limit the number of meta-edges scales as
\begin{equation}\label{eq:Mmeta_scaling}
M_\T{meta}\sim
\begin{cases}
N^{4-\alpha} & \T{for } \alpha\geq 1/2,\\
N^{11\alpha-2} & \T{for } \alpha< 1/2.\\
\end{cases}
\end{equation}
We obtain the scaling of the average meta-degree by combining Eqs.~(\ref{eq:Nmeta_scaling}) and (\ref{eq:Mmeta_scaling}):
\begin{equation}\label{eq:kmeta_scaling}
\avg{k_\T{meta}}\sim
\begin{cases}
N^{2-\alpha} & \T{for } \alpha\geq 1/2,\\
N^{5\alpha-1} & \T{for } \alpha< 1/2.\\
\end{cases}
\end{equation}

If, as in the main text, we only consider link-link interactions, the exponential cutoff in Eqs.~(\ref{eq:Nmeta_fixed_N}) and (\ref{eq:Mmeta_fixed_N}) disappears and relying on Eq.~(\ref{eq:Nmeta_Mmeta_fixed_N_link-link}) we obtain
\begin{equation}\label{eq:kmeta_fixed_N_link-link}
\avg{k_\T{max}}\sim N^{2-\alpha}.
\end{equation}

\subsection{Onset of physicality and the jammed state}

There are two central quantities that we study in the $N\rightarrow\infty$ limit: the onset of physicality and the jammed state.
In the following we define these quantities and describe their relation to similar concepts.

\subsubsection{Onset of physicality}\label{sec:Mphys_def}

In the initial steps of adding links to an LPN, links are unlikely to get rejected due to physical conflicts, while near the end of the process only a small portion are successfully added~(Fig.~\ref{fig:jamming_illust}a).
To characterize this transition between non-physical and physical stages, we define the the onset of physicality $M_\T{phys}$ as the number of links above which new links are rejected with finite probability.

This definition, however, is only useful in the $N\rightarrow\infty$ limit, since in finite systems link rejection always happens with finite probability; therefore we are interested in the scaling of $M_\T{phys}$ for large $N$.
In order to measure $M_\T{phys}$ in finite simulations, we measure $M_\T{phys}$ as the number of links above which at least $1-\phi$ fraction of the links are rejected at any time, i.e.,
\begin{equation}\label{eq:Mphys_finite_def}
M_\T{phys} = \max_M \left\{M:\ \frac{t-M(t)}{t}>1-\phi \quad\forall M(t) > M\right\},
\end{equation}
where $t=1,2,\dots,N(N-1)/2$ is the number of link candidates considered and $M(t)$ is the number of links successfully added upto time $t$.

\subsubsection{Jammed state}

We define the jammed state as an LPN with maximal number of links, i.e., a network where no more links can be added without violating physicality. 
The jammed state depends on the algorithm we use to add links to the network, in this paper we are interested in the jammed state reached by adding links in random order (Sec.~\ref{sec:generating_LPNs}). 
The number of links in this jammed state $M_\T{max}$ is not equal to the global maximum of $M$.
To find the global maximum, one can find the maximum independent set in the meta-graph.
However, the maximum independent set is an NP-complete problem in general graphs, suggesting that it is also difficult to characterize in meta-graphs~\cite{karp1972reducibility}.

Note that jammed state of LPNs is related to, but distinct from the jammed states studied in hard particle packings (HPP).
In the latter, jammed states refer to maximal packings, such that the particles touch, while maximal packings where there is a gap between the particles are called saturated.
Jammed and saturated states differ from each other in mechanical properties, such as rigidity and response to stress~\cite{torquato2010jammed}.
The generation of LPNs is similar to a classical hard rod packing problem with some important differences: (i)~the length of the links (or rods) is heterogeneous, (ii)~not all rod positions are considered, only links that connect a predefined set of nodes, and (iii)~the links  allow to overlap if they are connected to the same node.
In jammed LPNs, links typically do not touch unless they are connected to the same node, similar to the saturated state in HPP.
However, the fact that links are connected to each other at their endpoints makes jammed LPNs similar to jammed HPP with respect to some properties. 

\subsection{Behavior of LPNs}

\begin{figure}[h]
	\centering
	\includegraphics[width=1.\textwidth]{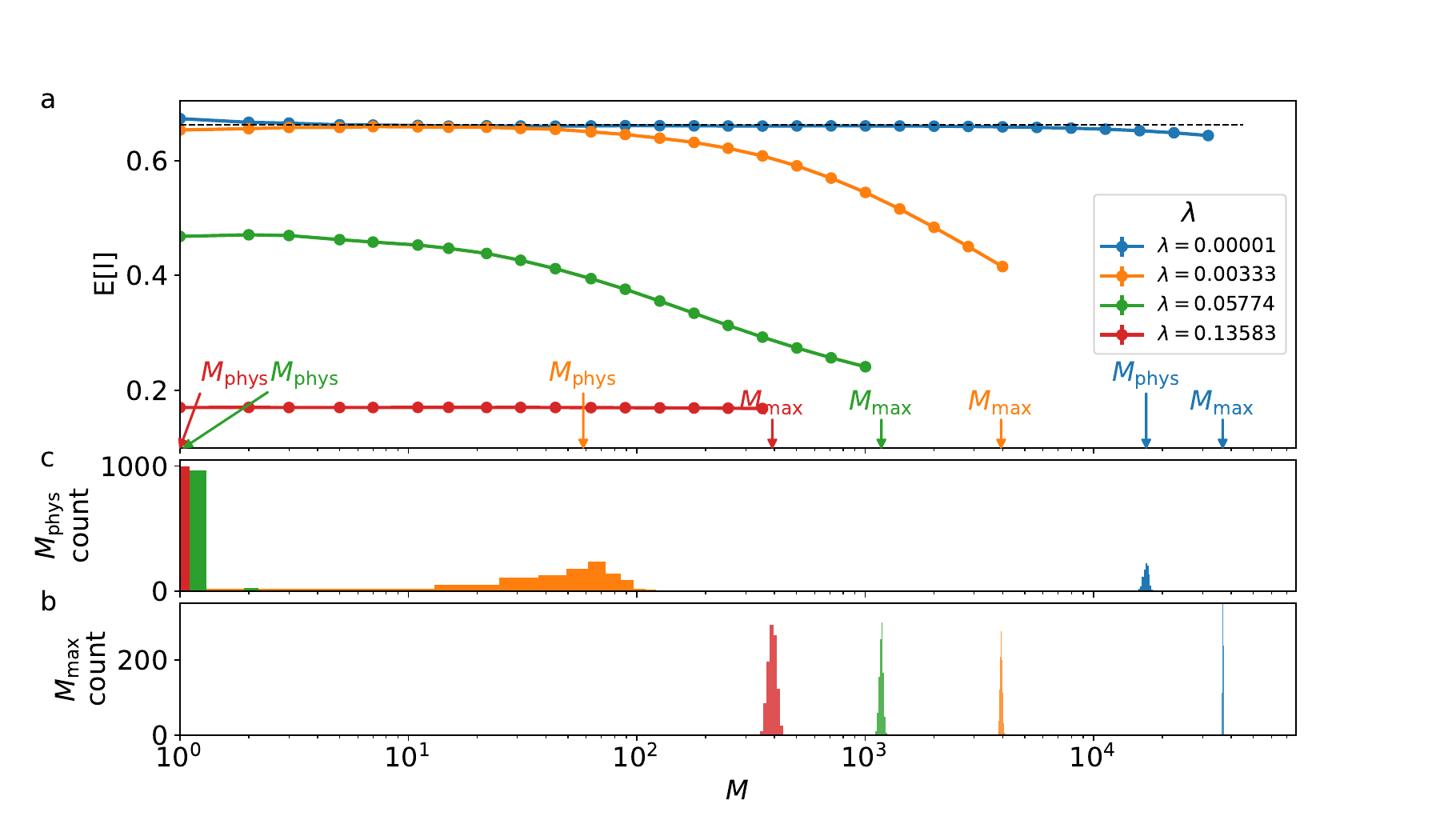}
	\caption{{ \bf Behavior LPNs with node-node, node-link and link-link interactions.} (a)~The expected link length $\avg{l(M)}$ after the addition of $M$ links for $\lambda = N^{-\alpha}$ where $\alpha=2$, $1$, $0.5$ and $0.35$. For $\alpha=2$ and $1$, initially $\avg{l(M)}$ is equal to the random expectation, i.e., the expected length of two randomly selected points in the unit cube (dashed line). At the onset of physicality $M_\T{phys}$ the expected link length decreases compared to this random expectation. For $\alpha=0.5$ and $0.35$, physicality even affects the first link due to node-link interactions. The process halts at the jammed state $M_\T{max}$, after which no more links can be added without violating physicality. We measure $\avg{l(M)}$ for networks with $N=300$ nodes, and data points represent an average of 1,000 independently generated networks. (b,c)~Histograms of $M_\T{phys}$ and $M_\T{max}$ for the 1,000 independent realizations. The onset of physicality $M_\T{phys}$ is calculated using Eq.~(\ref{eq:Mphys_finite_def}).}
	\label{fig:jamming_illust}
\end{figure}

To illustrate the process of generating an LPN, we measure the expected link length after the addition of $M$ links as
\begin{equation}
\avg{l(M)}= \frac{1}{M}\sum_{i\leq M} l_i,
\end{equation}
where $l_i$ is the length of $i$th the link added to the network.
Figure~\ref{fig:jamming_illust}a shows $\avg{l(M)}$ for $\lambda=N^{-2}$, $N^{-1}$, $N^{-0.5}$ and $N^{-0.35}$ for LPNs including node-node, node-link and link-link interactions, while Fig.~1g in the main text shows the same for LPNs with link-link interaction only.
Overall, we observe that $\avg{l(M)}$ decreases as we add links to the network: as more links are present the likelihood that long link candidates overlap with an existing link increases and therefore get rejected increases, descreasing $\avg{l(M)}$.
The process halts after the addition of $M_\T{max}$ links, after which no more links can be added.

The key difference between LPNs with all physical interactions and LPNs with link-link interactions that we observe is the onset of physicality $M_\T{phys}$, i.e., the $M$ value where $\avg{l(M)}$ differs from the non-physical expectation.
For LPNs with only link-link interactions, the initial links are not affected by physicality and $M_\T{phys}>1$ for any $\lambda$.
However, including node-link interactions can even affect the placement of the first link.
Without physicality, the typical link length of the first link is proportional to the side length of the cube, i.e., $l\sim 1$.
The probability that a link overlaps with a node is proportional to the volume of the link $\sim l \lambda^2$.
Assuming that there are $N$ nodes in the network and $\lambda\sim N^{-\alpha}$, the expected number of nodes the first link overlaps with is
\begin{equation}
\sim N \cdot N^{-2\alpha},
\end{equation}
meaning that if $\alpha>1/2$ physicality reduces the length of the first link in the $N\rightarrow \infty$ limit, i.e., $M_\T{phys} = 1$.
We investigate the scaling of $M_\T{phys}$ in more detail in Sec.~\ref{sec:rand_indep_set}.

\clearpage

\section{Rigorous lower bounds on $M_\T{max}$}\label{sec:Mmax_bounds}

The mapping  between $\lambda$-physical networks and the independent sets of the meta-graph introduced in Sec.~\ref{sec:meta-graph} allows us to formulate simple rigorous lower bounds for $M_\T{max}$.
A point set $\s P$ is in one-to-one correspondence with an independent set of $\s M(\s P, \lambda)$. 
Adding edges to the physical network in random order is equivalent to building independent sets using the following randomized greedy algorithm: Let $\s G=(\s V,\s E)$ be an abstract graph.
If an ordering $\sigma$ is given on the vertices of $\s G$ then there is a corresponding independent set $\s I_\sigma$ in $\s G$ which is produced in a recursive way.
The vertex with the smallest index is automatically included in $\s I_\sigma$. We make a decision for each vertex in the order $\sigma$. Assume that this decision has been made up to index $i$.
Then we include $v\in \s V$ with $\sigma(v)=i+1$ in  $\s I_\sigma$ if and only if no neighbor of $v$ with smaller index is included.

\begin{figure}[t!]
	\centering
	\includegraphics[width=1.\textwidth]{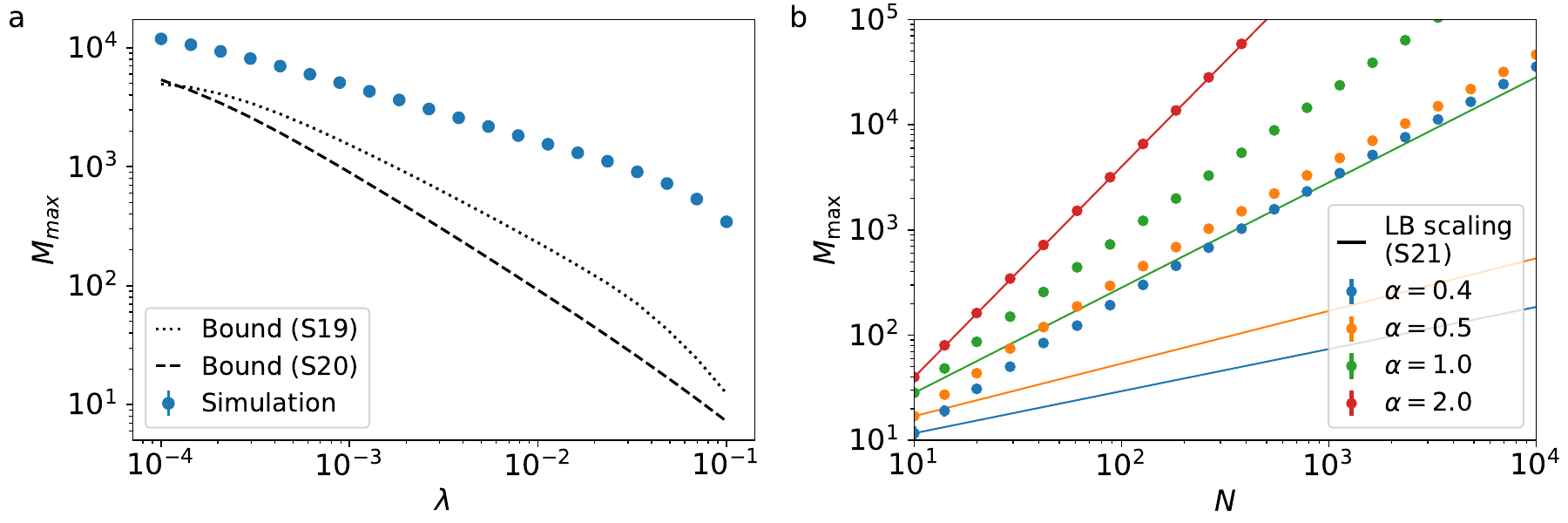}
	\caption{{ \bf Lower bounds for $M_\T{max}$.} (a)~We numerically generate LPNs in the jammed state for various $\lambda$ and we measure $M_\T{max}$ together with its lower bounds provided by Eqs.~(\ref{eq:lowerbound1}) and (\ref{eq:lowerbound2}). We find that the lower bounds typically underestimate the order of magnitude of $M_\T{max}$. Both bounds are the tightest for small $\lambda$. The markers represent the average of 10 independent LPNs with $N=200$ nodes. (b)~We compare the scaling of the lower bound provided by Eq.~(\ref{eq:lower_bound_scaling}) to the numerical measurements of $M_\T{max}$ with varying $N$ and setting $\lambda=N^{-\alpha}$. Finite size simulations suggest that, with the exception of $\alpha=2$, the lower bounds do not capture the asymptotic scaling of $M_\T{max}$ in the large network limit.}
	\label{fig:lower_bounds}
\end{figure}

It is known that for any graph the expected size of the independent set has the lower bound
\begin{equation}\label{eq:lowerbound_general}
\mathbb{E}(\lvert\mathcal I_\sigma\rvert)\geq\sum_{v\in \mathcal V}\frac{1}{k_v+1},
\end{equation}
where $k_v$ is the degree of $v$ and $\mathbb{\cdot}$ denotes expectation. 
The expected maximum number of physical links in $\mathcal G(\mathcal P,\lambda)$ is equal to the expected size of independent sets in $\mathcal M(\mathcal P,\lambda)$.
 Thus we have 
\begin{align}
\mathbb{E}(M_\T{max})\geq & \sum_{e\in \mathcal M(\lambda,\mathcal P)} \frac{1}{1+k_e}\geq \label{eq:lowerbound1}\\ 
\geq & N_\T{meta}\Bigl(1+\avg{k_\T{meta}}\Bigr)^{-1},\label{eq:lowerbound2}
\end{align}
where $k_e$ is the meta-degree of link candidate $e$, and the second inequality is Jensen's inequality.
Figure~\ref{fig:lower_bounds}a compares the actual $M_\T{max}$ to the lower bounds, all measured on numerically generated meta-graphs.
We have to restrict ourselves to networks of at most a few hundred nodes for which we can build the meta-graph explicitly.
For these networks we find that both lower bounds are the tightest for very dense and sparse physical networks; however, for intermediate values the bounds are rather poor, and both significantly contribute to underestimating $M_\T{max}$.

Does the lower bound become tight for large networks?
To find an answer we explore the scaling of the lower bound in the $N\rightarrow \infty$ limit setting $\lambda=N^{-\alpha}$.
In Sec.~\ref{sec:large_N}, we derived the asymptotic scaling of $N_\T{meta}$ and $\avg{k_\T{meta}}$, substituting into the lower bound~(\ref{eq:lowerbound2}) provides the scaling of $M_\T{max}$.
For both strict LPNs with node-link and link-link interactions and the more permissive LPNs with only link-link interactions, we obtain
\begin{equation}\label{eq:lower_bound_scaling}
\mathbb{E}(M_\T{max})\gtrsim \begin{cases}
N^{2} & \T{for } \alpha\geq 2,\\
N^{\alpha} & \T{for } \alpha< 2\\
\end{cases}.
\end{equation}
The fact that the asymptotic behavior of the lower bound does not depend on the presence of node-link interactions hints that such interactions become irrelevant to the jammed state in the large network limit.
To compare the predicted scaling of the lower bound, we generate networks with increasing $N$ while keeping $\lambda N^\alpha$ constant, this ensures that $\lambda$ remains on the order of $N^{-\alpha}$.
Figure~\ref{fig:lower_bounds}b shows that for $\alpha<2$ the bound is not tight and provides apparently inaccurate scaling.

\clearpage

\section{Random independent sets}\label{sec:rand_indep_set}

In this section we calculate $n_\T{IS}$ the expected fraction of vertices in the maximal greedy independent set (IS) for a class of random networks relying on a differential equation formalism~\cite{brightwell2017greedy,krivelevich2020greedy}, the fraction $n_\T{IS}=N_\T{IS}/N$ is also known as the expected greedy independence ratio or jamming ratio~\cite{krivelevich2020greedy, finch2003mathematical}. 

More specifically we consider networks with $N$ vertices such that each vertex pair $(u,v)$ is connected independently with probability
\begin{equation}
p(u,v)=w_uw_v,
\end{equation}
where $w_v$ is a prescribed weight of vertex $v$, either provided by a deterministic sequence or drawn from some distribution $p(w)$.
The expected degree of vertex $v$ is therefore $\mathbb{E}(k(v))=w_v\sum_{u\neq v} w_u$ or $\mathbb{E}(k(v))=w_v\int wp(w)dw$.

The randomized greedy construction of ISs works by placing the nodes in a random order $\sigma$, then sequentially in this order adding them to the IS whenever possible.
Meaning that we can add the $t$th node to the independent set if none of its neighbors have been added to the IS before $t$, the probability of this is
\begin{equation}
P(t\in \mathcal I_\sigma) = \prod_{s<t,s\in I_\sigma}(1-w_t w_s)\approx
 \exp\left(-w_t\sum_{s<t,s\in I_\sigma}w_s\right)=\exp(-w_t W_\T{IS}(t)),
\end{equation}
where the approximation follows from
\begin{equation}
\log P(t\in \mathcal I_\sigma) = \sum_{s<t,s\in I_\sigma}\log(1-w_t w_u) = -\sum_{s<t,s\in I_\sigma}w_t w_u + O\left((w_t w_u)^2\right),
\end{equation}
and we introduced $W_\T{IS}(t)$, the total weight of nodes in the IS before $t$.

To obtain the time evolution of $W_\T{IS}(t)$, we substitute it with its expectation value over all possible $\sigma$ orders:  
\begin{equation}\label{eq:disc_W_IS}
W_\T{IS}(t+1)=W_\T{IS}(t)+ \int w\exp(-w W_\T{IS}(t))p(w)dw,
\end{equation}
with initial condition $W_\T{IS}(0)=0$.
Similarly the expected number of nodes in the IS before $t$ is given by
\begin{equation}
N_\T{IS}(t+1)=N_\T{IS}(t)+ \int \exp(-w W_\T{IS}(t))p(w)dw,
\end{equation}
with initial condition $N_\T{IS}(0)=0$.
The final expected size of the IS is
\begin{equation}
\lvert\mathcal{I}_\sigma\rvert=N_\T{IS}(N).
\end{equation}
It can be useful to transform the equations by taking the continuous time limit using $\tau=t/N$:
\begin{align}
\frac{1}{N}\dot{W}_\T{IS}(\tau)&=\int w\exp(-w W_\T{IS}(\tau))p(w)dw,\label{eq:cont_W_IS}\\
\frac{1}{N}\dot N_\T{IS}(\tau)&=\int \exp(-w W_\T{IS}(\tau))p(w)dw.\label{eq:cont_N_IS}
\end{align}
These equations are further simplified as
\begin{align}
\frac{1}{N}\dot{W}_\T{IS}(\tau)&=F^\prime\left(-W_\T{IS}(\tau)\right),\label{eq:WIS_general}\\
\frac{1}{N}\dot N_\T{IS}(\tau)&=F\left(-W_\T{IS}(\tau)\right),\label{eq:NIS_general}
\end{align}
where $F(z)=\int_0^\infty \exp(-wz)p(w)dw$ is the moment generating function of the weight distribution $p(w)$.

\subsubsection{Example: Erd\H{o}s-R\'enyi model}

\begin{figure}[t!]
	\centering
	\includegraphics[width=1.\textwidth]{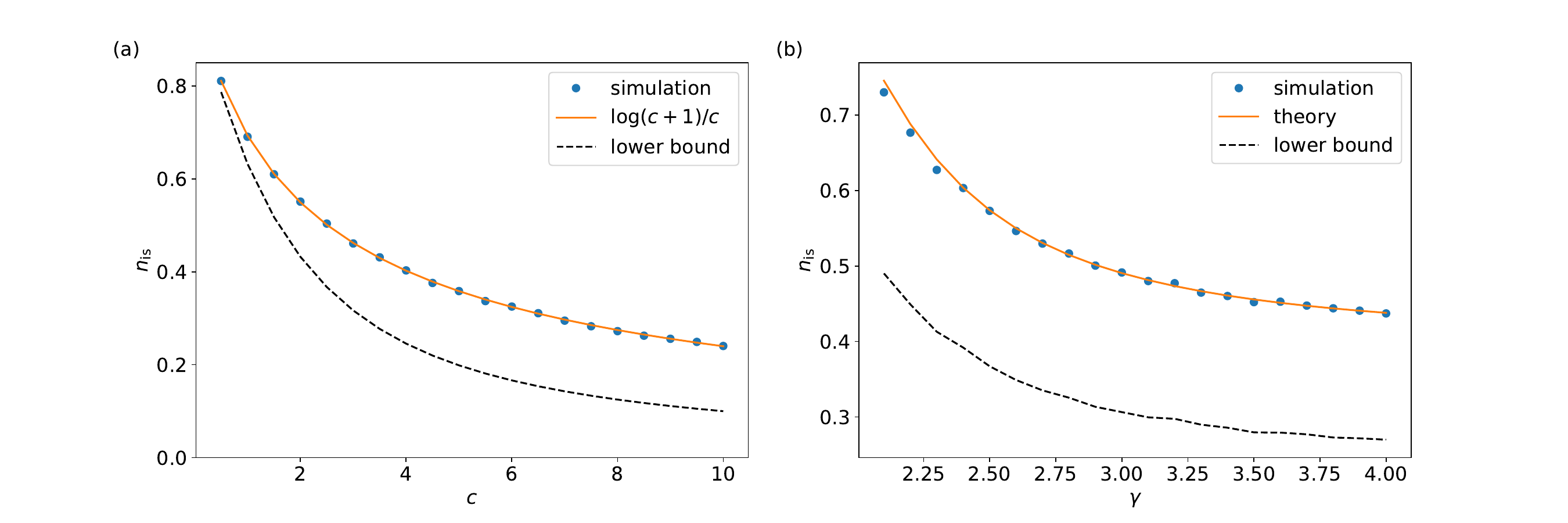}
	\caption{{ \bf Size of independent sets in random networks with fixed expected degree.} We compare the theoretical prediction of the fraction of nodes in the greedy maximal independent sets $n_\T{IS}=N_\T{IS}/N$ for random model networks finding an excellent agreement. (a)~Example 1: Erd\H os-R\'enyi networks. (b)~Scale-free random networks. Markers indicate simulations for networks with Erd\H os-R\'enyi $N=10,000$ nodes and scale-free random networks with $N=1,000$ nodes and average degree $c=2$. The continuous orange line is the solution of Eqs.~(\ref{eq:WIS_general}) and (\ref{eq:NIS_general}), and the dashed line is the lower bound~(\ref{eq:lowerbound_general}).  }
	\label{fig:model_indepset}
\end{figure}

Choosing $w_v\equiv \sqrt{p}$ we recover the Erd\H{o}s-R\'enyi model. Substituting to Eqs.~(\ref{eq:cont_W_IS}) and (\ref{eq:cont_N_IS}) we obtain
\begin{equation}\label{eq:ER_IS}
\dot{n}^\T{ER}_\T{IS}(\tau)=\exp(-c n^\T{ER}_\T{IS}(\tau)),
\end{equation}
where $n_\T{IS}(\tau)=N_\T{IS}(\tau)/N$ and $c=Np$ is the average degree of the network. Together with the initial condition $n_\T{IS}(0)=0$, the solution of Eq.~(\ref{eq:ER_IS}) is
\begin{equation}\label{eq:ER_nIS}
n^\T{ER}_\T{IS} = \frac{\log(c+1)}{c}.
\end{equation}
Figure~\ref{fig:model_indepset}a compares this solution to simulations showing excellent agreement.
It is interesting to compare the solution to the lower bound obtained from Eq.~(\ref{eq:lowerbound2})
\begin{equation}
n^\T{ER}_\T{lb} = \sum_{k\geq 0}\frac{1}{k+1}\frac{c^k}{k!}e^{-c} = \frac{1}{c}(1-e^{-c}).
\end{equation}
The solution in Eq.~(\ref{eq:ER_nIS}) and the lower bound have the same asymptotic behavior for large $c$ (for example, if $c\sim N^\alpha$); however, this behavior might not be visible in simulations due to the logarithmic correction in Eq.~(\ref{eq:ER_nIS}).

\subsubsection{Example: power law degree distribution}

Choosing 
\begin{equation}
w_i\sim i^{-\alpha}
\end{equation}
($i=1,2,\ldots,N$) generates a network that has a degree distribution with a power law tail $p(k)\sim k^{-\gamma}$ where $\gamma=1+\frac{1}{\alpha}$; and normalizing $w_i$ such that $\sum_i w_i = \sqrt{cN}$ sets the average degree to $c$~\cite{aiello2001random}.
Substituting $p(w)=\frac{1}{N}\sum_i\delta(w_i)$ into Eq.~(\ref{eq:disc_W_IS}) and numerically solving it we obtain the heuristic estimate for $n^\T{pl}_\T{IS}$.
Figure~\ref{fig:model_indepset}b compares this prediction to simulations again showing excellent agreement.

\subsection{Approximating the meta-graph}

\begin{figure}[h]
	\centering
	\includegraphics[width=1.\textwidth]{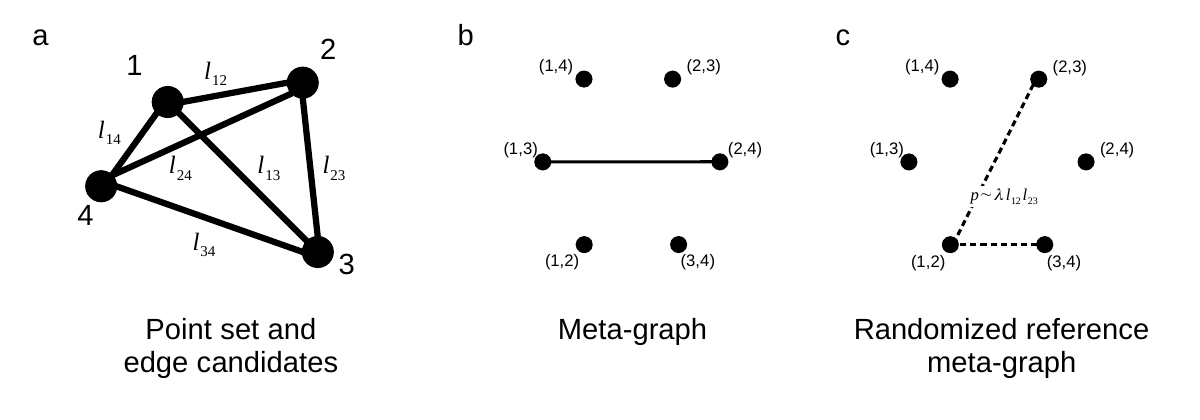}
	\caption{{ \bf Randomized reference meta-graph.} The randomized reference graph can be interpreted as a meanfield version of the original meta-graph such that the length of the links are preserved. The reference graph is defined on the same set of nodes as the meta-graph but two nodes are connected with the probability equal to the probability of two randomly placed links intersecting with the same length.}
	\label{fig:approx_meta}
\end{figure}

The equations derived in this section allow us to analytically estimate $M_\T{max}$, the maximum number of edges that we can randomly add to a $\lambda$-physical network.
The approach is that we approximately model the meta-graph with a random graph where the probability that two meta-nodes are connected is equal to the probability that two random segments of the same length are at most $\lambda$ distance from each other; therefore, the connection probability only depends on the length of the corresponding link candidates (Fig.~\ref{fig:approx_meta}).

The probability that two randomly chosen segments in the unit cube are at most $\lambda$ distance from each other is approximately
\begin{equation}
\frac{\pi}{2}\lambda l_1 l_2,
\end{equation}
where $l_1$ and $l_2$ are the lengths of the segments and the approximation ignores boundary effects and the inhomogeneity of the unit cube.
Therefore, to approximate the meta-graph, we choose
\begin{equation}\label{eq:metanode_weight}
w_v = l_v\sqrt{\frac{\pi}{2}\lambda}.
\end{equation}
To calculate $M_\T{max}=N_\T{IS}$, we need to know $p(w)$ or equivalently the length distribution of the link candidates $p_\T{LC}(l)$.
The distribution $p_\T{LC}(l)$ for large $N$ converges to the distribution of the distance between two randomly selected points from unit cube, sometimes called cube line picking.
The cube line picking has a known explicit, although complicated form, and its mean is $\avg l\approx 0.662$ is the Robbins constant.

For LPNs with node-link interactions, a point pair is included as a node in the meta-graph, if the segment connecting the two points is at least $\lambda$ distance away from all other points.
Equation~(\ref{eq:node_link_intersect_prob}) shows that the probability that a segment with length $l$ is at least $\lambda$ distance away from $N-2\approx N$ random points in the unit cube is approximately $\exp(-\pi l \lambda^2 N)$.
Using these observations we adopt Eq.~(\ref{eq:cont_W_IS}) to calculate $L_\T{IS}(\tau)$, the total length of the meta-nodes in the independent set:
\begin{equation}\label{eq:Lis}
\dot L_\T{IS}(\tau) = \frac{N^2}{2}\int^{\sqrt{3}}_0 l\exp\left[-\frac{\pi}{2}\lambda L_\T{IS}(\tau)l-\pi\lambda^2Nl\right]p_\T{LC}(l)dl,
\end{equation}
where $\tau=2t/N^2$.
Note that the first term in the exponent corresponds to the link-link interaction and the second term to the node-link interaction.
Similarly, we obtain the number of meta-nodes in the independent set $N_\T{IS}(\tau)$ by integrating
\begin{equation}\label{eq:Nis}
\dot N_\T{IS}(\tau) = \frac{N^2}{2}\int^{\sqrt{3}}_0 \exp\left[-\frac{\pi}{2}\lambda L_\T{IS}(\tau)l-\pi\lambda^2Nl\right]p_\T{LC}(l)dl.
\end{equation}
For LPNs with only link-link interactions, the second term in the exponent of Eqs.~(\ref{eq:Lis}) and (\ref{eq:Nis}) is left out.

\begin{figure}[t!]
	\centering
	\includegraphics[width=1.\textwidth]{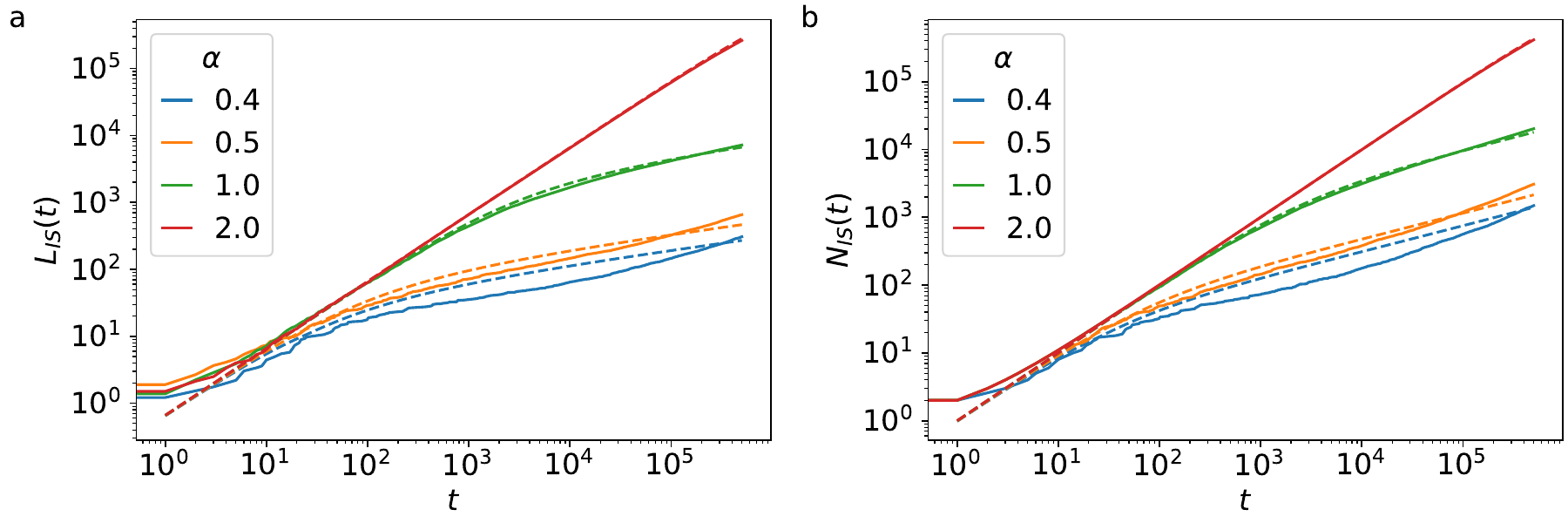}
	\caption{{ \bf Time evolution of }$L_\T{IS}(t)${\bf and }$N_\T{IS}(t)${\bf .} We numerically generate a single instance of LPN with only link-link interactions for various $\alpha$ and $N=1000$ and we measure the total link length $L_\T{IS}(t)$ and the number of links $N_\T{IS}=M(t)$ in the independent set over time. We compare $L_\T{IS}(t)$ and $N_\T{IS}(t)$ to the theoretical prediction of Eqs.~(\ref{eq:Lis}) and (\ref{eq:Nis}), and we find that despite the random approximation of the meta-graph, the theory describes the simulations well on a log scale.}
	\label{fig:theory_vs_simluation}
\end{figure}

To test the accuracy of the theoretical approximation, we simulate LPNs with only link-link interactions with $N=1000$ nodes and compare the numerically observed time evolution of $L_\T{IS}(t)$ and $N_\T{IS}(t)$ to the predictions of Eqs.~(\ref{eq:Lis}) and (\ref{eq:Nis}).
Figure~\ref{fig:theory_vs_simluation} shows that despite the random approximation of the meta-graph, the theory well approximates $L_\T{IS}(t)$ and $N_\T{IS}(t)$ and accurately captures their order of magnitude.

In the following we extract the asymptotic scaling of the onset of physicality $M_\T{phys}$ and the jammed state $M_\T{max}$ in the $N\rightarrow \infty$ limit while setting $\lambda = N^{-\alpha}$.

\subsubsection{Onset of physicality}

In Sec.~\ref{sec:Mphys_def}, we defined $M_\T{phys}$ in the large network limit as the number of links above which physical links get rejected with finite probability.
As long as $M(t)<M_\T{phys}$ all links get accepted, meaning that $M(t)=t$ (where $t=\tau N^2/2$) and the typical link length is of the order of the size of the unit cube, i.e., $\avg {l(t)}\sim 1$.
For links to get rejected with finite probability, the exponent in Eq.~(\ref{eq:Lis}) must not converge to zero at $M(t)=M_\T{phys}$ as $N\rightarrow\infty$.
The exponent has two terms, the first $\sim \lambda L_\T{IS}l$ corresponds to link-link interactions and the second $\sim \lambda^2 N l$ to node-link interactions.
The longest links are the most likely to get rejected; therefore we investigate the convergence of these two terms for $l\sim 1$.

If $\alpha\leq 1/2$, the node-link interaction term
\begin{equation}
\lambda^2 N\sim N^{1-2\alpha}
\end{equation}
becomes non-zero in the large network limit.
This means that long links are rejected due to overlap with isolated nodes already at the beginning of the generation of the LPN, and $M_\T{max}=1$.

If $\alpha> 1/2$, the node-link interaction term becomes irrelevant and the link-link interaction term
\begin{equation}
\lambda L_\T{IS}\sim N^{-\alpha}M_\T{max}
\end{equation}
dominates, which becomes non-zero in the large network limit after the addition of $M_\T{max}\sim N^\alpha$ links.

In summary, for LPNs with both node-link and link-link interactions the onset of physicality happens at
\begin{equation}\label{eq:Mphys_scaling_nodelink}
M_\T{phys}\sim
\begin{cases}
1 & \T{for } \alpha\leq 1/2\\
N^\alpha & \T{for } \alpha> 1/2.
\end{cases}
\end{equation}
For LPNs with only link-link interactions only, the node-link interaction term is missing and the onset of physicality happens at
\begin{equation}
M_\T{phys}\sim N^\alpha
\end{equation}
for any $\alpha$.

To test the predicted scaling, we generate LPNs with increasing $N$ while setting $\lambda=N^{-\alpha}$ and we measure $M_\T{phys}(N)$ using the definition (\ref{eq:Mphys_finite_def}).
Figure~\ref{fig:nodelink_scaling}a compares the numerical measurements of $M_\T{phys}(N)$ to the predicted scaling for LPNs with both node-link and link-link interactions, while Fig.~2d in the main text compares it for LPNs with only link-link interactions.
For both cases, we find that simulations are consistent with the predictions.

\subsubsection{The jammed state}\label{sec:jammed_state_LIS_Mmax}

We reach the jammed state after we exhausted all link candidates corresponding to time $\tau=1$.
Therefore, to calculate the asymptotic scaling of number of links in the jammed state $M_\T{max}$, we need to solve Eqs.~(\ref{eq:Lis}) and (\ref{eq:Nis}) at $\tau=1$ for large $N$.
First we assume that for large $N$
\begin{equation}\label{eq:assump}
\frac{\pi}{2}\lambda L_\T{IS}(\tau)\gg\pi \lambda^2N=\pi N^{1-2\alpha},
\end{equation}
this clearly holds for $\alpha>1/2$, for $\alpha\leq 1/2$ we have to check the results for consistency. 
Our second assumption is that as $N$ increases $L_\T{IS}(\tau)\rightarrow \infty$ also; and therefore the typical length $l$ of the segments that are added to the independent set tend to zero over time, this is useful since as $l\rightarrow 0$ the segment length distribution becomes $p_\T{LC}(l)\rightarrow 4\pi l^2$.
Substituting this into Eq.~(\ref{eq:Lis}), we obtain
\begin{equation}\label{eq:LIS_near_jammed}
\dot L_\T{IS}(\tau) = 4\pi\frac{N^2}{2}\int^{\sqrt{3}}_0 l^3\exp\left[-\frac{\pi}{2}\lambda L_\T{IS}(\tau)l\right]dl.
\end{equation}
The integrand is sharply peaked at low $l$; we can, therefore, extend the upper bound of integration to $\infty$. With this step together with the change of variable $x=\frac{\pi}{2}\lambda L_\T{IS}(\tau)l$, we get
\begin{equation}
\dot L_\T{IS}(\tau) = 4\pi\frac{N^2}{2}\left(\frac{\pi}{2}\lambda L_\T{IS}(\tau)\right)^{-4}\int^{\infty}_0 x^3\exp\left[-x\right]dx=\frac{32}{3\pi^3}\frac{N^2}{\lambda^4L_\T{IS}(\tau)^4}\Gamma(4).
\end{equation}
By integration we obtain the solution
\begin{equation}\label{eq:LIS_solution_near_jamming}
L_\T{IS}(\tau) = \left(\frac{3\cdot 5\cdot 64}{\pi^3}\lambda^{-4}N^2 \tau \right)^{1/5};
\end{equation}
therefore the total length of the independent set at $\tau=1$ is
\begin{equation}\label{eq:LIS_scaling}
L_\T{IS}=L_\T{IS}(\tau=1)\sim \left(\lambda^{-4}N^2\right)^{1/5}=N^{\frac{4\alpha+2}{5}}.
\end{equation}
Note that for large $N$ the inequality
\begin{equation}
\lambda L_\T{IS}\sim N^{(2-\alpha)/5}>N^{1-2\alpha}
\end{equation}
 holds for $\alpha>1/3$; therefore the result is consistent with our initial assumptions for all possible $\alpha$ values.
This means that node-link interactions become irrelevant compared to link-link interactions in the large network limit $N\rightarrow\infty$.

Similarly the number of meta-nodes in the independent set $N_\T{IS}$ (or equivalently the maximal number of links in the physical network $M_\T{max}$) is
\begin{equation}\label{eq:Mmax_scaling}
N_\T{IS}\sim N^{\frac{3\alpha+4}{5}}.
\end{equation}

\begin{figure}[h!]
	\centering
	\includegraphics[width=.4\textwidth]{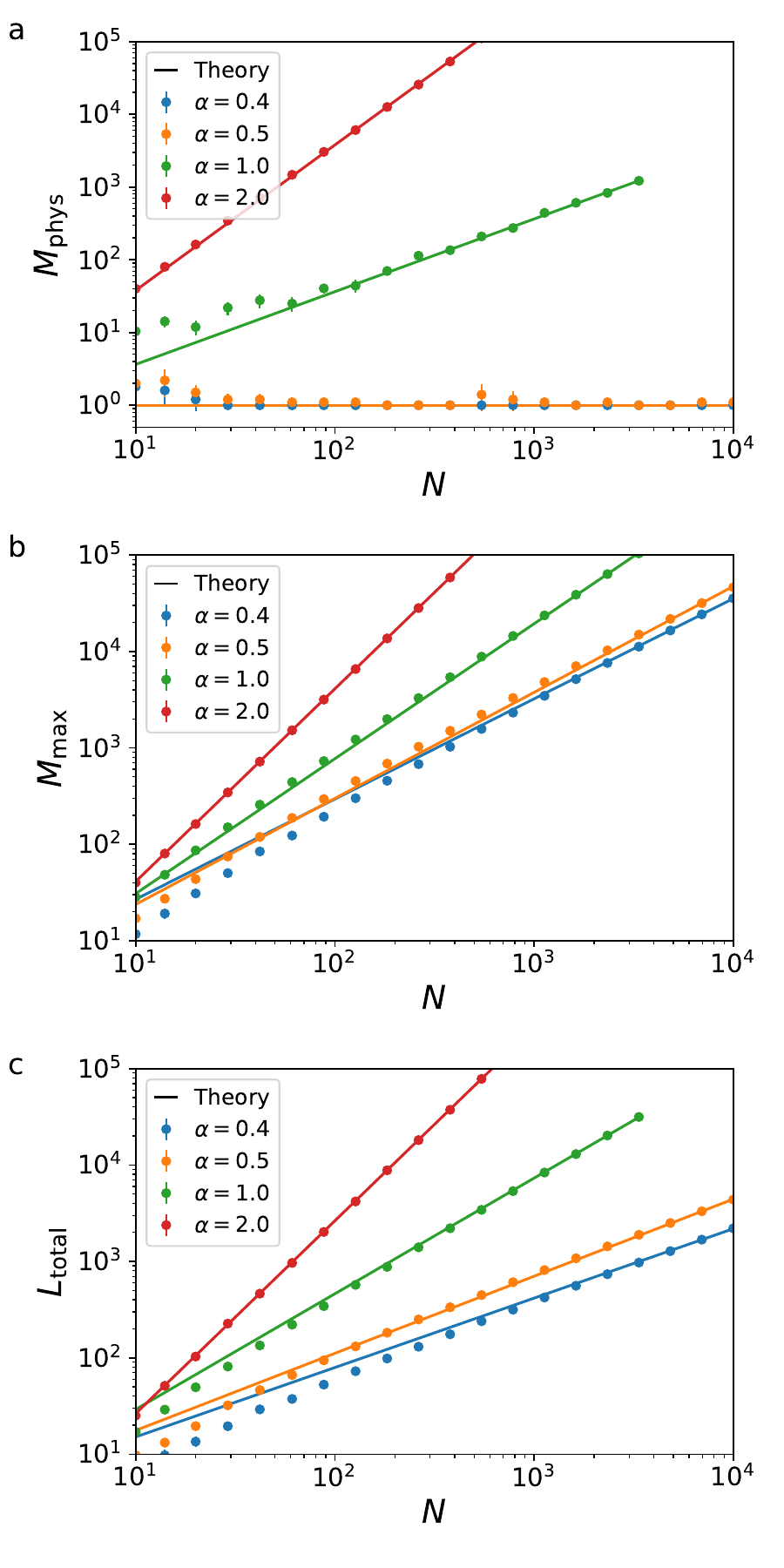}
	\caption{{ \bf Asymptotic scaling in LPNs with node-link and link-link interactions interactions.} We find that numerical measurements of $M_\T{phys}(N)$, $M_\T{max}(N)$ and $L_\T{total}(N)$ are consistent with their scaling predicted by theory in the large network limit. (a)~For the onset of physicality, Equation~(\ref{eq:Mphys_scaling_nodelink}) predicts that for $\alpha\leq 1/2$ node-link interactions affect the placement of the first links, i.e., $M_\T{max}=1$. While for $\alpha>1/2$, link-link interactions dominate and we predict the same scaling as in LPNs with only link-link interactions. (b,c)~Theory predicts that in the jammed state link-link interactions are the dominant physical interactions for any $\alpha>1/3$, and we predict the same scaling as in LPNs with only link-link interactions. Markers represent the average of 10 independently generated networks and the errorbars indicate the standard error of the mean. The slope of the solid lines corresponds to the predicted scaling exponent and the intercept is chosen by fitting the predicted scaling to the final 20\% of the data points.}
	\label{fig:nodelink_scaling}
\end{figure}

To test the predicted scaling, we generate jammed LPNs with increasing $N$ while setting $\lambda=N^{-\alpha}$ and we measure the number of links $N_\T{IS}=M_\T{max}$ and the total length of the links $L_\T{IS}=L_\T{total}$.
Figures~\ref{fig:nodelink_scaling}b,c compare the numerical measurements to the predicted scaling for LPNs with both node-link and link-link interactions, while Figs.~2e,f in the main text compare it for LPNs with only link-link interactions.
For both cases, we find that simulations are consistent with the predictions.

\begin{figure}[h]
	\centering
	\includegraphics[width=1.\textwidth]{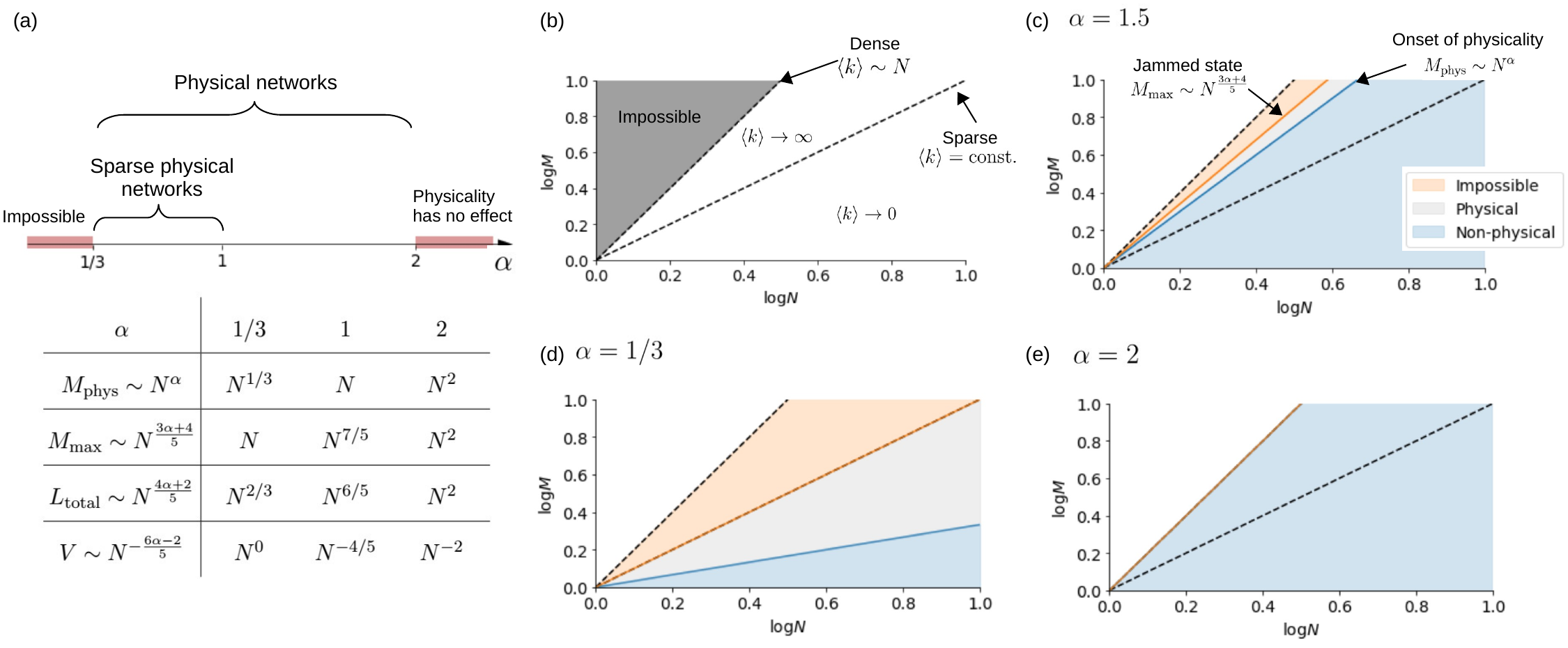}
	\caption{{ \bf Phase diagram.} (b)~The average degree $\avg k = 2M/N$ of an abstract network in the large $N\rightarrow \infty$ is determined by the scaling $M\sim N^\beta$, where $0\geq\beta\geq 2$. For example, if $M\sim N$ the average degree remains constant and the network is sparse, for $\beta>1$ the average degree diverges, and $M\sim N^2$ represents the dense limit, where $\avg k \sim N$. For $\beta<1$, the average degree tends to 0, meaning that in the  $N\rightarrow \infty$ limit almost all nodes are isolated. (c)~The effect of physicality depends on the physicality parameter $\lambda = N^{-\alpha}$ and the number of links added to the network. If $M\lesssim M_\T{phys}\sim N^\alpha$, than link-link physical interactions do not effect the evolution of the network, and physicality affects the global properties of the network only above $M_\T{phys}$ (blue area). We reach the jammed state at $M_\T{max}\sim N^{\frac{3\alpha+4}{5}}$, meaning that we cannot generate physical networks with $M\gtrsim M_\T{max}$ (orange area). (d)~If $\alpha=1/3$ the jammed state is sparse and networks with even $\avg k\rightarrow 0$ are affected by physicality. (e)~If $\alpha=2$, the scaling of both $M_\T{phys}$ and $M_\T{max}$ coincide with the dense network limit. }
	\label{fig:schematic_phases}
\end{figure}

Figure~\ref{fig:schematic_phases} summaries the possible asymptotic behavior of the LPNs with only link-link interactions.

\clearpage

\subsection{Fluctuations of $M_\T{phys}$, $M_\T{max}$ and $L_\T{total}$}

The quantities $M_\T{max}$, $M_\T{phys}$ and $L_\T{total}$ are affected by multiple random processes, such as the random positions of the $\mathcal P$ point set and the random order which we select the links for inclusion, and this complex dependence on the history of the network may to lead to non-self-averaging properties.
The analytical solution developed in Sec.~\ref{sec:rand_indep_set} relies on substituting $L_\T{total}$ with its expectation. The fact that the analytical predictions well approximate the numerical results suggests that $M_\T{phys}$, $M_\T{max}$ and $L_\T{total}$ are concentrated on their expectation values.
To provide further support for this, we derive the scaling of the variances of these quantities with network size $N$ for LPNs with link-link interactions.

\subsubsection{Onset of physicality}

Section~\ref{sec:Mphys_def} defines the onset of physicality $M_\T{phys}$ for finite size networks as the number of links above which at least $1-\phi$ fraction of the link candidates where rejected.
Here we estimate the scaling of $\sigma^2(M_\T{phys})$, i.e., the scaling of the variance of the numerically measured $M_\T{phys}$.
The probability of a link candidate with length $l$ to get accepted at time $t$ is
\begin{equation}\label{eq:acceptance_prob}
p(l,t)=\exp\left[-\frac{\pi}{2}\lambda L_\T{IS}(t)l\right].
\end{equation}
Before the onset of physicality all link candidates are accepted, meaning that the number of accepted links is equal to $t$ (i.e., $M(t)=t$) and the total length is $L_\T{IS}(t)=\avg l \cdot M(t)$, where $\avg l$ is the average length of a random segment in the unit cube.
The typical length of links near the onset of physicality is on the scale of the available volume; therefore probability of accepting a link candidate at $t$ becomes approximately
\begin{equation}\label{eq:acceptance_prob2}
p(M)\approx\exp[-c\lambda M],
\end{equation}
where $c$ is a constant and $M$ is the number of accepted links.
When we measure $M_\T{phys}$ we are in effect measuring $p(M)$, and $M_\T{phys}$ is found if the estimated $\bar p(M)$ falls below a predefined threshold $\phi$.
However, we can only estimate $p(M)$, i.e., we can only determine it up to an error $\bar p(M)= p(M)\pm\sigma(p)$.
The onset of physicality is then determined by
\begin{equation}
\phi = \exp[-c\lambda M_\T{phys}]\pm \sigma_p.
\end{equation}
Solving for $M_\T{phys}$ and keeping only first order terms of $\sigma_p$, we obtain
\begin{equation}\label{eq:Mphys_error}
M_\T{phys}=\frac{\ln(\phi^{-1})}{c}\cdot \lambda^{-1}\pm\frac{1}{c\phi}\lambda^{-1}\sigma_p.
\end{equation}
The error of numerically estimating the binomial proportion $p(M)$ is
\begin{equation}
\sigma_p \sim \sqrt{\frac{p(1-p)}{n}},
\end{equation}
where $n$ is the sample size.
In our case the sample size is on the scale of the number of links added to the network, i.e., $n\sim M_\T{phys}\sim N^\alpha$; therefore
\begin{equation}\label{eq:sigmap_scaling}
\sigma_p \sim N^{-\alpha/2}.
\end{equation}
From Eqs.~(\ref{eq:Mphys_error}) and (\ref{eq:sigmap_scaling}) we obtain that the variance of $M_\T{phys}$ scales as
\begin{equation}\label{eq:var_Mphys_scaling}
\sigma^2(M_\T{phys})\sim \left(\lambda^{-1}\sigma_p\right)^2\sim N^\alpha. 
\end{equation}
Figure~\ref{fig:MphysMmaxLtot_var}a compares the predicted scaling to numerical simulations.

\subsubsection{Jammed state}

The total link length $L_\T{total}$ is the sum of the length of the links $l(t)$ added at each time step,
where $l(t)$ is a random variable depending on $L_\T{total}(t)$, and $l(t)=0$ indicates that no link was added at $t$.
Throughout our derivation we assume that consecutive $l(t)$ variables only depend on $L_\T{total}(t)$ and are otherwise independent of each other.
Under the same assumption the variance of the sum of the link lengths $\sigma^2\left(L_\T{total}\right)$ is equal to the sum of the variances of $l(t)$.
We separate the contribution of the link candidates to $\sigma^2\left(L_\T{total}\right)$ into two terms
\begin{equation}
\sigma^2\left(L_\T{total}\right)=\sigma^2_1\left(L_\T{total}\right)+\sigma^2_2\left(L_\T{total}\right),
\end{equation}
corresponding to the evolution of the network before and after the onset of physicality, respectively.
Up to the onset of physicality $M_\T{phys}\sim N^\alpha$, all link candidates are accepted, hence the variance of $l(t)$ is a constant equal to the variance of $p_\T{LC}(l)$; therefore
\begin{equation}\label{eq:varL_prephys_contribution}
\sigma^2_1\left(L_\T{total}\right)\sim N^\alpha
\end{equation}

We showed in Eq.~(\ref{eq:LIS_solution_near_jamming}) that near the jammed state $L_\T{IS}(t)\sim \lambda^{-4/5}t^{1/5}$; therefore the probability of accepting a link with length $l$ becomes
\begin{equation}
\sigma^2_2\left(L_\T{total}\right)\approx\exp\left[-c \lambda \cdot \lambda^{-4/5}t^{1/5}l\right]=\exp\left[-c (\lambda t)^{1/5}l\right],
\end{equation}
where $c$ is a constant.
The contribution of the link candidates to $\sigma^2\left(L_\T{total}\right)$ after the onset of physicality is approximately
\begin{equation}\label{eq:varL_jam_contribution}
\begin{split}
\sigma^2_2\left(L_\T{total}\right)\approx \int_{N^\alpha}^{N^2/2} dt\left[ \left(\int l^2 \exp\left[-c (\lambda t)^{1/5}l\right] p_\T{LC}(l)dl\right) \right. - \\
\left. -  \left(\int l \exp\left[-c (\lambda t)^{1/5}l\right] p_\T{LC}(l)dl\right)^2 \right].
\end{split}
\end{equation}
The probability of accepting a link is positive only for short links in the large network limit, and for small $l$ the length distribution of link candidates is $p_\T{LC}(l)\sim l^2$ (Sec.~\ref{sec:jammed_state_LIS_Mmax}).
Substituting this into Eq.~(\ref{eq:varL_jam_contribution}) and introducing the new integration variable $z\sim  (\lambda t)^{1/5} l$, we obtain
\begin{equation}\label{eq:varL_jam_contribution2}
\sigma^2_2\left(L_\T{total}\right)\approx \int_{N^\alpha}^{N^2/2} dt c_1 (\lambda t)^{-1}-c_2\left[(\lambda t)^{-4/5}\right]^2,
\end{equation}
where $c_1$ and $c_2$ are constants.
Performing the integration, we find that $\sigma^2_2\left(L_\T{total}\right)\sim N^\alpha\ln N$ in leading order.
Together with Eq.~(\ref{eq:varL_prephys_contribution}) this predicts that the variance of the total link length at the jammed state scales as
\begin{equation}\label{eq:var_Ltotal_scaling}
\sigma^2\left(L_\T{total}\right)\sim N^\alpha\ln N.
\end{equation}

\begin{figure}[h]
	\centering
	\includegraphics[width=1.\textwidth]{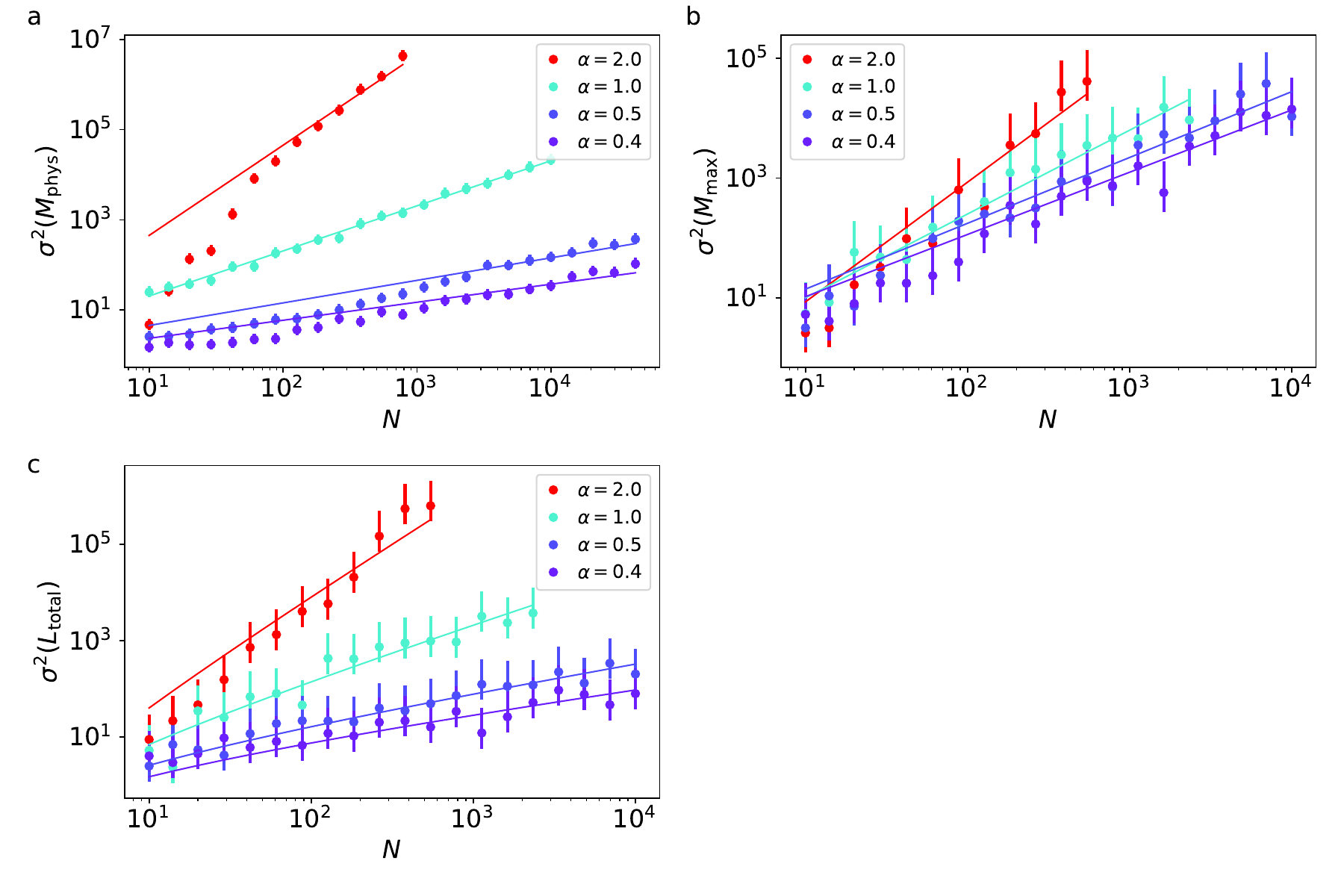}
	\caption{{ \bf Fluctuations of $M_\T{phys}$, $M_\T{max}$ and $L_\T{total}$.} We compare the numerically measured variances of $M_\T{phys}$, $M_\T{max}$ and $L_\T{total}$ (circles) to their predicted scaling (lines). The slope of the lines are provided by Eqs.~(\ref{eq:var_Mphys_scaling}), (\ref{eq:var_Mmax_scaling}) and (\ref{eq:var_Ltotal_scaling}), respectively. The markers indicate the estimated variances relying on (a)~100 and (b,c)~10 independent network realizations. The error bars indicate the 95\% confidence intervals.}
	\label{fig:MphysMmaxLtot_var}
\end{figure}

To calculate the variance of the number of links in the jammed state $\sigma^2(M_\T{max})$, we introduce a binary random variable $b(t)$ for each time step, $b(t)=1$ indicating that we successfully added a link candidate and $b(t)=0$ indicating that we rejected it.
Similarly to the derivation of $\sigma^2\left(L_\T{total}\right)$, we can obtain $\sigma^2(M_\T{max})$ as the sum of the variances of $b(t)$.
Before the onset of physicality, all link candidates are accepted; therefore the contribution of this stage of the evolution of the network to $\sigma^2(M_\T{max})$ is zero.
Approaching the jammed state, we can write the sum of the variance of $b(t)$ as
\begin{equation}\label{eq:varMmax_jam_contribution}
\begin{split}
\sigma^2_2\left(M_\T{max}\right)\approx \int_{N^\alpha}^{N^2/2} dt\left[ \left(\int \exp\left[-c (\lambda t)^{1/5}l\right] p_\T{LC}(l)dl\right) \right. - \\
\left. -  \left(\int \exp\left[-c (\lambda t)^{1/5}l\right] p_\T{LC}(l)dl\right)^2 \right].
\end{split}.
\end{equation}
We again use $p_\T{LC}(l)\sim l^2$ and introduce the new integration variable $z\sim  (\lambda t)^{1/5} l$, and keeping only the leading order term we obtain that the scaling of $\sigma^2_2\left(M_\T{max}\right)$ in the large network limit is 
\begin{equation}\label{eq:var_Mmax_scaling}
\sigma^2_2\left(M_\T{max}\right)\sim N^{\frac{3\alpha+4}{5}}.
\end{equation}
Figures~\ref{fig:MphysMmaxLtot_var}b and c compare the predicted scaling of $\sigma^2_2\left(M_\T{max}\right)$ and $\sigma^2_2\left(L_\T{total}\right)$ to numerical simulations.

\clearpage

\section{Link length distribution in the jammed state}\label{sec:link_length}

In this section we analytically characterize the length distribution of a randomly selected link in jammed LPNs, we rely on the differential equations developed in Sec.~\ref{sec:rand_indep_set}. 
The link length distribution $p(l)$ is shaped by two factors: the length distribution of link candidates and the probability that a link candidate gets accepted.
The length distribution of link candidates $p_\T{LC}(l)$ is given by the so-called cube line picking distribution which is the distribution of the distance between two random points selected uniformly from the unit cube.
The important properties of $p_\T{LC}(l)$ for our derivation are: 
(i)~that the support of $p_\T{LC}(l)$ is $[0,\sqrt{3}]$, meaning that the system size will induce an upper cutoff in $p(l)$ and 
(ii)~that for small $l$ the effect of boundaries becomes negligible and $p_\T{LC}(l)$ is equal to the surface of a sphere with radius $l$, i.e., $p_\T{LC}(l)\approx 4\pi l^2$.

Equation~(\ref{eq:Lis}) indicates that the probability of accepting a link of length $l$ at time $t$ is the exponentially decaying function $\exp\left[-\frac{\pi}{2}\lambda L_\T{IS}(t)l\right]$, where $L_\T{IS}(t)$ the total link length increases over time, meaning that the length of links accepted decreases over time (see Fig.~1e in the main text).
Before the onset of physicality, even the longest links are accepted; therefore the upper cutoff of $p(l)$ is determined by $p_\T{LC}(l)$.
On the other hand, very short links $l\ll (\lambda L_\T{IS}(t=N^2/2))^{-1}$ always get accepted, where $L_\T{IS}(t=N^2/2)~\sim N^\frac{4\alpha+2}{5}$ is the total link length in the jammed state.
This means that for small enough $l\ll l^*\sim N^{-\frac{2-\alpha}{5}}$ the link length distribution is again determined by $p_\T{LC}(l)$ and we get
\begin{equation}
p(l)\sim l^2.
\end{equation}
Note that $l^*\sim L_\T{IS}/M_\T{max}$ provides the scaling of the typical link length.
To characterize $p(l)$ for intermediate $l\sim l^*$ values, we write the link length distribution up to normalization as
\begin{equation}\label{eq:pl}
p(l)\sim \int_0^{N/2}dt \exp\left[-\frac{\pi}{2}\lambda L_\T{IS}(t)l\right]p_\T{LC}(l).
\end{equation}
Since $l^*\rightarrow 0$ for $N\rightarrow\infty$ for all $\alpha>1/3$, we can substitute Eq.~(\ref{eq:LIS_scaling}) for $L_\T{IS}(t)$ and $l^2$ for $p_\T{LC}(l)$, obtaining
\begin{equation}
p(l)\sim \int_0^{N/2}dt \exp\left[-c(\lambda t)^{1/5}l\right]l^2,
\end{equation}
where $c$ is a constant.
Introducing the new integration variable $z=(\lambda t )^1/5$ we get
\begin{equation}
p(\bar{l})\sim \bar{l}^{-3}\int_0^{c\bar{l}}dz z^4e^{-z}=\bar{l}^{-3}\gamma(5,c\bar{l}),
\end{equation}
where $\bar{l}=l/l^*$ is the link length measured in units of the typical link length and $\gamma(a,x)$ is the lower incomplete gamma function.
For large $\bar{l}$, we get $\gamma(5,c\bar{l})\approx \Gamma(5) - (c\bar{l})^4 e^{-c\bar{l}}$, keeping only the zeroth order term we get
\begin{equation}
p(\bar{l})\sim \bar{l}^{-3}.
\end{equation}

\begin{figure}[h]
	\centering
	\includegraphics[width=1.\textwidth]{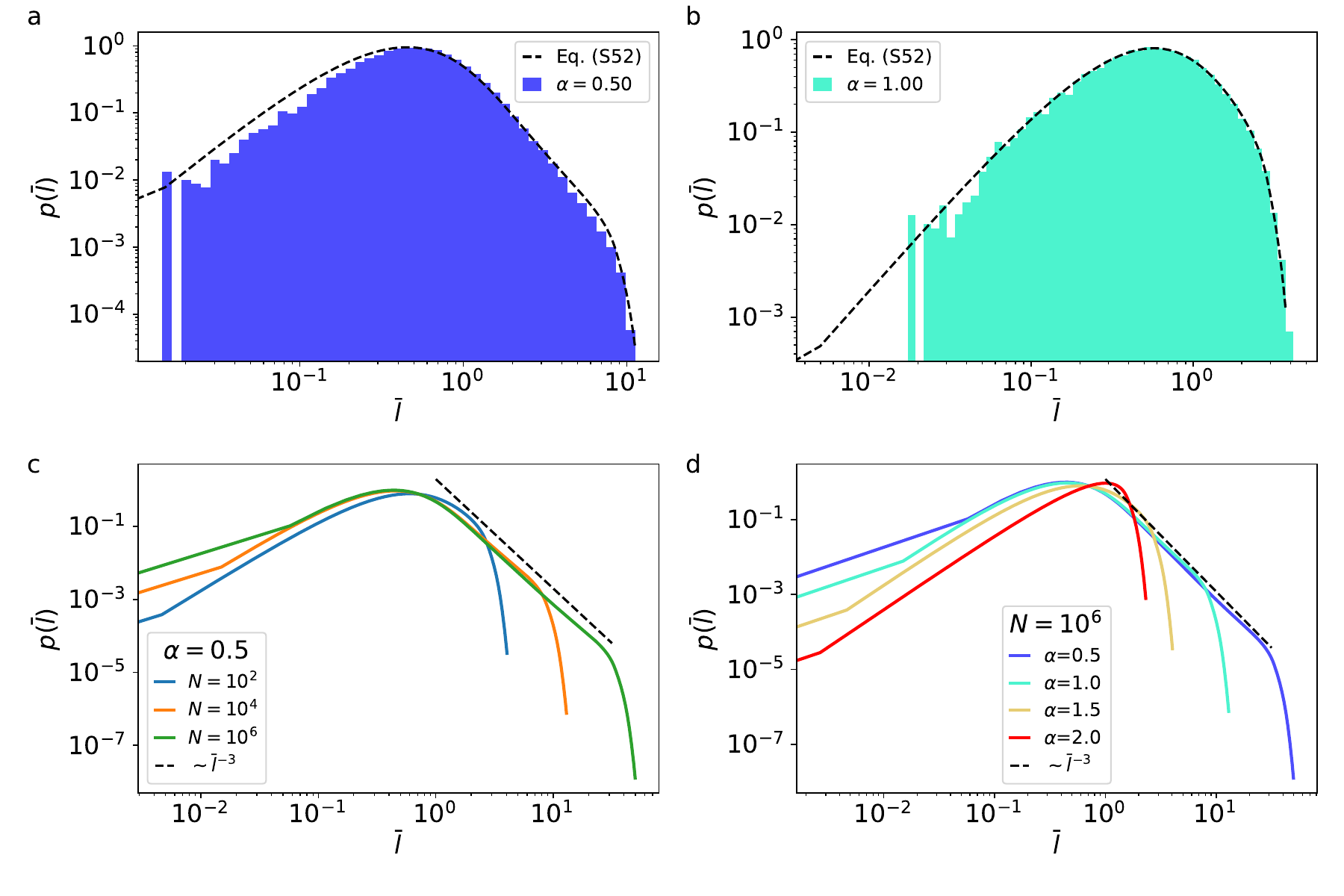}
	\caption{{ \bf Link length distribution.} (a,b)~Comparing the predicted link length distribution obtained by numerically integrating Eq.~(\ref{eq:pl}) to simulated jammed LPNs with (a)~$\alpha=1/2$, $N=10,000$ and (b)~$\alpha=1$, $N=1623$. We showed that the theoretical $p(\bar{l})$ has a power law tail with a cutoff; however, the power law section of the distribution remains narrow for the network sizes easily accessible by simulations. (c,d)~Numerically integrating Eq.~(\ref{eq:pl}) reveals that the power law regime of $p(\bar{l})$ becomes more pronounced with increasing $N$ and decreasing $\alpha$.}
	\label{fig:linklengthdist}
\end{figure}

\clearpage

\section{Degree distribution and clustering in the jammed state}\label{sec:degvar_clust}

Previously we showed that the length of the links in LPNs is reduced by volume exclusion.
Link length is a physical property of the system; volume exclusion, however, also affects network properties, such as the degree distribution and the abundance of triangles.
As we build a LPN by sequentially adding links, some of the link candidates are discarded due to physical conflicts.
If we turn off physicality, however, all links are allowed and the process generates an Erd\H os-R\'enyi network (ER).
Therefore, any difference in the properties of an LPN and an ER network with the same number of nodes and links is a consequence of physicality.
To demonstrate the effect of physicality on network properties, we generate jammed LPNs with only link-link interactions with fixed $N$ as a function of $\lambda$ together with an ER counterpart with $N$ nodes and $M_\T{max}$ links, and we compare their degree variances and clustering coefficients.

\subsection{Degree variance}

Figure~\ref{fig:jammed_prop}a shows  the variance of the degree distribution of LPNs ($\sigma_\T{LPN}^2$) and the corresponding ER networks ($\sigma^2_\T{ER}$) as a function of $\lambda$.
For small $\lambda$ physicality has no effect and the jammed state is a fully connected network where each node has $k=N-1$ degree; therefore for both the LPNs and ER networks $\sigma^2$ tends to zero as $\lambda\rightarrow 0$.
As $\lambda$ increases, the node degrees are no longer uniform and initially both $\sigma_\T{LPN}^2$ and $\sigma_\T{ER}^2$ increases.
Also with increasing $\lambda$, the jammed state becomes more sparse, i.e., $M_\T{max}(\lambda)$ decreases.
The node degree in ER networks follows a binomial distribution with mean $\avg k _\T{ER}=2M_\T{max}(\lambda)/N$ and variance
\begin{equation}
\sigma^2_\T{ER}= N\left(1-\frac{2M_\T{max}(\lambda)}{N^2}\right)\frac{2M_\T{max}(\lambda)}{N^2},
\end{equation} 
meaning that $\sigma^2_\T{ER}$ peaks when the jammed state contains half of all possible links.
Figure~\ref{fig:jammed_prop}a shows that the degree variance $\sigma^2_\T{LPN}$ also peaks at intermediate $\lambda$ values. However, $\sigma^2_\T{LPN}$ is higher than $\sigma^2_\T{ER}$ for the entire range of $\lambda$, indicating that volume exclusion increases degree heterogeneity in LPNs.

\subsection{Clustering coefficient}

\begin{figure}[h]
	\centering
	\includegraphics[width=1.\textwidth]{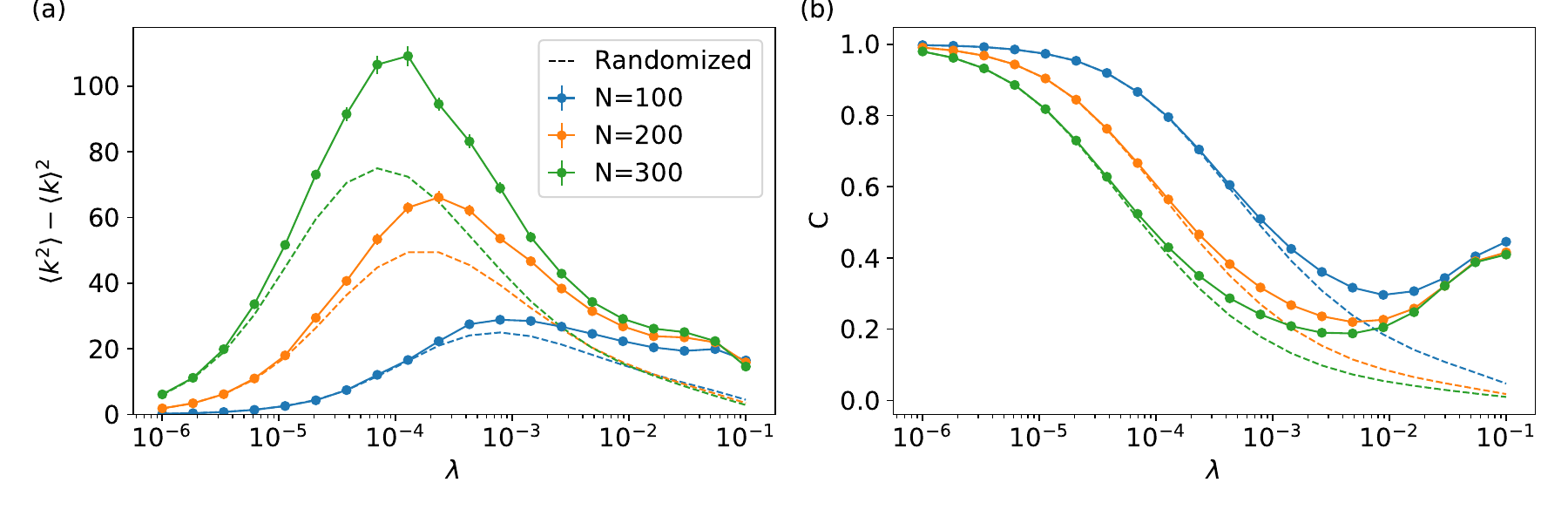}
	\caption{{ \bf Properties of the jammed network.} (a)~Variance of the degree distribution of jammed physical networks compared to the expected degree variance of the randomized reference networks with the same number of nodes and edges ($ \sigma^2_\T{rand}=Np(1-p)$, where $p=2M/N^2$). (b)~Average local clustering coefficient of jammed networks compared to the randomized reference networks ($C_\T{rand}=p$). The data points represent an average of 50 independent networks, the error bars, often smaller than the symbols, indicate the 95\% confidence interval.}
	\label{fig:jammed_prop}
\end{figure}

The local clustering coefficient of node $i$ is
\begin{equation}
C_i = \frac{n_{\triangle,i}}{k_i(k_i-1)/2},
\end{equation}
where $n_{\triangle,i}$ is the number of triangles node $i$ participates in and $k_i$ is degree of node $i$; $C_i$ is undefined for nodes with degree less than 2.
Figure~\ref{fig:jammed_prop}b shows the average local clustering coefficient for both the LPNs ($C_\T{LPN}$) and their ER counterpart ($C_\T{ER}$). 
For small $\lambda$, the jammed state corresponds to the fully connected network, hence we have both $C_\T{LPN}=1$ and $C_\T{ER}=1$ for $\lambda\rightarrow 0$ and fixed $N$.
As $\lambda$ increases, $M_\T{max}(\lambda)$ decreases resulting in connected wedges formed by two adjacent links that are not closed by a third link to form a triangle.
For ER networks, all node pairs are connected with the same probability $p=2 M_\T{max}(\lambda)/N(N-1)$; therefore the probability of three nodes forming a triangle is $p^3$ and three nodes forming a connected triplet is $p^2$, leading to
\begin{equation}
C_\T{ER}=p=\frac{2M_\T{max}(\lambda)}{N(N-1)}.
\end{equation} 
Therefore $C_\T{ER}$ monotonically decreases as $\lambda$ increases.

Figure~\ref{fig:jamming_illust} showed that as we increase the link thickness $\lambda$, the typical link length in the jammed state decreases.
Such shorter links imply that for high $\lambda$ nodes tend to connect to nodes in their physical vicinity increasing the chance of forming triangles.
Indeed, Figure~\ref{fig:jammed_prop}b shows that for large $\lambda$ the clustering coefficient $C_\T{LPN}$ increases compared to the random expectation $C_\T{ER}$, confirming that volume exclusion increases the density of triangles in LPNs.

\subsection{Stuck nodes for LPNs with only link-link interactions}\label{sec:stucknodes}

\begin{figure}[h]
	\centering
	\includegraphics[width=.75\textwidth]{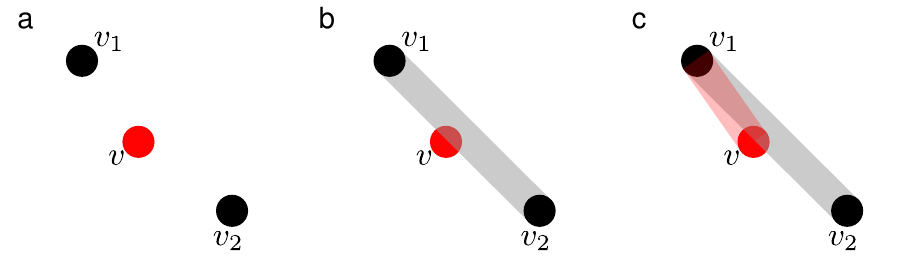}
	\caption{{ \bf Stuck nodes in LPNs with only link-link interactions.} (a)~During the generation of an LPN, consider a physical node $v$ that is still isolated. (b)~Since node-link interaction is not taken into account a newly added link $(v_1,v_2)$ may overlap with node $v$. As we continue the generation of the LPN, any link that connects $v$ to nodes other than $v_1$ or $v_2$ would generate a physical conflict; and therefore is forbidden. We call such nodes stuck. (c)~ However, node $v$ may still be able to connect to nodes $v_1$ and $v_2$, since overlap is allowed between links that share an endpoint. Therefore stuck nodes have degree at most two.}
	\label{fig:stuck_nodes}
\end{figure}

In LPNs with node-link interactions, physical nodes and links cannot overlap by definition, unless the link is adjacent to the node.
During the generation of an LPN with only link-link interactions, however, it might happen that a newly added link $e=(v_1,v_2)$ overlaps with isolated node $v$.
If this happens, node $v$ becomes stuck in the sense that as we add further links to the network, node $v$ may only be connected nodes $v_1$ and $v_2$, any other link would generate a conflict with link $e$ (Fig.~\ref{fig:stuck_nodes}).

The presence of such nodes may alter the network properties of the LPN, for example, it can contribute to the increased degree variance of the network.
This raises the question: how does the number of stuck nodes $N_\T s$ depend on $\lambda$?
To provide an upper bound for $N_\T s$ in the large network limit $N\rightarrow \infty$, we return to the analytical description of LPNs introduced in Sec.~\ref{sec:rand_indep_set}.
Equation~(\ref{eq:LIS_scaling}) predicts that the total link length in the jammed state scales as
\begin{equation}
L_\T{total}\sim N^{\frac{4\alpha+2}{5}},
\end{equation}
which in turn provides the scaling of the total volume of the links
\begin{equation}\label{eq:Vtotal_scaling}
V_\T{total}\sim \lambda^2 L_\T{total}\sim N^{-\frac{6\alpha-2}{5}},
\end{equation}
The probability that a random node overlaps with any of the links is at most to $V_\T{total}$.
This is an upper bound, because a newly added link $e=(v_1,v_2)$ can only overlap with node $v$ if $v$ is not yet connected to any node other node than $v_1$ and $v_2$.
Once $v$ gains more than two connections, link-link interactions make it impossible for node $v$ to become stuck.

A consequence of the upper bound~(\ref{eq:Vtotal_scaling}) is that the probability of a node to become stuck converges to zero in the large network limit for any $\alpha>1/3$, suggesting that the effect of stuck nodes becomes irrelevant for most network properties.
Note, however, that the number of stuck nodes $N_\T s \lesssim N\cdot V_\T{total}$ may diverge for $1/3<\alpha<1/2$, although sub-linearly.

\clearpage

\section{Spectra of linear physical networks}\label{sec:spectra}

In this section, we examine the eigenvalues and eigenvectors of the adjacency matrix of LPNs.
Volume exclusion determines which links can and cannot be added to physical networks, hence it induces a correlation between the physical layout and the abstract network structure of LPNs.
In the following, we use numerical simulations and analytical considerations to show the emergence of eigenvectors of the adjacency matrix that only depend on the position of physical nodes, which connects the abstract network structure captured by the adjacency matrix to the physical structure captured by the node positions.

\subsection{Eigenspectrum of Erd\H{o}s-R\'enyi networks}

If volume exclusion has no effect on network evolution, e.g., $M<M_\T{phys}$, all link condidates are accepted and the resulting network is equivalent to an Erd\H{o}s-R\'enyi (ER) network.
Any difference observed between the eigenspectrum of a LPN and an ER network of the same size is an affect of volume exclusion.

The distribution of the eigenvalues $\mu$ of an ER network in the large network limit with diverging average degree (i.e., $\avg k\rightarrow \infty$ as $N\rightarrow\infty$) is given by Wigner's semicircle law~\cite{van2010graph}.
\begin{equation}\label{eq:wigner}
p(\mu)=\frac{\sqrt{4Np(1-p)-(\mu+p)^2}}{2\pi Np(1-p)},
\end{equation}
for $\lvert\mu\rvert\leq 2p(1-p)\sqrt{N}$, where $p$ is the probability that two nodes are connected in the ER network.
The largest eigenvalue $\mu_1$ separated from the bulk and corresponds to the average degree:
\begin{equation}
\mu_1=(N-2)p+1+O(N^{-1/2}).
\end{equation}

\subsection{Simulations}\label{sec:eig_sim}

\begin{figure}[h]
	\centering
	\includegraphics[width=1.\textwidth]{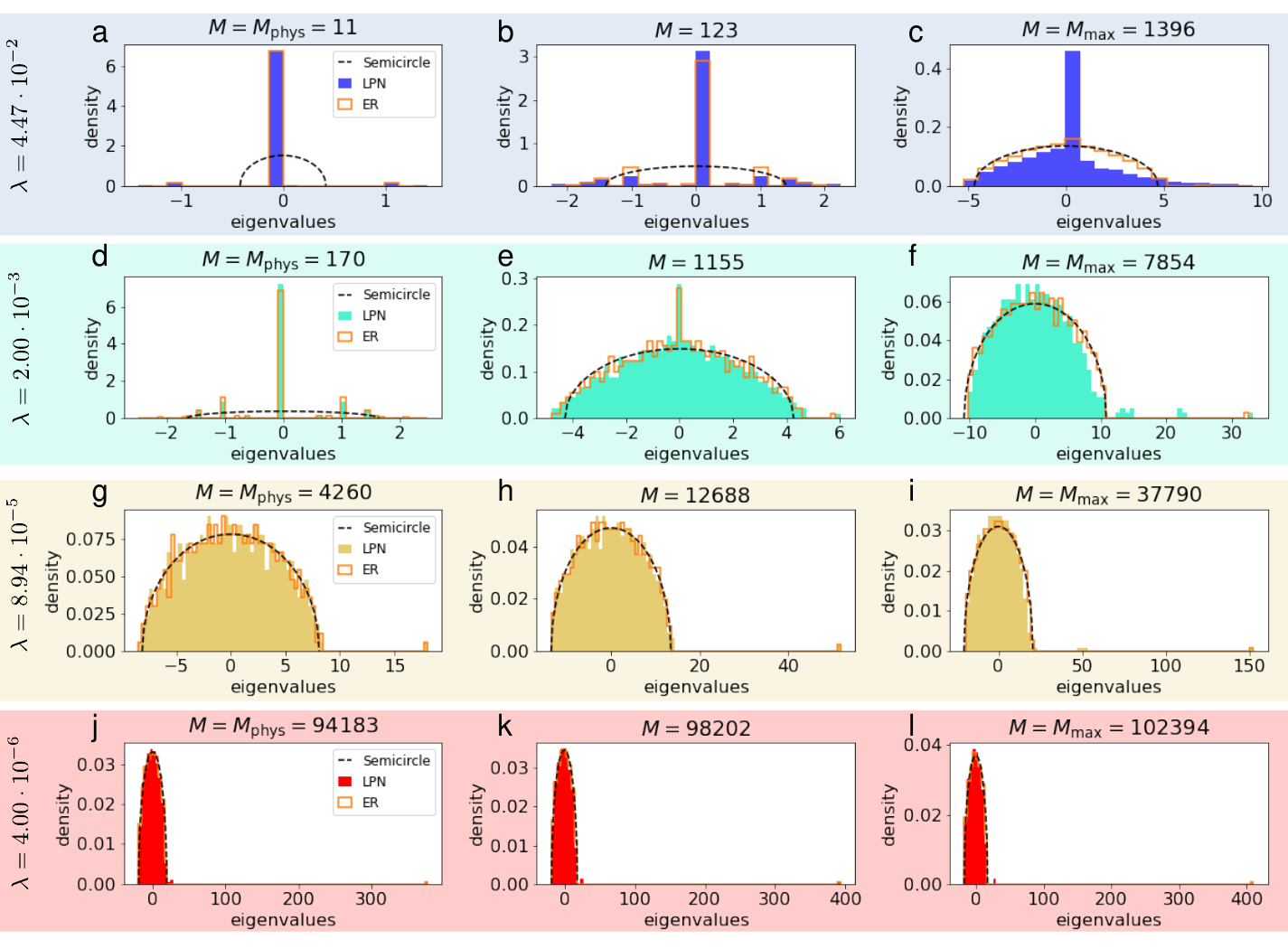}
	\caption{{ \bf Eigenspectra of LPNs.} We show the spectral density of the adjacency matrix of LPNs with $N=500$ node and varying $\lambda$ link thickness. (a,d,g,j)~At the onset of physicality, the spectral density of an LPN is consistent with the spectral density of ER networks with the same number of nodes and links, indicating that physicality has no effect on the evolution of the abstract network for $M<M_\T{phys}$.
	(c,f,i,l)~In the jammed state the spectrum of an LPN differs from the spectrum of its ER counterpart in two main ways: (i)~the bulk of the distribution becomes right skewed, and (ii)~a group of three eigenvalues ($\mu_2$, $\mu_3$ and $\mu_4$)  separate from the bulk. These differences are a consequence of physicality.
	(c)~For $\lambda$ values where the jammed state corresponds to a sparse abstract network, the separation of the three eigenvalues from the bulk is less pronounced.
	(f)~For certain $\lambda$ values, we observe that additional groups of eigenvalues separate from the bulk.
	Each subplot show the spectral density of a single LPN with $N=500$ nodes. The orange outline is the spectral density of an ER network with the same number of nodes and links as the corresponding LPN, and the dashed line is Wigner semicircle law.
	}
	\label{fig:eigspectra}
\end{figure}

We generate LPNs with $N=500$ nodes and varying link width $\lambda$, and we calculate the eigenvalues of their adjacency matrix at different points $M\leq M_\T{max}$ of the network evolution. 
Figure~\ref{fig:eigspectra} compares the spectra of LPNs with only link-link interactions to ER networks with the same number of nodes and links.
We observe that at the onset of physicality ($M=M_\T{phys}$) the eigenspectrum of an LPN is consistent with spectrum of its ER network counterpart, which in turn for a dense enough network is well approximated by Wigner's semicircle law Eq.~(\ref{eq:wigner}) (Fig.~\ref{fig:eigspectra}a,d,g,j).
In the jammed state ($M=M_\T{max}$), LPNs differ for the ER networks in two main ways: (i)~the bulk of the eigenvalue distribution becomes right skewed, and (ii)~a group of three eigenvalues $\mu_2$, $\mu_3$ and $\mu_4$ become separated from the bulk (Fig.~\ref{fig:eigspectra}c,f,i,l).
We observe two deviations from this general picture: (i)~For large $\lambda$, LPNs reach the jammed state while remaining sparse. For such networks, the skewness of the eigenspectrum is prominently observable in the numerical simulations and the separation of the three leading eigenvalues from the bulk is less clear (Fig.~\ref{fig:eigspectra}c).
(ii)~For some values of $N$ and $\lambda$, we observe that in addition to $\mu_2$, $\mu_3$ and $\mu_4$ further groups of eigenvalues separate from the bulk.
We examine these two deviations in more detail in Secs.~\ref{sec:higher_eig} and \ref{sec:eig_separation}.

\begin{figure}[h]
	\centering
	\includegraphics[width=1.\textwidth]{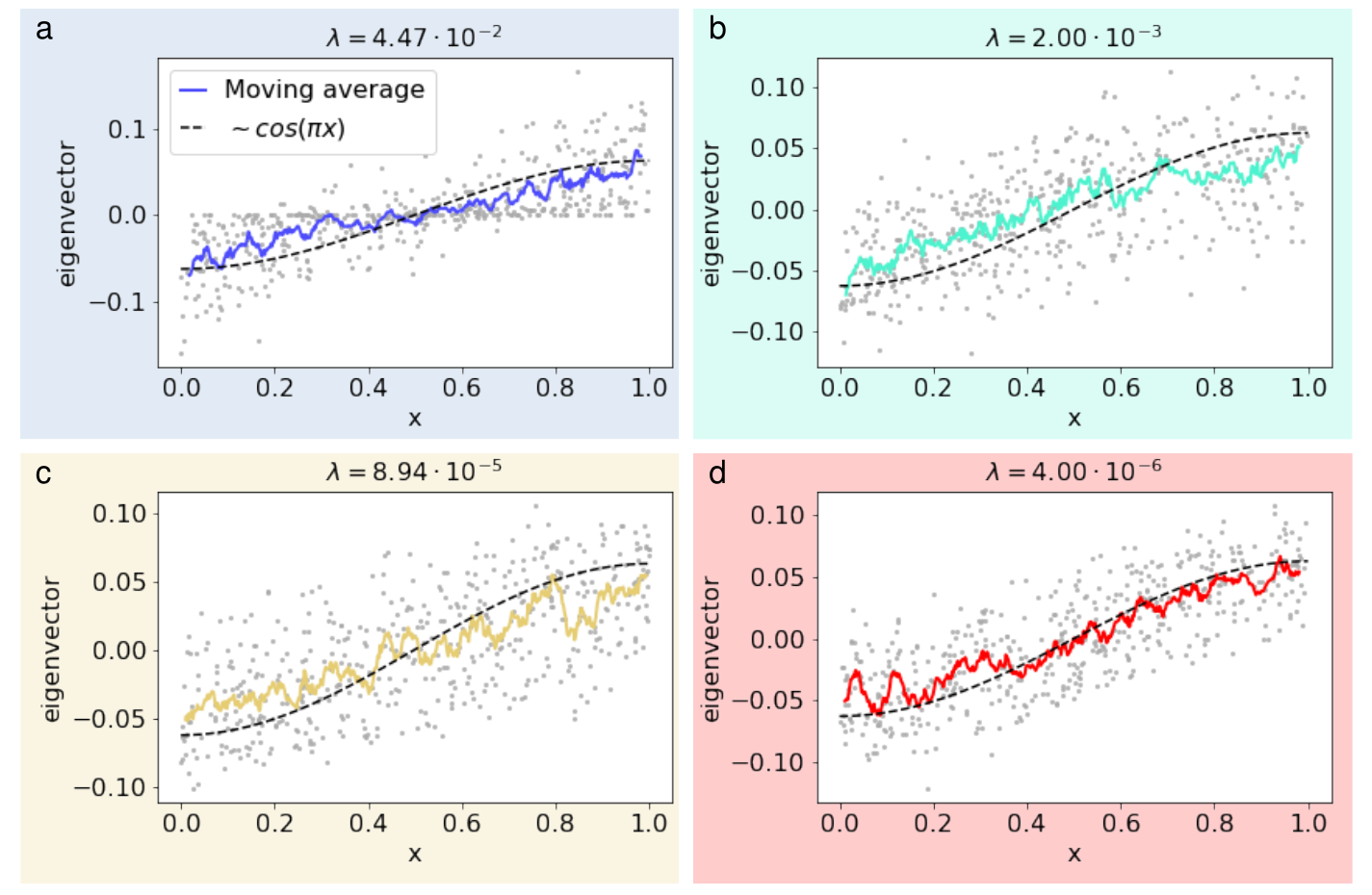}
	\caption{{ \bf Eigenvectors of LPNs.} In the jammed state the eigenvectors $\V v^{(2)}$, $\V v^{(3)}$ and $\V v^{(4)}$ capture the physical structure of the network and are largely determined by the node positions. To demonstrate this we pick the eigenvector $\V v$ that correlates the most with the $x$ coordinate of the nodes and plot $v_i$, the value of the eigenvector at node $i$, as a function of $x_i$, the position of node $i$.
	The gray markers correspond to the individual nodes of the networks, the solid line shows a moving average of 15 nodes. The dashed line shows the theoretical prediction.
	The subplots correspond to the networks in Fig.~\ref{fig:eigspectra}, we use matching colors.
	 (a)~Figure~\ref{fig:eigspectra}c showed that $\mu_2$, $\mu_3$ and $\mu_4$ do not separate clearly from the bulk for parameters $N=500$ and $\lambda=4.47\cdot 10^{-2}$. We observe, however, that the corresponding eigenvectors still correlate with the node position, despite apparent lack of separation. 
	 For LPNs with only link-link interactions with high $\lambda$, we observe stuck nodes, i.e., isolated nodes that overlap with links due to the lack of node-link volume exclusion. The value of the eigenvector $\V v$ is zero at nodes that remain isolated in the jammed state, hence the increased number of nodes with $v_i=0$.}
	\label{fig:eigvecs}
\end{figure}

The eigenvectors $\V v^{(2)}$, $\V v^{(3)}$ and $\V v^{(4)}$ corresponding to  $\mu_2$, $\mu_3$ and $\mu_4$, respectively, capture the large-scale physical structure of the network and are largely independent of the details of the wiring of the network.
In other words, the value of the eigenvector at node $i$ is largely determined by the position of node $i$ and does not depend on exactly which nodes $i$ is connected to, i.e., $v^{(j)}_i\approx v(\V r_i)$ for $j=2,3,4$.
To show this we first select the eigenvector $\V v$ out of the group of $\V v^{(2)}$,$\V v^{(3)}$ and $\V v^{(4)}$ which correlates the most with the $x$ coordinate of the nodes. 
We then plot $v_i$ as a function of $x_i$ which shows a clear correlation between the eigenvector and the node position (Fig.~\ref{fig:eigvecs}).

\clearpage

\subsubsection{Groups of separated eigenvalues in the jammed state}
\label{sec:higher_eig}

\begin{figure}[h]
	\centering
	\includegraphics[width=1.\textwidth]{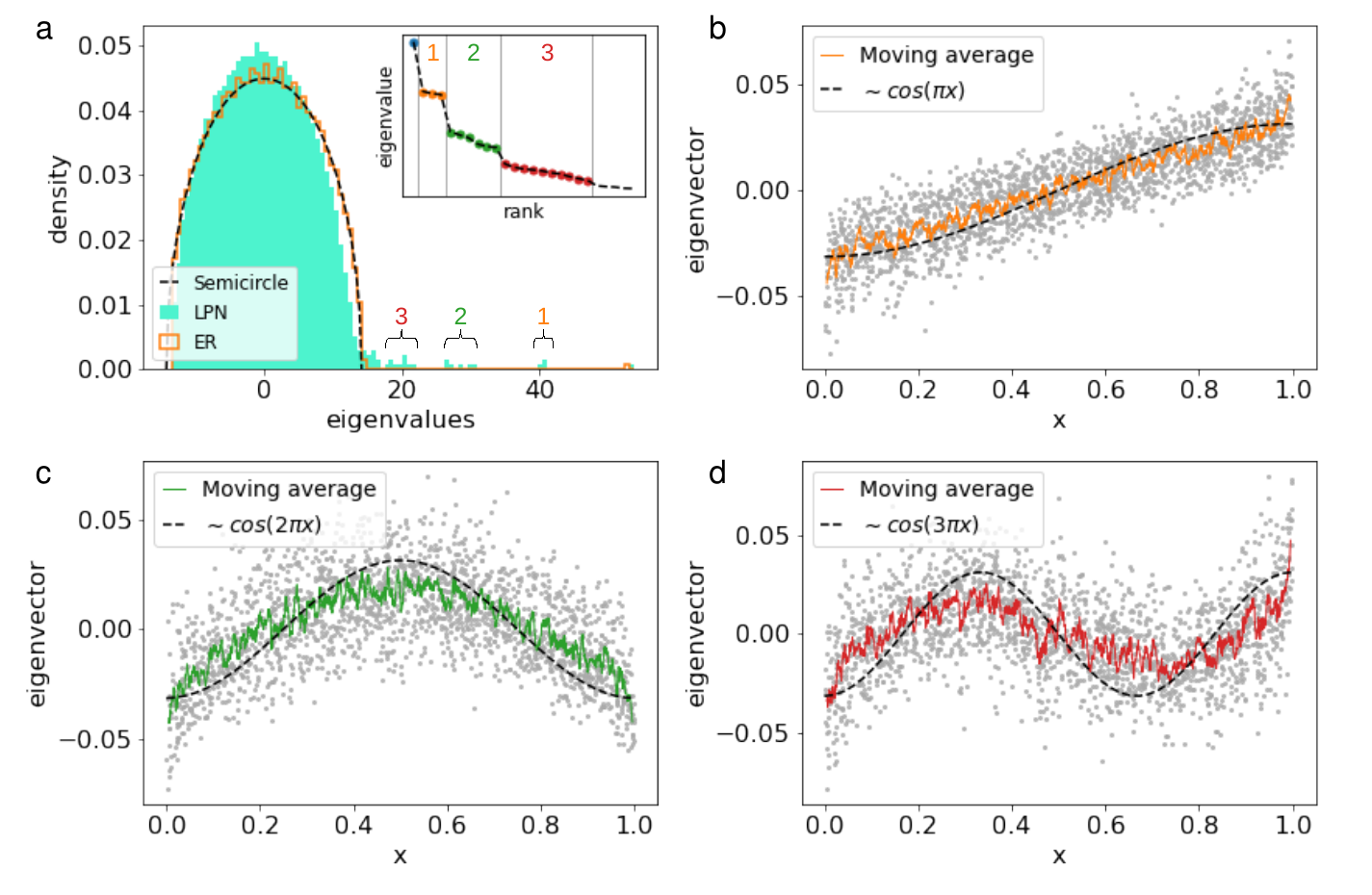}
	\caption{{ \bf Groups of separated eigenvalues in the jammed state.} (a)~The spectral density for an LPN with only link-link interactions and with parameters $N=2000$ and $\alpha=1$. Three groups of eigenvalues containing 3, 6 and 10 eigenvalues, respectively, separate from the bulk of the distribution. The inset shows that rank-plot of the eigenvalues. (b-d)~We plot an eigenvector from each group that depends the most on the $x$ node coordinate.
	The gray markers correspond to the individual nodes of the networks, the solid line shows a moving average of 15 nodes. The dashed line shows the theoretical prediction. }
	\label{fig:eig-higher}
\end{figure}

Figure~\ref{fig:eigspectra}f shows the eigenvalue distribution of an LPN with $N=500$ and $\lambda=N^{-1}=2\cdot 10^{-2}$, indicating that the group of three eigenvalues that separate from the bulk is followed by an additional group eigenvalues that also separate.
To better understand this behavior, we generate a larger LPN with $N=2000$ and $\lambda=N^{-1}$ and calculate the spectrum of its adjacency matrix.
Figure~\ref{fig:eig-higher}a indicates that for these parameter values there are three separated groups of eigenvalues: the initial group containing three eigenvalues, followed by a group of six and a group of ten eigenvalues.
On Figs.~\ref{fig:eig-higher}b-d, we plot an eigenvector from each group that has the highest dependence on the $x$ node coordinate.
We find that eigenvector in the first group is monotonic function of  $x$ (Fig.~\ref{fig:eig-higher}b), while the eigenvector from the second group has a maximum at $x=1/2$ (Fig.~\ref{fig:eig-higher}c) and the eigenvector in the third group has a maximum at $x=1/3$ and $x=2/3$ (Fig.~\ref{fig:eig-higher}d).
This indicates that the second and third group of eigenvalues also capture physical structure, and we observe that these eigenvectors are well-approximated by sinusoidal functions.

\clearpage

\subsubsection{Separation of $\mu_2$, $\mu_3$ and $\mu_4$}\label{sec:eig_separation}

In the jammed state of an LPN, the top three eigenvalues following the largest separate from the bulk (Fig.~\ref{fig:eigspectra}) and the corresponding eigenvectors capture the physical structure of the LPN (Fig.~\ref{fig:eigvecs}).
We observed, however, for parameters $N=500$ and $\lambda =N^{-1/2}=4.47\cdot 10^{-2}$ that the separation is less clear (Fig.~\ref{fig:eigspectra}c); yet, similarly to the case where the separation is pronounced, the corresponding eigenvectors capture physical layout of the network.
To better understand this behavior, we calculated the spectra of LPNs with increasing size $N$ while setting $\lambda = N^{-1/2}$, i.e., choosing parameter $\alpha=1/2$.
Note that according to Eq.~(\ref{eq:Mmax_scaling}), with increasing network size the average degree of the abstract network in the jammed state increases, albeit slowly, as $\avg k \sim N^{\frac{3\alpha-1}{5}}=N^{0.1}$.
We find that the spectra of LPNs for $N=500$, $1000$, $2000$ and $4000$ are similar and that the numerical results do not decisively determine whether the three leading eigenvalues become bounded from the bulk in the large network limit $N\rightarrow \infty$ or not.
The lack of numerical evidence is likely explained by the slow divergence of the average degree in the jammed state.

Figure~\ref{fig:eigvecs-N} shows the eigenvector that correlates the most with the $x$ node coordinate for each $N$.
We find for all network sizes that the eigenvectors $\V v^{(2)}$, $\V v^{(3)}$ and $\V v^{(4)}$ capture the physical layout of the network.
However, the role of stuck nodes becomes more pronounced with increasing $N$.

\begin{figure}[h]
	\centering
	\includegraphics[width=1.\textwidth]{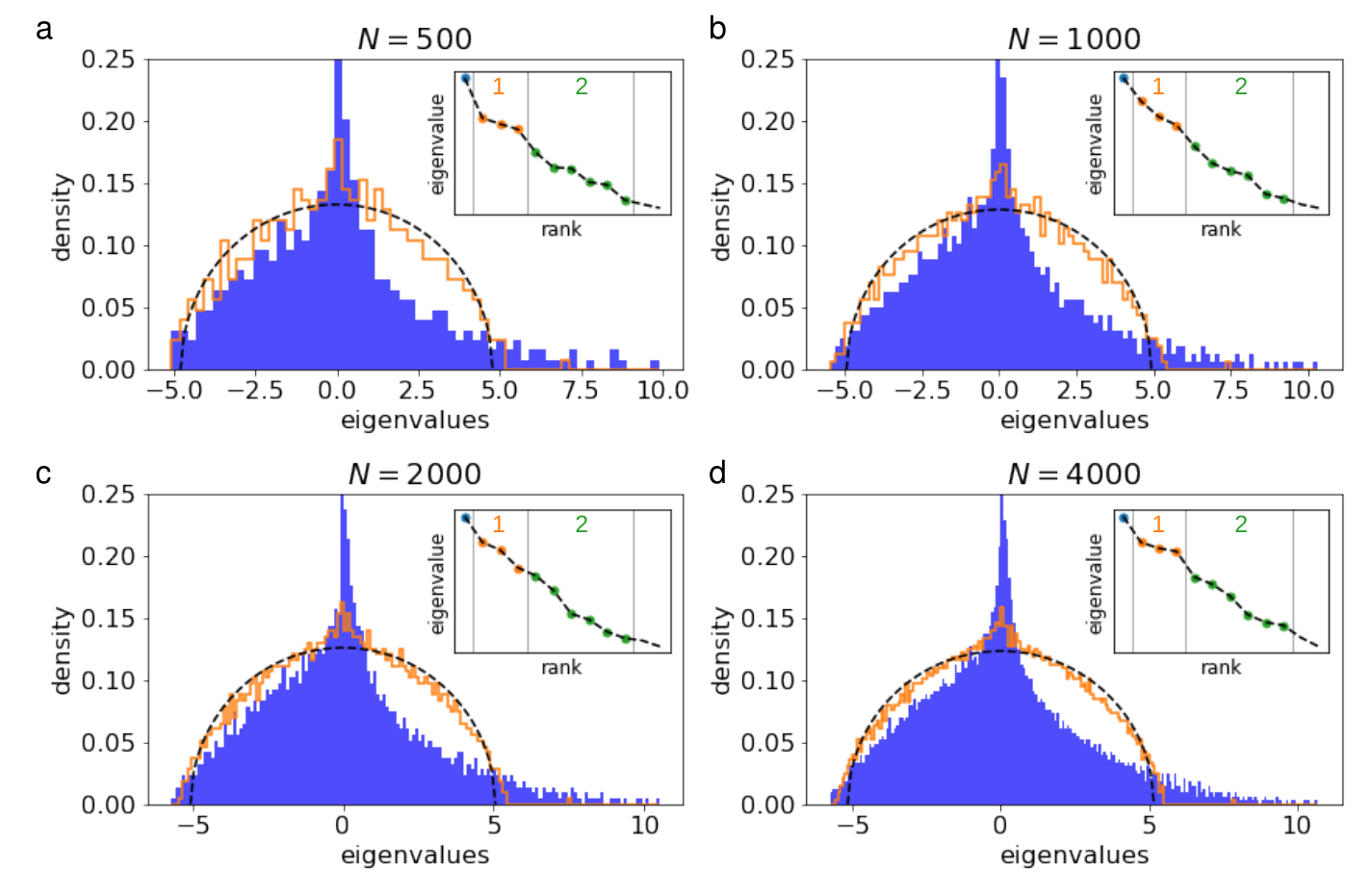}
	\caption{{ \bf Separation of eigenvalues for large $\lambda$.} The spectral density of LPNs with only link-link interactions and with increasing size while setting $\lambda=N^{-1/2}$, which corresponds to setting the exponent $\alpha=1/2$.
	 The numerical simulations do not provide decisive evidence whether the group of $\mu_2$, $\mu_3$ and $\mu_4$ eigenvalues become bounded from the bulk in the $N\rightarrow \infty$ limit or not.
	Networks with $\alpha\leq 1/2$ contain a diverging number, but zero fraction, of stuck nodes (Sec.~\ref{sec:stucknodes}). Stuck nodes have maximum degree 2, resulting in the sharp peak around $\mu=0$; to increase legibility, we only show the spectral density upto $0.25$.
	 The subplots show the spectral density of a single LPN and the insets show the rank-plot of the first 10 eigenvalues. The orange outline is the spectral density of an ER network with the same number of nodes and links as the corresponding LPN, and the dashed line is Wigner's semicircle law.}
	\label{fig:eigspectra-N}
\end{figure}

\begin{figure}[h]
	\centering
	\includegraphics[width=1.\textwidth]{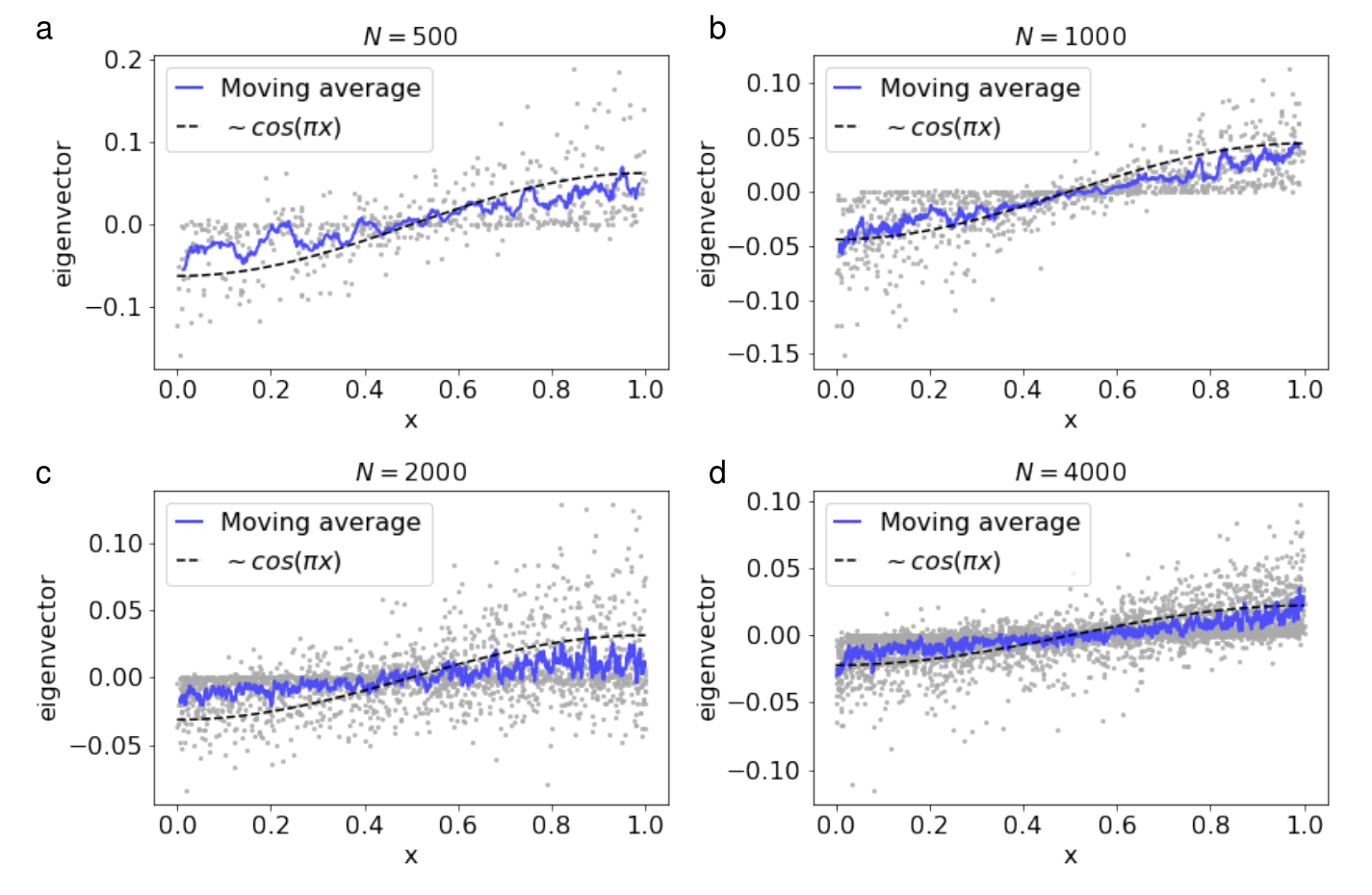}
	\caption{{ \bf Leading eigenvectors for LPNs with $\alpha=1/2$.} 
From eigenvectors $\V v^{(2)}$, $\V v^{(3)}$ and $\V v^{(4)}$ we select the eigenvector $\V v$ that correlates the most with the $x$ node coordinate and we plot $v_i$, the value of the eigenvector at node $i$, as a function of $x_i$, the $x$ coordinate of node $i$, for varying $N$ while setting $\lambda=N^{-1/2}$.
We find for all $N$ values that there is a positive correlation between $v_i$ and $x_i$, although the correlation is weakened by the presence of stuck nodes.
	The gray markers correspond to the individual nodes of the networks, the solid line shows a moving average of 15 nodes. The dashed line shows the theoretical prediction. }
	\label{fig:eigvecs-N}
\end{figure}

\clearpage

\subsection{Theory}\label{sec:eig_theory}

Let $\mu_i$ be the $i$th eigenvalue of the adjacency matrix and $\V v^{(i)}$ the corresponding eigenvector such that $\mu_1\geq\mu_2\geq\ldots\geq\mu_N$.
Numerical simulations indicate that the eigenvectors $\V v^{(2)}$, $\V v^{(3)}$ and $\V v^{(4)}$ correspond to the spatial locations of the nodes in the jammed state of the physical network (Sec.~\ref{sec:eig_sim}).
In this section, we show that these eigenvectors are given by the first few Fourier basis and the corresponding eigenvalues are equal to the Fourier coefficients of $P(l)$, the probability that two nodes at distance $l$ are connected.
We rely on an approximation similar to the high density approximation of the Euclidean random matrix theory~\cite{mezard1999spectra,goetschy2013euclidean}.

We start by searching for eigenvectors $\V v$ that only depend on the position of the nodes, i.e., $v_{i}=v(\V r^{(i)})$, where $v_i$ is the value of the eigenvector $\V v$ at node $i$ and $\V r^{(i)}$ is the location of node $i$.
The characteristic equation determining $\mu$ and $\V v$ is
\begin{equation}\label{eq:char_eq}
\mu v(\V r^{(i)}) = \sum_{j=1}^N A_{ij} v(\V r^{(j)}),
\end{equation}
we aim to approximate this equation such that it does not depend on the details of $\M A$, only on its average behavior.

As a first step, we show that in the jammed state physical nodes are densely connected to other nodes in their spatial neighborhood.  
The characteristic link length in the jammed state is
\begin{equation}
l^* \sim \frac{L_\T{tot}}{M_\T{max}}\sim N^{-\frac{2-\alpha}{5}},
\end{equation}
meaning that for $\alpha<2$ the characteristic scale $l^*$ tends to zero in the large network limit.
The typical number of other nodes a node $i$ can potentially connect to is proportional to the number of nodes in a ball with radius $l^*$ centered around $i$
\begin{equation}
N^* \sim N\cdot N^{-3\frac{2-\alpha}{5}}=N^{\frac{3\alpha-1}{5}},
\end{equation}
which diverges as $N\rightarrow\infty$ for $\alpha>1/3$.
The typical degree of a node also scales as
\begin{equation}
\avg{k}=2\frac{M_\T{max}}{N}\sim N^{\frac{3\alpha-1}{5}},
\end{equation}
meaning that indeed nodes are densely connected within their neighborhood, i.e., they connect to finite fraction of the nodes within the characteristic length $l^*$.
In other words, for $1/3<\alpha<2$ in the large network limit, $l^*$ goes to zero, yet $N^*$ diverges, and nodes are densely connected with other nodes within their neighborhood.

This observation allows us to substitute the adjacency matrix $\M A$ with its expectation in the characteristic equation~(\ref{eq:char_eq}), leading to
\begin{equation}
 \mu v(\V r^{(i)}) \approx \sum_{j=1}^N P(\left\lvert \V r^{(i)}-\V r^{(j)}\right\rvert) v(\V r^{(j)}),
\end{equation}
where  $P(l)$ the probability that two nodes at distance $l$ are connected.
Assuming that the eigenvector $\V v(\V r)$ changes much slower than the typical distance between neighboring nodes $N^{-1/3}$ and the probability that two nodes connect only depends on their relative position (i.e., boundary effects are negligible) allows us to substitute the summation with an integration:
\begin{equation}
\mu v(\V r) \approx N \int_{[0,1]^3} d^3\V s P(\lvert \V s -\V r\rvert) v(\V s),
\end{equation}
where the integration runs over the unit cube.
Since the characteristic link length $l^*$ tends to zero, we can expand the limits of integration to infinity wiht out introducing an error in the $N\rightarrow \infty$ limit:
\begin{equation}\label{eq:meanfield_char_eq}
\mu v(\V r) \approx N \int_{-\infty}^\infty d^3\V s P(\lvert \V s -\V r\rvert) v(\V s),
\end{equation}

To continue, we search for eigenvectors of the form
\begin{equation}\label{eq:v}
v(\V r) \sim e^{-i\V k \V r},
\end{equation}
where $\V k\in \mathbb{R}^3$ is a wavevector.
Substituting into Eq.~(\ref{eq:meanfield_char_eq}) we get
\begin{equation}
\mu e^{-i\V k \V r} \approx N \int_{-\infty}^\infty d^3\V s P(\lvert \V s -\V r\rvert) e^{-i\V k \V s}.
\end{equation} 
Multiplying both sides with $e^{i\V k \V r}$ leads to
\begin{equation}\label{eq:mu}
\mu \approx N \int_{-\infty}^\infty d^3\V w P(\lvert \V w \V\rvert) e^{-i\V k \V w},
\end{equation} 
 where we introduced the new integration variable $\V w = \V s -\V r$.
 Equation~(\ref{eq:mu}) indicates that under the assumptions we made, $v(\V r) = e^{-i\V k \V r}$ is indeed an eigenvector of the adjacency matrix with eigenvalue provided by Eq.~(\ref{eq:mu}).
 
For numerical simulations, we observed that in the jammed state a group of three eigenvalues are separated from the bulk of the spectral density and the corresponding eigenvectors correlate with the position of the nodes.

Theory predicts the emergence of a group of three eigenvectors with the same eigenvalue corresponding to wavevectors $\V k = \pi(1,0,0)$, $\pi(0,1,0)$ and $\pi(0,0,1)$, consistent with simulations (Fig.~\ref{fig:eigvecs}).
Theory further predicts a group of six eigenvectors corresponding to all possible combinations of second order sinusoidal functions and a group of ten eigenvectors corresponding to all possible combinations of third order sinusoidal functions.
Indeed, we find for sufficiently large LPNs (e.g., for $N=2000$ and $\lambda=N^{-1}$) such separated eigenvalue groups can emerge (\ref{fig:eig-higher}).

\subsection{Predicting node position}

In the jammed state, the eigenvectors $\V v_2$, $\V v_3$ and $\V v_4$ of an LPN are largely determined by its physical layout.
This allows us to predict the node positions relying on the structure of the abstract network only, i.e., relying on the adjacency matrix.
Theory and symmetry considerations predict that all three eigenvectors have the same eigenvalue, i.e., $\mu_2=\mu_3=\mu_4$,
hence any linear combination of $\V v_2$, $\V v_3$ and $\V v_4$ is also an eigenvector.
This means that we cannot in general assign an eigenvector to each of the axis of the unit cube to predict the position of nodes. (Although in numerical simulations we find that the eigenvectors tend to be close to parallel with the axes due to inhomogeneities of the cube.)
Therefore, to quantify the predictive power of the eigenvectors, we first find a linear transformation that best aligns the eigenvectors $\V v^{(2)}$, $\V v^{(3)}$ and $\V v^{(4)}$ with the node coordinates.
Specifically, we find matrix $\M a\in \mathbb{R}^{3\times 3}$ and vector $\V b\in \mathbb{R}^3$ that minimizes
\begin{equation}
C = \sum_{i=1}^N \left\lvert\M a \V w^{(i)}+\V b-\V r^{(i)} \right\rvert^2,
\end{equation}
where $\V w^{(i)}=[v^{(2)}_i,v^{(3)}_i,v^{(4)}_i]$ is a three dimensional vector with elements corresponding to the value of the eigenvectors at node $i$, and $\V r^{(i)}$ is the position of node $i$.
Then, we calculate the coefficient of determination $R^2$, where $R^2=1$ indicates perfect alignment, and $R^2=0$ corresponds to guessing the center of the unit cube as the position of each node.

\clearpage

\section{Real physical networks and the generalized meta-graph}\label{app:realmeta}

Real physical networks are typically not linear, meaning that they are not solely composed of spheres and straight rods.
If we add a new non-straight physical link to a network, we can route it infinitely many ways; therefore it is not possible to keep track of emerging physical conflicts relying on the original definition of the meta-graph.
Despite this limitation, we show that it is possible to define a generalized version of the meta-graph that is useful to characterize the physical structure of any existing physical network.

\subsection{Skeletonized representation of physical networks}

Most real physical networks, from neural or vascular networks to rock fissures, are obtained as volumetric data from experiments.
Volumetric representation of a physical network means that the three-dimensional space is divided into voxels, the three-dimensional equivalent of pixels, and the voxels are labeled to be inside or outside the physical network.
While such representation provides the most accurate description of the shape of a physical network that is available, it is both computational and analytically demanding to analyze.
Therefore volumetric data is routinely approximated by skeletonization, capturing less details, but providing a more concise description.

The skeleton of a physical network is in fact a variant of a linear physical network: a skeletonization algorithm approximates the shape of a physical network with vertices and straight segments inside the physical network and associates a radius to each vertex.
Multiple segments in the skeleton can correspond to what is considered a separate entity in the original network, e.g., a single neuron in a neural network or a non-branching section of a vessel in the vascular network is represented by a collection of straight segments in the skeleton.
Therefore it is common to associate a label with each segment connecting it the original object it represents.
Altogether, a skeleton representation for our purposes must have the following properties
\begin{definition}
A skeleton representation  $\s S$ is a graph with vertex set $\s V$ and edge set $\s E$ together with
\begin{itemize}
\item a position  $\V r: \s V\rightarrow \mathbb R ^3$ and a radius $\rho: \s V\rightarrow \mathbb R^+$ associated to each vertex,
\item and a label $\sigma: \s E \rightarrow \mathbb Z$ associated to each edge.
\end{itemize}
\end{definition}

Note that the skeleton $\s S$ is a physical realization of the abstract network $\s G$, where the node set of $\s G$ is the set of labels in $\s S$.

To recover an approximate volume of a physical network from a skeleton, we take the union of spheres centered at $\V r(v)$ with radius $\rho(v)$ for each vertex $v\in \s V$, and truncated cones that have axis corresponding to the segment $(\V r(v),\V r(w))$ and parallel faces with radii $\rho(v)$ and $\rho(w)$ for each edge $(v,w)\in \s E$.
Alternatively, a less accurate but simpler approximate volume can be obtained by substituting each edge by a cylinder with axis $(\V r(v),\V r(w))$ and radius $(\rho(v)+\rho(w))/2$.

The quality of the approximation can be controlled by the number of vertices in the skeleton.
There is, however, no single definition of the cost function that characterizes how good an approximation is, and there are a large number of skeletonization algorithms available and used in various scientific disciplines~\cite{saha2016survey}.
We obtained the data that we work with already in a skeleton representation, unless otherwise noted.

\subsection{The generalized meta-graph}

The goal of the generalized meta-graph is to characterize a given physical network by identify components that are in a physically confined space.
We define the generalized meta-graph $\s M_\T g$ for a skeleton representation of a physical network.
\begin{definition}
Given a skeleton representation $\s S$ and a parameter $\Delta\lambda$, the associated generalized meta-graph $\s M_\T g(\Delta\lambda,\s S)$ is a graph with vertex set corresponding to the edge labels of $\s S$.
We increase the radius of each skeleton-vertex by $\Delta \lambda$, and meta-vertices $l_1$ and $l_2$ are connected
 if the approximate volume corresponding to the labels $l_1$ and $l_2$ now overlap.
\end{definition}

The labels $l_i$ of a skeleton representation $\s S$ correspond to separate nodes or links of a physical network, for example, a label can identify the skeleton of a single neuron in a neural network or a vessel segment in a vascular network.
The degree of a neuron or vessel in $\s M_\T g(\lambda,\s S)$ quantifies its physical confinement: it counts the neurons or vessels that surround it in space.
A skeleton $\s S$ is a physical realization of an abstract network $\s G$, such as the synaptic network for neurons or the network of vessel segments connected together by junction points.
There are, however, many alternative skeletons $\s S^\prime$ that realize the same abstract network, prompting the question: What are the properties of $\s S$ that are common in all realizations and what are differences?
Two neurons that are connected in the synaptic network or two vessels bound together at a junction point are necessarily adjacent in physical space, hence they become connected in the generalized meta-graph for low $\Delta \lambda$ for any possible physical layout $\s S^\prime$ realizing the abstract network.
To measure the excess confinement of a neuron, i.e., the confinement beyond what is necessitated by the synaptic network, we define the restricted meta-graph, where we exclude all edges in the generalized meta-graph that are between neurons that are synaptic partners or vessel segments bound together.

\begin{definition}
Given a generalized meta-graph $\s M_\T g(\Delta\lambda,\s S)$ and a corresponding abstract network $\s G$, the restricted meta-graph $\s M_\T r(\Delta\lambda,\s S)$ is obtained by removing each edge $e=(l_1,l_2)$ from $\s M_\T g(\Delta\lambda,\s S)$ if $l_1$ and $l_2$ are adjacent in $\s G$.
\end{definition}

\begin{figure}[h]
	\centering
	\includegraphics[width=.8\textwidth]{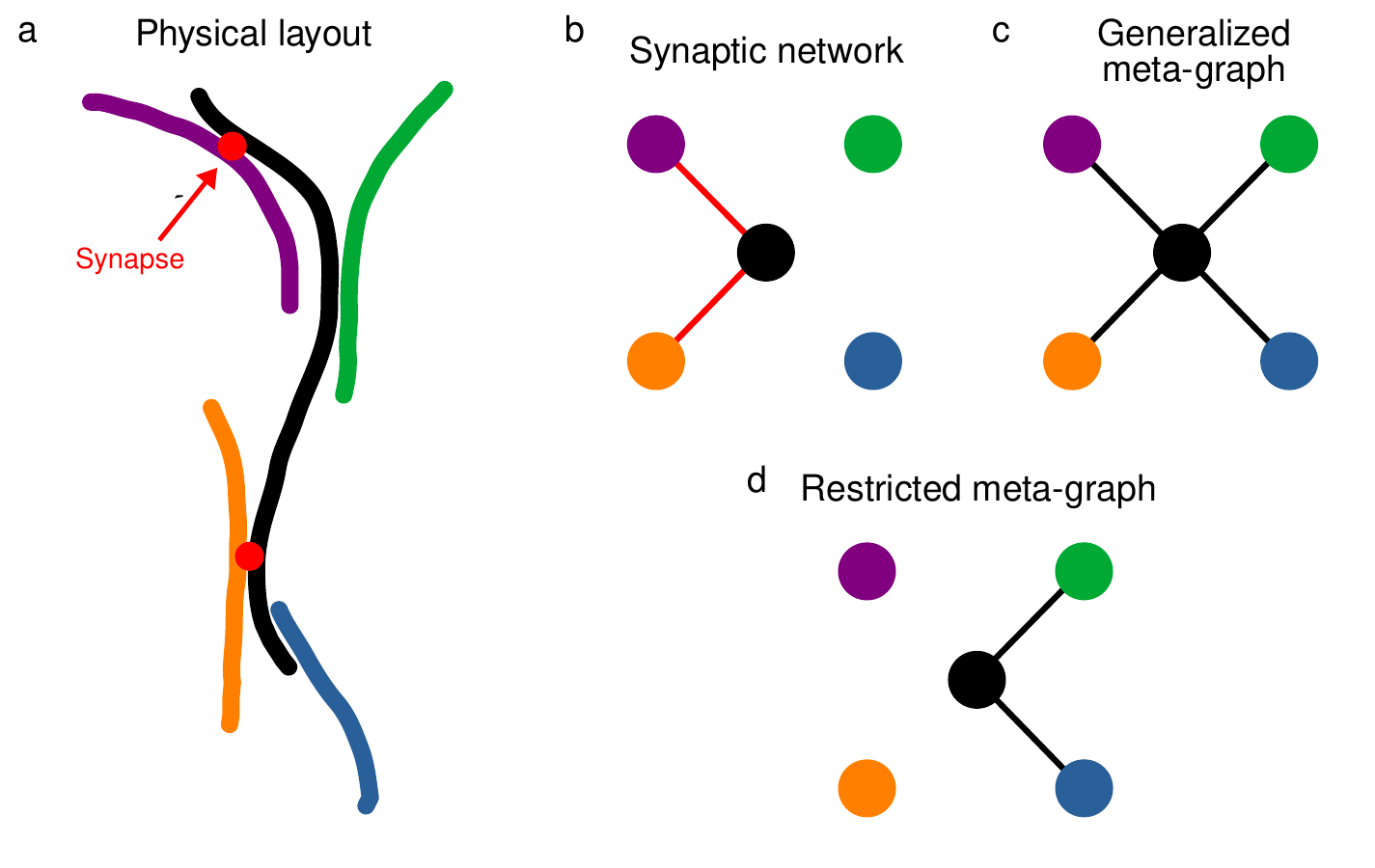}
	\caption{{ \bf The generalized and the restricted meta-graph.}
	(a)~The schematic layout of five neurons. The black neuron must be physically adjacent with its synaptic partners (purple and orange) and depending on the layout may be confined by other neurons it does not synapse with (green and blue).
	(b)~In the synaptic network realized by the physical layout (a), the black neuron has degree two.
	(c)~In the generalized meta-graph a neuron is connected with all neurons it overlaps with after thickening its branches by $\Delta\lambda$; therefore the generalized meta-degree of the black neuron is four.
	(d)~In the restricted meta-graph, we remove edges representing an overlap between synaptic partners; therefore, the restricted meta-degree of the black neuron is two.}
	\label{fig:gen_vs_rest}
\end{figure}

Figure~\ref{fig:gen_vs_rest} shows an example comparing the generalized and the restricted meta-graph.

Note that the original meta-graph $\s M(\lambda,\s P)$ is a special case of the restricted meta-graph.
We start from a skeleton $\s S$ corresponding to a complete graph on $\s P$ with uniform link thickness $0$, and labeling each link uniquely.
The restricted meta-graph $\s M_\T r(\lambda,\s S)$ obtained by thickening each link by $\lambda$ is equivalent to the original meta-graph $\s M(\lambda,\s P)$.

\clearpage

\subsection{Data sets}

\subsubsection{Fruit fly brain}

Relying on automated imaging techniques, a recent project mapped out a large fraction of the brain of the fruit fly {\it Drosophila melanogaster} containing the three-dimensional map of approximately 25,000 neurons and the location 20 million synapses~\cite{scheffer2020connectome}.
We downloaded the skeletonized data describing the shape of each neuron through the publicly available NeuPrint API~\cite{clements2020neuprint}.
To reduce the computational complexity, we focus on analyzing the Medula brain region (labeled ME(R), Fig.~\ref{fig:realmeta}a)~\cite{ito2014systematic}, which contains 2,979 neurons and 1,464,000 segments, making it a computationally difficult task to identify collisions between neurons exactly.
To overcome this difficulty, we substitute each neuron skeleton by a point cloud and use an efficient k-d tree implementation to query minimum distances between them.

In the skeletonized data set, each segment is labeled by the neuron that it belongs to, hence the meta-vertices represent neurons.
The skeleton together with the location of the synapse is a physical realization of the synaptic network.

\subsubsection{Vascular network}

The vascular network data set describes the vasculature found in a $600\times 600 \times 662 \;\mu \T{m}$ sample of a mouse cortex (Fig.~\ref{fig:realmeta}b)~\cite{gagnon2015quantifying}.
The data is provided as a skeleton including radii at the skeleton vertices.
We uniquely label non-branching vessel sections, i.e., each path connecting a pair of skeleton vertices with degree $k\neq 2$ receives a unique identifier.
The skeleton is a physical realization of an abstract network, where nodes are vessel segments, and two nodes are connected if the corresponding vessels are bound together at a junction point.
We construct the generalized meta-graph such that the meta-vertices represent the labeled vessel segments.

\subsubsection{Mitochondrial network}

The mitochondrial network data represents the mitochondrial reticulum of yeast cells (Fig.~\ref{fig:realmeta}c)~\cite{viana2020mitochondrial}.
The data set is available both as a skeleton and as a mesh representing the surface.
There is no radii provided with skeleton vertices; therefore we extracted a radius for each skeleton vertex based on the surface mesh using the skeletor python package~\cite{skeletor}.
Similarly to the vascular network, we uniquely label non-branching sections of the skeleton, and we construct the generalized meta-graph such that the meta-vertices represent these labeled sections.

\subsubsection{Root system}

The root network describes the root system of a {\it Cryptomeria japonica} tree (Fig.~\ref{fig:realmeta}d)~\cite{ohashi2019reconstruction}. 
The data is provided as a skeleton including radii at the skeleton vertices.
We uniquely label non-branching root sections, i.e., each path connecting a pair of skeleton vertices with degree $k\neq 2$ receives a unique identifier. 
We construct the generalized meta-graph such that the meta-vertices represent these labeled sections.

\clearpage

\subsection{The restricted meta-graph of real networks}

\begin{figure}[h]
	\centering
	\includegraphics[width=1.\textwidth]{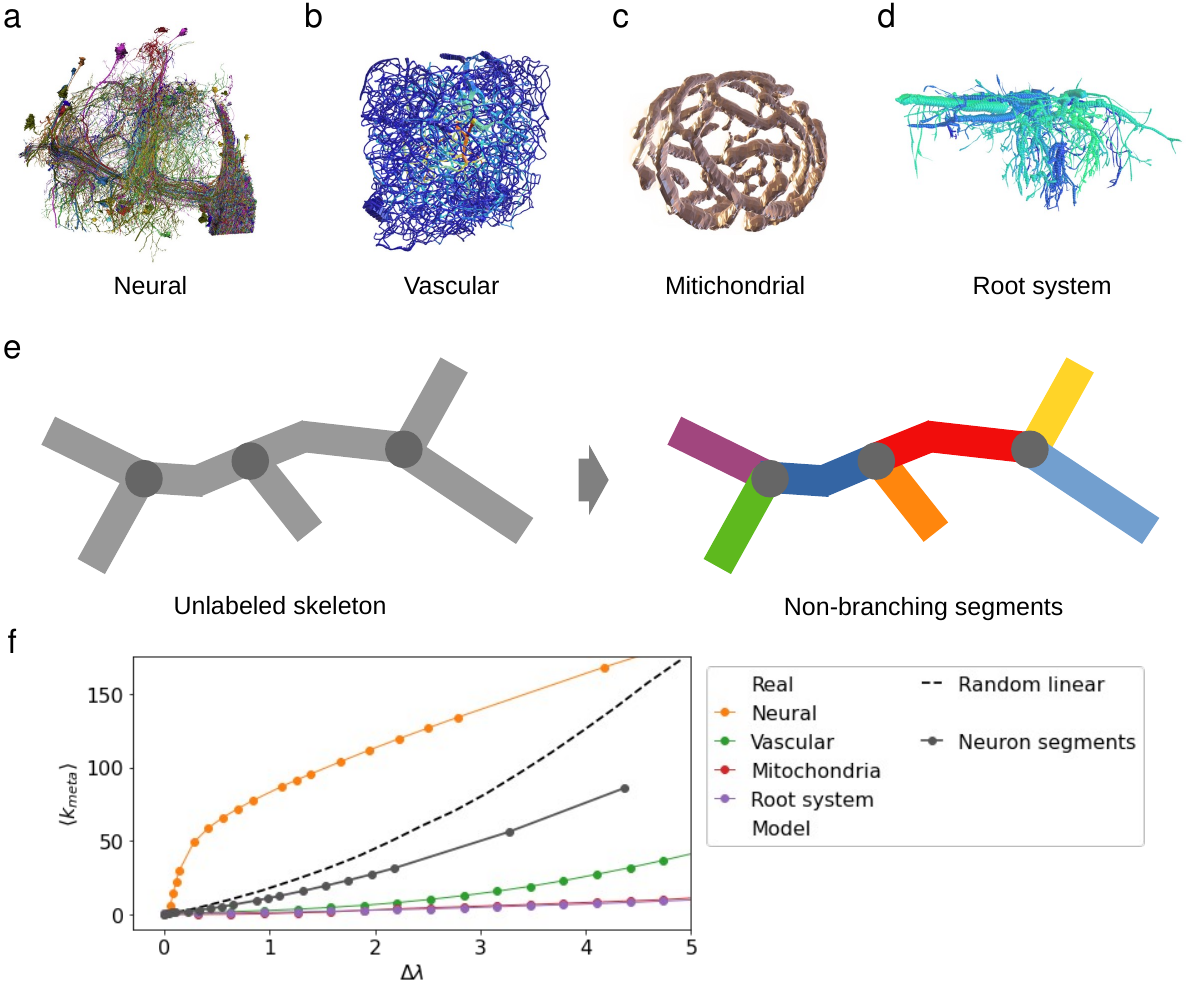}
	\caption{{ \bf The restricted meta-graph of real networks.} (a-d)~Three-dimensional rendering of the skeletonized description of the four real physical networks.
	(e)~For the vascular, mitochonrial and root system networks, we label non-branching sections uniquely, i.e., paths connecting skeleton vertices with degree not equal to two (color coded sections). In the generalized and restricted meta-graph the vertices represent these non-branching sections.
	(f)~Average degree of the restricted meta-graph $\avg {k_\T{meta}}$ as the thickness of network is increased by $\Delta \lambda$, where $\Delta \lambda$ is measured in units equal to the average radius of the original physical network.
	Dots represent real physical networks and the dashed line represents a linear physical network with $N=300$ nodes and $\lambda=N^{-1/2}$. To show that the complex shape of the neurons is responsible for the shape of $\avg {k_\T{meta}}(\Delta \lambda)$ we divide the neurons into smaller non-branching sections and we calculate $\avg {k_\T{meta}}$ treating these sections as the vertices of the restricted meta-graph.}
	\label{fig:realmeta}
\end{figure}

We calculate the restricted meta-graph for the four real physical networks as a function of $\Delta \lambda$, where we measure $\Delta \lambda$ in units of the average radius of the original network.
As a reference, we also generate a jammed random linear physical network with $N=300$ nodes and $\lambda=N^{-1/2}$ and calculate its restricted meta-graph by thickening the links present in the jammed state.
Figure~\ref{fig:realmeta}f shows the average meta-degree $\langle k_\T{meta}\rangle$ as a function of $\Delta \lambda$ for each network, revealing two distinct patterns: 
for the brain network we observe an initial fast increase in the average meta-degree followed by a slower, steady growth.
Such rapid growth is absent in the vascular, mitochondrial and root system networks, and is also absent in random linear networks.

This different behavior represents the differences in the building blocks: the connectome consists of highly intertwined neurons with complex shapes, while the other three networks consist of tube-like components, such as vessels, molecular chains and roots.
Indeed, if we subdivide each neuron into smaller non-branching segments before constructing the meta-graph, we recover the superlinear behavior without the initial rapid growth of of $\langle k_\T{meta}\rangle$ (Fig.~\ref{fig:realmeta}e).

\clearpage

\subsection{The generalized and the restricted meta-graph of the fruit fly brain network}

\begin{figure}[h]
	\centering
	\includegraphics[width=.5\textwidth]{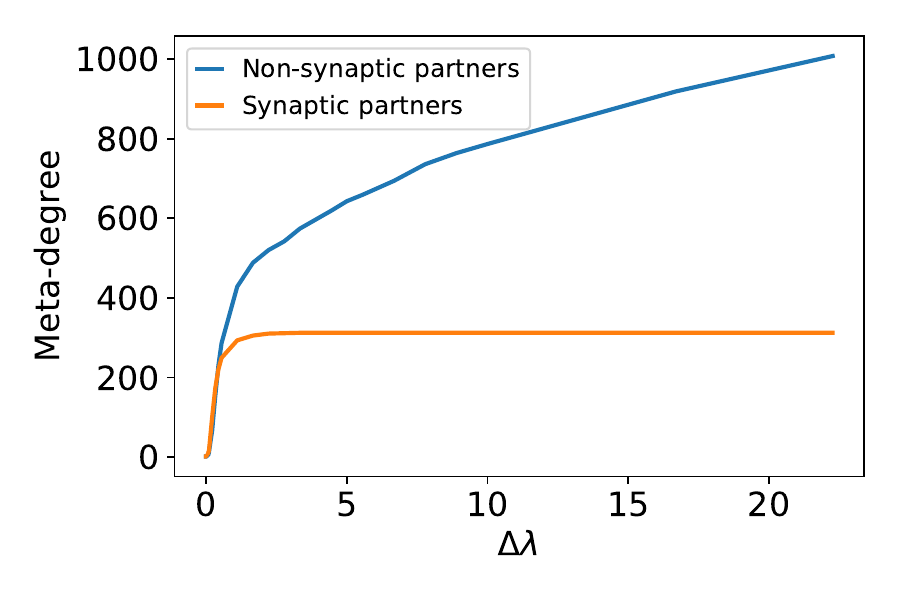}
	\caption{{ \bf Conflicts with synaptic partners.}
	To form synapses neurons must be in close vicinity of each other; therefore, after thickening the neurons, synaptic partners will be in physical conflict.
	We show the meta-degree (i.e., the number of conflicts) of the neuron highlighted in Fig.~4 of the main text with non-synaptic (blue) and synaptic (orange) partners as a function of $\Delta \lambda$.
	Initially, for $\Delta \lambda=0$ the neuron has no conflicts. After increasing the thickness of the neural branches all 312 synaptic partners become quickly in conflict with the neuron.
	The number of conflicts with non-synaptic partners outnumbers the conflicts with synaptic partners, for example, at $\Delta \lambda\approx 3.34$ the neuron has 312 conflicts with synaptic partners and 574 conflicts with non-synaptic partners.
	}
	\label{fig:synaptic_conflict}
\end{figure}

Neurons can only form synapses if their axons and dendrites are in close physical proximity; therefore, if we increase the diameter of neuron branches, neurons will quickly overlap with synaptic partners (Fig.~\ref{fig:synaptic_conflict}).

\begin{figure}[h]
	\centering
	\includegraphics[width=1.\textwidth]{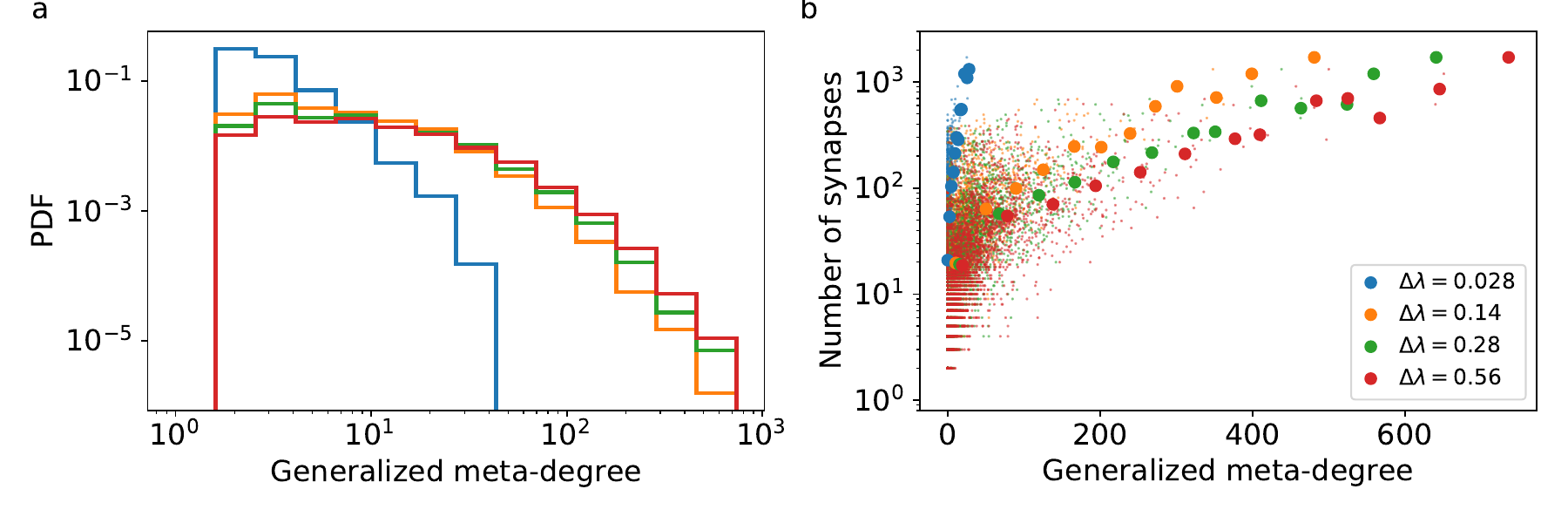}
	\caption{{ \bf Generalized meta-graph of the fruit fly brain network.}
	(a)~The generalized meta-degree distribution for varying $\Delta \lambda$ values. Physically confined neurons have high meta-degree.
	(b)~Two neurons that are connected by a synapse are necessarily adjacent in physical space, hence are connected in the generalized meta-graph. Hence we expect to find a positive correlation between the generalized meta-degree and number of synapses.}
	\label{fig:generalized-metacorr}
\end{figure}

In the generalized meta-graph $\s M_\T g(\lambda)$ of the fruit fly brain network, vertices represent neurons and edges indicate conflicts between pairs of neurons -- both between pairs that are connected via synapses and pairs that are not.
Since synaptic partners necessarily overlap, we expect a positive correlation between the number of synapses a neuron has and the generalized meta-degree of the neuron, this expectation is indeed confirmed by numerical measurements (Fig.~\ref{fig:generalized-metacorr}).

\begin{figure}[h]
	\centering
	\includegraphics[width=1.\textwidth]{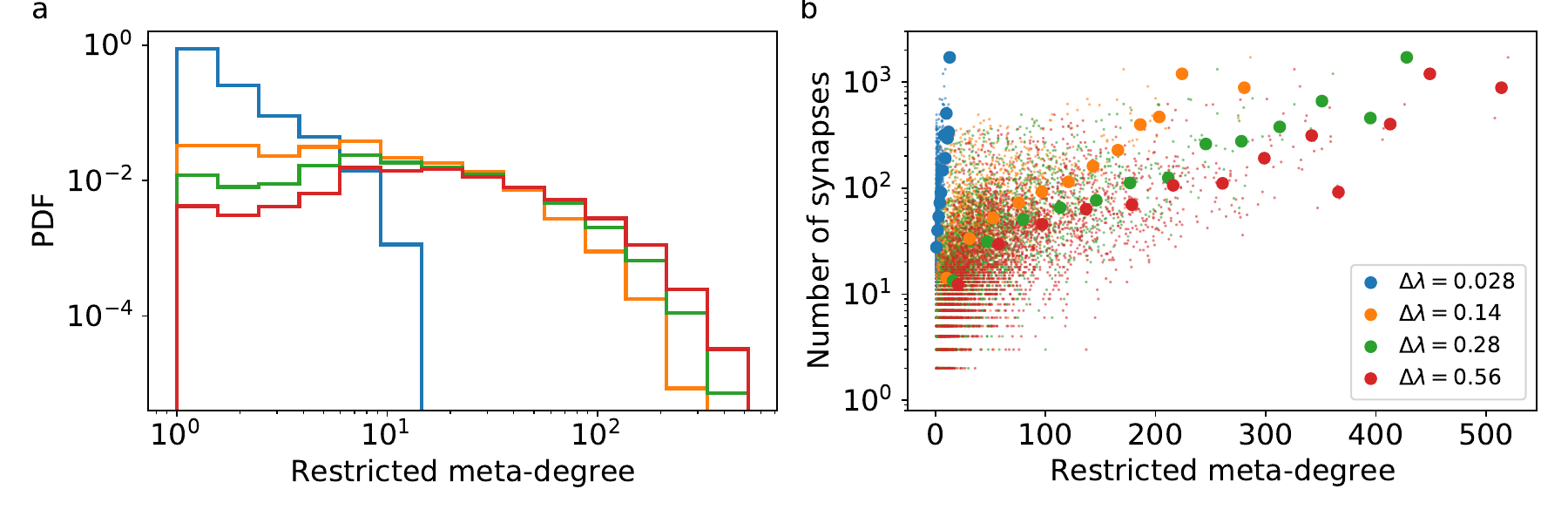}
	\caption{{ \bf Restricted meta-graph of the fruit fly brain network.}
	(a)~The restricted meta-degree excludes edges between neurons that are synaptic partners, capturing the excess confinement of a neuron that is not necessitated by the synaptic network. For illustration purposes, in the main text we focused on value of $\Delta \lambda$ (blue) that produced a sparse restricted meta-graph. Increasing $\Delta \lambda$ may significantly increase the average meta-degree.
	(b)~We again observe a positive correlation between the restricted meta-degree and number of synapses for all tested $\Delta \lambda$ values.}
	\label{fig:resrticted-metacorr}
\end{figure}

\begin{figure}[h]
	\centering
	\includegraphics[width=1.\textwidth]{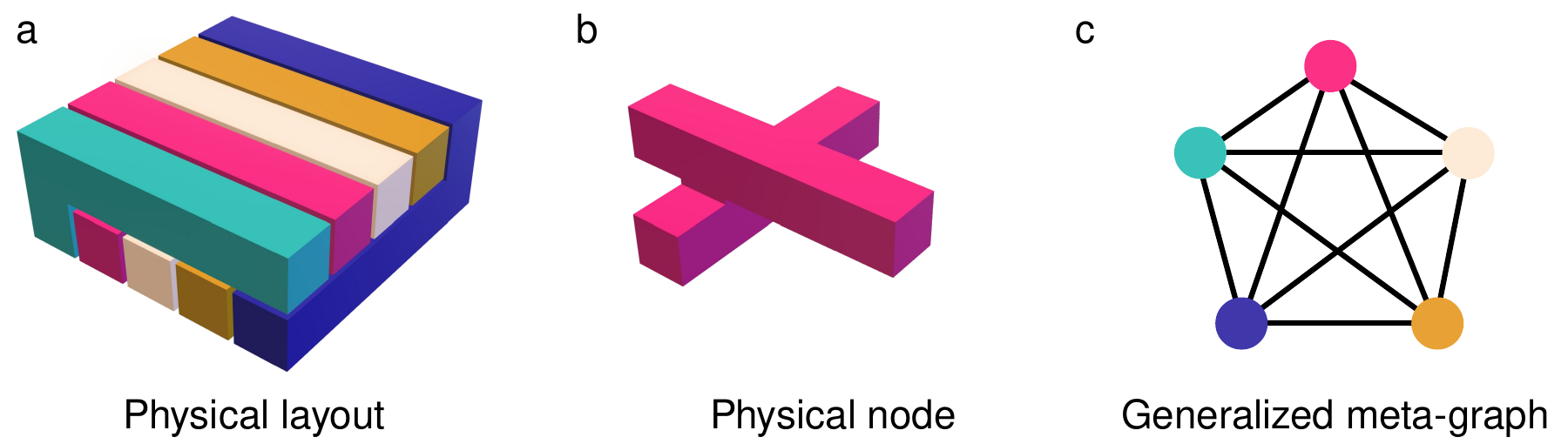}
	\caption{
{ \bf Physical layout with negative correlation between restricted meta-degree and abstract network degree.
(a)~A physical network with $N=5$ nodes, each color corresponds to a separate node. We construct the network by placing five parallel logs in one layer, then we place another five parallel logs on top of them rotated by 90 degrees.
(b)~Logs of the same color are bound together, meaning that each physical node is a cross.
(c)~In this construction, each physical node touches every other node; therefore, the corresponding generalized meta-graph is fully connected, each node has generalized meta-degree 4.
Since all physical nodes are adjacent in space, this physical layout can realize any abstract network. 
For example, if nodes represent neurons, the physical layout can support any synaptic network.
If neuron $i$ in the synaptic network is connected to $k_i$ other neurons, its restricted meta-degree is $N-k_i$, meaning that there is a perfect negative correlation between the degree in the synaptic network and the restricted meta-degree.}}
	\label{fig:full_meta}
\end{figure}

On the other hand, edges of the restricted meta-graph $\s M_\T r(\lambda)$ represent physical conflicts only between neuron pairs that are not connected by synapses.
In other words, $\s M_\T r(\lambda)$ focuses on conflicts that are not necessitated by the synaptic network.
Figure~\ref{fig:resrticted-metacorr} shows that we again find a positive correlation between the number of synapses and the restricted meta-degree of a neuron, indicating that neurons central in the synaptic network are also tightly confined by other neurons.
This is non-obvious, as we can construct physical networks that have a negative correlation between the number of synapses and the restricted meta-degree (Fig.~\ref{fig:full_meta}).

\clearpage

\subsection{LPNs as null models for real physical networks}

Random LPNs may serve as null models for analyzing real physical networks, similarly to the role of the Erd\H{o}s-R\'enyi model and the configuration model in analyzing abstract networks.
To demonstrate this application, we constructed a null model for the vascluar network data set by the following steps: 
\begin{itemize}
\item We identified the location of branching points in the vascular network. 
\item We constructed an abstract network where nodes represent branching points and the branching points are connected if they are connected by a vessel in the physical network.
\item To construct the null model, we generated a random LPN with the same number of nodes and links, by keeping the coordinates of the branching points and the diameter lambda to match the average diameter in the original network.   
\end{itemize}

\begin{figure}[h]
	\centering
	\includegraphics[width=1.\textwidth]{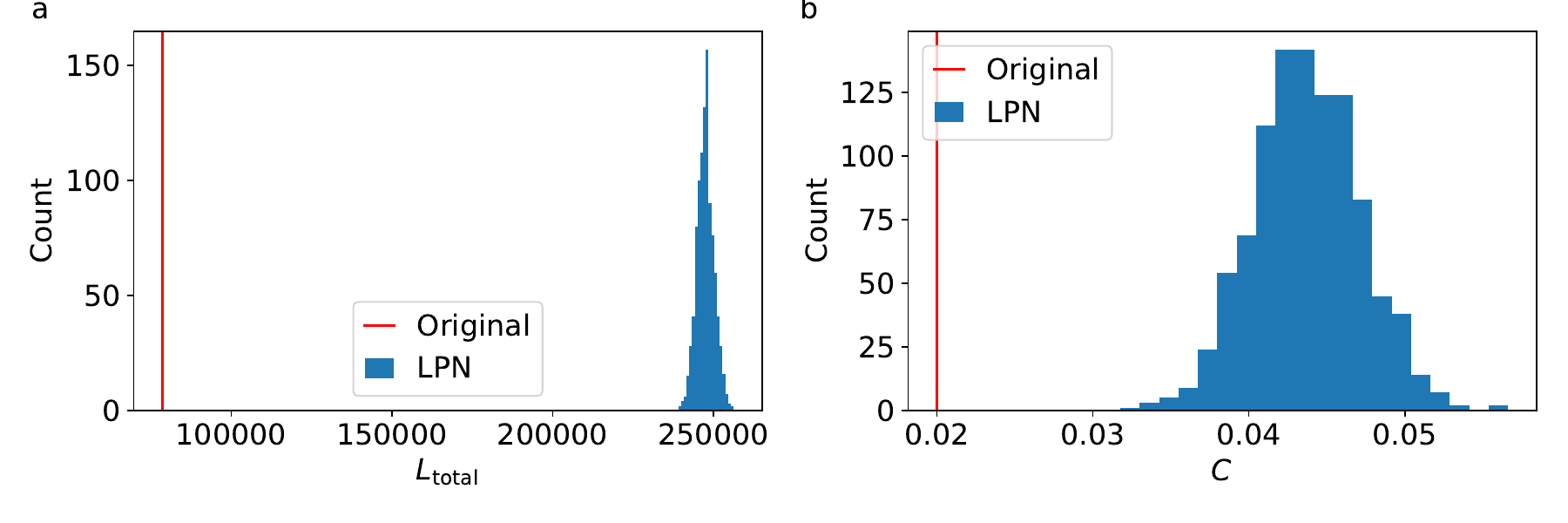}
	\caption{{ \bf LPNs as null models.} We compare the original vascular network to a random LPN generated with the same number of nodes and links, such that we keep the coordinates of the nodes fixed and we set the diameter of the links to be equal to the average diameter of the original network. 
	(a)~The total link length $L_\T{total}$ of the original network is significantly smaller than for random LPNs. For the original network, we calculated $L_\T{total}$ as the sum of the pairwise distances between connected node pairs. 
	(b)~The global clustering coefficient $C$ of the original network is also significantly lower than for random LPNs.
	The histograms show 1000 independent random LPNs.}
	\label{fig:nullmodel}
\end{figure}

Figure~\ref{fig:nullmodel} compares a physical property (total link length $L_\T{total}$) and an abstract network property (clustering coefficient $C$) of the original network to the null model.
We found that $L_\T{total}$ is much longer in the corresponding LPN than in the vascular network, hence volume exclusion is not sufficient to create the locality observed in the real data.
This could be explained by the fact that in a tissue vessels need to fight for empty space not only with the other blood vessels, but also with other structural elements.
Surprisingly, we also found that triangles are exceedingly rare in the real vascular network, even compared to random LPNs.

\clearpage
